\documentclass[12pt]{article}

\usepackage{amsfonts,dsfont}
\usepackage{xcolor}
\usepackage{hyperref}
\hypersetup{
	colorlinks = true,
	citecolor = cyan,
	linkcolor = magenta,
	linktocpage
}
\usepackage{color}
\usepackage{cite}
\usepackage{amsmath, amssymb}
\usepackage{amsthm}
\usepackage{bm}
\usepackage{comment}
\usepackage{fancyhdr}
\usepackage{geometry}
\usepackage{mathtools}
\usepackage{tikz}
\allowdisplaybreaks

\input otex.sty
\counterwithin*{equation}{section}
\numberwithin{Theorem}{section}

\newcommand{\bH}{\textbf{H}}

\newcommand{\D}{\mathcal{D}}
\newcommand{\G}{\mathcal{G}}
\newcommand{\cH}{\mathcal{H}}
\newcommand{\cI}{\mathcal{I}}

\newcommand{\cQ}{\mathcal{Q}}

\newcommand{\Dt}{\mathrm{D}}

\newcommand{\bp}{\bm{p}}

\newcommand{\bs}{\bm{\sigma}}
\newcommand{\bt}{\bm{\tau}}

\newcommand{\ii}{^{(\infty)}}
\newcommand{\mK}{\mathcal{K}}
\newcommand{\n}{^{(n)}}
\renewcommand{\k}{^{(k)}}

\newcommand{\ind}{\mathcal{I}}
\newcommand{\indK}{\cI_{\K}}
\newcommand{\indKn}{\indK^{(n)}}
\newcommand{\indKa}{\indK^\alpha }
\newcommand{\lc}{\bm{\ell}}

\newcommand{\pp}[2]{\frac{\partial #1}{\partial #2}}

\newcommand{\vv}{\mathbf{v}}
\newcommand{\vvn}{\widetilde{\vv}}

\newcommand{\vV}{\mathbf{V}}

\newcommand{\mD}{\mathbb{D}}

\newcommand{\mZ}{\mathcal{Z}}
\newcommand{\oz}{\overline{z}}
\newcommand{\ow}{\overline{w}}
\newcommand{\oZ}{\overline{Z}}

\def\i{\,{\rm i}\,}

\newcommand{\tD}{\text{D}}

\newcommand{\E}{\mathcal{E}}
\newcommand{\mrH}{\mathcal{\r{H}}}
\newcommand{\mI}{\mathcal{I}}
\newcommand{\K}{\mathcal{K}}

\newcommand{\mT}{\mathcal{T}}
\newcommand{\mU}{\mathcal{U}}

\newcommand{\mX}{\mathcal{X}}

\def\r#1{\overline{#1}{}}
\newcommand{\rd}{\r{\d}}
\newcommand{\rDiff}{\r{\D}}
\newcommand{\rD}{\r{D}}

\newcommand{\rH}{\r{\mathfrak{U}}}

\newcommand{\rl}{\r{l}}
\newcommand{\rL}{\r{L}}

\newcommand{\rr}{\r{\rk}}

\newcommand{\rt}{\r{\t}}

\newcommand{\rU}{\r{U}}

\newcommand{\rX}{\r{X}}

\newcommand{\rY}{\r{Y}}
\newcommand{\rZ}{\r{Z}}
\newcommand{\reta}{\r{\eta}}
\newcommand{\rphi}{\r{\phi}}
\newcommand{\rvarphi}{\r{\varphi}}
\newcommand{\rpi}{\r{\pi}}
\newcommand{\rpsi}{\r{\psi}}
\newcommand{\rrho}{\r{\rho}}
\newcommand{\rxi}{\r{\xi}}
\newcommand{\rzeta}{\r{\zeta}}
\newcommand{\rns}{\r{n}_\star}

\newcommand{\tU}{\widehat{U}}
\newcommand{\tV}{\widehat{V}}

\newcommand{\ns}{n_\star}

\newcommand{\nEQ}{n_\natural}
\newcommand{\nf}{n_f}

\newcommand{\de}{\mathcal{R}}
\newcommand{\rde}{\r{\G}}
\newcommand{\rG}{\r{\G}}
\newcommand{\rDelta}{\r{\Delta}}
\newcommand{\LR}[1]{\mathcal{L}_{\de}^{(#1)}}
\newcommand{\LRn}{\LR{n}}
\newcommand{\LG}[1]{\mathcal{L}_{\G}^{(#1)}}
\newcommand{\LGn}{\LG{n}}
\newcommand{\LrG}[1]{\mathcal{L}_{\rG}^{(#1)}}
\newcommand{\LrGn}{\LrG{n}}

\newcommand{\nde}{\mathcal{N}}
\newcommand{\LN}[1]{\mathcal{L}_{\nde}^{(#1)}}
\newcommand{\LNn}{\LN{n}}

\newcommand{\sym}{\Sigma}
\newcommand{\symR}[1]{\sym_{\de}^{#1}}
\newcommand{\symRn}{\symR{n}}
\newcommand{\symG}[1]{\sym_{\G}^{#1}}
\newcommand{\symGn}{\symG{n}}
\newcommand{\symrG}[1]{\sym_{\rG}^{#1}}
\newcommand{\symrGn}{\symrG{n}}

\newcommand{\symN}[1]{\sym_{\nde}^{#1}}
\newcommand{\symNn}{\symN{n}}

\newcommand{\vsnf}{\Psi}
\newcommand{\symM}{M}
\newcommand{\symMR}[1]{\symM_{\de}^{#1}}
\newcommand{\symMRn}{\symMR{n}}
\newcommand{\symMRnREF}{\symM_{\de,\text{REF}}^n}

\newcommand{\symMrGn}{\symM_{\rG}^{n}}
\newcommand{\symMN}[1]{\symM_{\nde}^{#1}}
\newcommand{\symMNn}{\symMN{n}}

\newcommand{\REF}{_\text{REF}}

\newcommand{\disym}{\Upsilon}

\def\uaJ{u^\alpha _J}

\def\zn{z^{(n)}}

\def\mTn{\mT^{(n)}}
\def\Dn{\D^{(n)}}
\def\En{\E^{(n)}}
\def\Gn{\G^{(n)}}
\def\Hn{\cH^{(n)}}

\def\cQn{\cQ^{(n)}}

\def\Di{\D^{(\infty)}}
\def\Ai{\mathcal{A}^{(\infty)}}
\def\Gi{\G^{(\infty)}}

\def\sa{s }

\def\style{\mathsf}
\def\a{\style{a}}
\def\rk{\style{r}}
\def\d{\style{d}}
\def\indexletter{\style{b}}
\def\index#1#2{{\dsty \indexletter_{#1}^{(#2)}}}
\def\indexzero{\indexletter_0}
\def\sindex#1#2{\mathsf{\scriptstyle b}_{#1}^{(#2)}}
\def\rindex#1#2{{\dsty \r{\indexletter}_{#1}^{(#2)}}}
\def\cartanletter{\style{c}}
\def\cartan#1#2{{\dsty \cartanletter_{#1}^{(#2)}}}
\def\rcartan#1#2{{\dsty \r{\cartanletter}_{#1}^{(#2)}}}
\def\tcartan#1#2{{\dsty \widetilde{\cartanletter}_{#1}^{(#2)}}}
\def\t{\style{t}}
\def\total#1#2{{\dsty \t_{#1}^{(#2)}}}
\def\rtotal#1#2{{\dsty \r{\t}_{#1}^{(#2)}}}

\def\rmap{\mathfrak{r}}
\def\rmapn{\rmap\n}

\def\prin{\text{prin}}
\def\para{\text{par}}

\operator{cls}

\def\cone{\mathcal{C}}
\def\chain{\bm{\mathcal{C}}}

\usepackage{tocloft}
\begin{document}
%%%%%

\thispagestyle{fancy}
\fancyhead{}
\fancyfoot{}
\renewcommand{\headrulewidth}{0pt}
\cfoot{\thepage}
\rfoot{\today}

\begin{center}
\begin{Large}
{\bf Convergence of Normal Form Power Series \sstrut{15}\\for Infinite-Dimensional Lie Pseudo-Group Actions}
\end{Large}
\end{center}
\vskip .4cm
\begin{center}
\textsc{
Peter J.\ Olver\textsuperscript{$\dagger$}
\qquad
Masoud Sabzevari\textsuperscript{$\ddagger$}
\qquad
Francis Valiquette\textsuperscript{$\star$}}
\end{center}
\vspace{-1.25cm}

\Abstract
We prove the convergence of normal form power series for suitably nonsingular analytic submanifolds under a broad class of infinite-dimensional Lie pseudo-group actions.   Our theorem is illustrated by a number of examples, and includes, as a particular case, Chern and Moser’s celebrated convergence theorem for normal forms of real hypersurfaces. The construction of normal forms relies on the equivariant moving frame method, while the convergence proof is based on the realization that the normal form can be recovered as part of the solution to an initial value problem for an involutive system of differential equations, whose analyticity is guaranteed by the Cartan--K\"ahler Theorem.

\vskip .25cm
\noindent {\bf Keywords}:   Involutive system of differential equations, Lie pseudo-group, moving frame, normal form power series, differential invariant, chain.

\vskip 0.25cm\noindent
{\bf 2020 Mathematics subject classification}:  22F05, 53A55, 58K50

\tableofcontents

%%%%%
\Section{Introduction} Introduction.
%%%%%

In general, a \is{normal form}, also known as a \is{canonical form}, is defined as a simple representative element chosen from an equivalence class of objects.   The identification of a normal form serves to simplify the treatment of such objects, and also solves the equivalence problem; namely, two objects are equivalent if and only if they have the same normal form.   A simple algebraic example is the Jordan canonical form, which represents the similarity class of a square matrix, \rf{OS}.  In dynamical systems, \rf{Murdock,Nayfeh}, normal forms are extensively used to study dynamics, including bifurcations, classification of singular points, and determining the behavior of solutions.

In the present paper, we focus on the problem of constructing normal forms of analytic $p$-dimensional submanifolds under the action of a Lie pseudo-group, which includes the case of Lie group actions.  Such problems arise in a wide range of applications, including classical invariant theory, \rf I, ordinary differential equations, \rf{F1995,KLS1985,SY1998}, partial differential equations, \rf{BGH1995,M2006}, differential operators, \rf{KO1989-2}, the calculus of variations, \rf{KO1989,KO1992,KOivb}, control theory, \rf{GS1987}, nonholonomic geometry, \rf{E2002}, image processing, \rf{BrMaMc,COSTH,HOpuzzle}, and many more. Normal forms can be algorithmically constructed using the method of equivariant moving frames, \rf{FO1999,M2012,O2018,OP08}, which produces formal power series whose non-constant Taylor coefficients provide a complete set of independent differential invariants of the pseudo-group action on submanifolds. Our main result is a theorem guaranteeing the convergence of such formal power series under rather general conditions on the Lie pseudo-group action in the infinite-dimensional case --- since convergence in the case of finite-dimensional Lie group actions is immediate --- and on the cross-section used for the moving frame construction.

Our results were inspired by the seminal paper of Chern and Moser, \cite{CM74}, that constructed normal form power series for nonsingular analytic real hypersurfaces in complex manifolds, and then proved their convergence.  This particular equivalence problem was first studied by Poincar\'e in \rf{P1907}, who gave two heuristic counting arguments that indicated that two real hypersurfaces in $\C^2$ are, in general, biholomorphically inequivalent, and raised the question of finding the invariants that distinguish them. This question was then solved in the two-dimensional case by Cartan, \rf{C1932}, and, subsequently, in higher dimensions by Chern and Moser, \rf{CM74}. Their analysis was based on an ingenious combination of Cartan's equivalence method and an innovative convergence proof, based on the method of chains that they introduced, which relies on the observation that the successive transformations mapping a regular hypersurface to its normal form can be characterized as solutions to ordinary differential equations, and are therefore analytic.  On the other hand, Kol\'{a}\v{r}, \rf{Kolar}, produced examples of singular hypersurfaces whose normal form  power series are divergent, thus indicating the subtlety of the convergence issue. The normal form analysis promoted by Chern and Moser has inspired  many developments in CR geometry, \rf{B1980,LS2020,KKZ2017,W1978}, and has also been applied to differential equations, \rf{FHM2023,KZ2018,MN2024,R2006}, control systems, \rf{RT2006}, and the geometry of submanifolds, \rf{BG1984,CG2000,HM2024,HM2024-2,O2018}. In the authors' previous paper \rf{OSV}, Chern and Moser's analysis was extended to construct normal form  power series for singular hypersurfaces by applying the equivariant method of moving frames  for Lie pseudo-groups, \rf{OP08}.  However, the convergence of the resulting power series continued to rely on the Chern--Moser chain-based arguments that only apply to a limited range of problems; see, for instance,  \cite{ES1996,FHM2023, KKZ2017, KZ2015, KZ2018, LS2020,M2020}. The present paper grew out of our inability at the time to provide an independent proof of convergence.

The aim of this paper is to establish a new theorem establishing convergence of normal form power series, that applies in great generality, and includes Chern and Moser's convergence theorem.   Our result is based on characterizing the normal form as part of the solution to an initial value problem for a certain involutive system of partial differential equations, whose solutions are analytic as a consequence of the Cartan--K\"ahler existence theorem. Since the theory of involutive differential equations is at the heart of our  proof, \sects {sec: jet} {sec: involutivity} summarize the general theory, as developed in \rf{FesSei,S}.  \is{Note\/}: We will use a purely partial differential equation version of the Cartan--K\"ahler theorem, which circumvents all the differential form constructions that appear in most other treatments, \eg \rf{BCGGG91,O95}.

There are four key steps in our argument.  The first is to recall in \sect{pg} that every analytic Lie pseudo-group  is characterized by an involutive system of partial differential equations known as its determining equations, \rf{OP05,Stormark}.  This means that every local diffeomorphism belonging to the pseudo-group is a solution to the determining equations and conversely.  A new contribution of the paper is the formulation in \sect{rde} of a system of partial differential equations that are satisfied by the restriction of the pseudo-group transformations to a prescribed $p$-dimensional submanifold, which we call the \is{reduced determining equations}. A pseudo-group is called \emph{reducible} if there is a one-to-one correspondence between the solutions of the pseudo-group determining equations and its reduced determining system; see \df{reducible} for the precise statement.  In \th{freered} we show that every Lie pseudo-group that eventually acts freely on an open subset of the submanifold jets space is reducible; such pseudo-groups are exactly those for which the method of equivariant moving frames can be applied, \rf{OP08,OP12}.  Using the Cartan--Kuranishi Prolongation Theorem, \rf{S}, we show in \th{reducibility thm} that the reduced determining equations of a reducible pseudo-group remain involutive and that furthermore, the Cartan characters of the reduced determining equations are equal to the first $p$ Cartan characters of the pseudo-group determining system.

The next step, carried out in \sect{nfde}, is to rewrite the reduced determining equations in an equivalent form which have the property that part of their solution, to a specified initial value problem, is the normal form of the submanifold upon which we based the reduction.  The resulting equations are therefore called the \is{normal form determining equations}.  Since the rewriting amounts to a simple algebraic change of variables, involutivity of the normal form determining equations is assured.  The final step is to apply the method of equivariant moving frames described in \sect{m} to prescribe formally well-posed initial conditions, cf. \cite[Section 9]{S}, for the normal form determining equations. These initial conditions are given by what we call a \is{well-posed cross-section} that serves to define the moving frame.  The well-posedness condition on the cross-section is a refinement of the notion of algebraic cross-section that was introduced in \rf{OP09}, the key difference being that a well-posed cross-section is determined with respect to a Pommaret basis, while an algebraic cross-section relies on a Gr\"obner basis construction.  To accomplish this, we show in \sect{imf} that once the reduced pseudo-group action becomes free at order $\nf$, the moving frame construction, and thus the prescribed initial conditions, is compatible with the involutivity of the normal form determining equations starting at order $\nf+1$. %\pjoc{I wonder if the next sentence should be deleted?}  This is achieved by introducing the \is{vertical symbol} of the normal form determining equations and the \is{prolonged annihilator symbol} of the pseudo-group action and showing in \th{Psi=Gamma} that the two symbols are equal once the order $\geq \nf+1$.  
In \th{well-posedK}, a simple algebraic test, based on the existence of a Rees decomposition for the cross-section at order $\geq \nf+1$ is established, which guarantees that a cross-section of minimal order will be well-posed.   Given well-posed initial conditions, an application of the Cartan--K\"ahler theorem immediately implies that the solution to the initial value problem for the normal form determining equations is analytic, which, in particular, yields the analyticity and hence convergence of the so-constructed normal form.  This leads to the main result of the paper, which is stated in \th{main}, that a well-posed cross-section to the prolonged pseudo-group action whose cross-section-based normalization constants define analytic functions determines a convergent normal form power series for any reducible analytic submanifold.

To the best of our knowledge, our convergence theorem provides the most general result available in the literature, which can be applied to an extremely broad range of Lie pseudo-group actions. All related works on the subject, \eg \rf{ES1996, FHM2023, KKZ2017, KZ2015, KZ2018}, prove the convergence of normal form power series within a specific context. In CR geometry, one of the most general results recently appeared in the work of Lamel and Stolovitch, \rf{LS2020}, who proved convergence of normal form power series for a class of nondegenerate CR submanifolds subject to certain constraints on the normal form.  
%Their results, however, only apply to a very particular family of pseudo-group actions.  
Moreover, as we argue in \sect{chain}, our main theorem sheds new light on, and generalizations of Chern and Moser's notion of chains that they used to prove the convergence of the normal forms constructed within their paper, \rf{CM74}; see also \cite{ES1996,M2020}.

The equivariant approach to moving frames developed in \rf{FO1999,M2012,OP08}, which underlies the final stage of our construction, is a generalization of the classical method, \rf{C1935,C2017,Gug}, that can be systematically and algorithmically applied to general Lie group actions as well as a wide range of Lie pseudo-groups.  In \rf{O2018}, the equivariant moving frame construction was reinterpreted as the specification of a normal form for a submanifold under a pseudo-group action, and \rf{OP08} ended with two explicit examples for relatively simple infinite-dimensional Lie pseudo-group actions. Being concerned with the algebraic formulation of the method, the resulting power series were only formal, and the question of convergence was not considered.  We remark that the implementation employed here differs from the original version introduced in \rf{FO1999} for general Lie group actions and \rf{OP08,OP09} for infinite-dimensional Lie pseudo-groups, in that it is based on the action of the reduced pseudo-group instead of the original pseudo-group.  That said, both implementations yield exactly the same differential invariants, invariant differential operators, etc.   Finally, we note that Arnaldsson \rf{Arnaldsson1,Arnaldsson2} has recently combined the equivariant moving frames with Cartan’s equivalence method for solving equivalence problems, basing his method on involutive bases for polynomial ideals.

\Rmk{1r} Our results are illustrated by a running example that was considered in the original moving frame papers \rf{OP05,OP08,OP09}.  In \sect{xx} we present a number of further examples illustrating our methods and results, including revisiting the Chern--Moser example of nonsingular real hypersurfaces in $\mathbb{C}^2$.  Applications to additional and more substantial Lie pseudo-group actions will be the subject of subsequent papers.

%%%%%
\Section{sec: jet} Jet Bundles and Differential Equations.
%%%%%

In this section we review the standard geometric language of jet spaces for studying systems of differential equations, and present the basic operations of prolongation and projection. While many of our initial considerations hold in more general contexts, we work in the analytic category throughout since we will ultimately rely on the Cartan--K\"ahler Theorem to prove the convergence of normal form power series.

%%%%%
\Subsection{sec: jet bundles} Jet Bundles.
%%%%%

Let $\mX$ be an analytic $p$-dimensional manifold, and $\pi\colon M \to \mX$ an analytic fiber bundle with $q$-dimensional fibers.  Locally, the total space $M$ is isomorphic to the Cartesian product $\mX \times \mU \subset \Rp \times \Rx q$; since all considerations are local, we do not lose any generality by working in the latter context.  Accordingly, we introduce the local coordinates $z = (x,u) \in M$ with $x=(x^1,\ldots,x^p) \in \mX$ parametrizing the base space, and so will play the role of independent variables, while $u=(u^1,\ldots,u^q) \in \mU$ parametrize the fibers, and will play the role of dependent variables in our system of differential equations.  In the following, we let $m = p+q$ denote the dimension of the total space $M$.

In general, given two analytic manifolds, say $\mX$ and $\mU$, and an integer $0 \leq n < \infty$, we let $\Jn = \Jn(\mX,\mU)$ denote the $n$-th order \emph{jet space}, whose points (jets) represent equivalence classes of locally defined functions $u \colon \mX \to \mU$ up to $n$-th order contact, or, equivalently, possessing the same order $n$ Taylor series at the base point $x$, \cite{O95}.  In particular $\J0 = \J0 M = M$. In the above framework, we can identify such functions with local sections of $M \to \mX$, and $\Jn(\mX,\mU) \subset \Jn M $ is an open subset (coordinate chart) of the jet bundle $\Jn M $ of sections of the fiber bundle.   Even more generally, the graphs of sections form $p$-dimensional submanifolds of $M$ that are transverse to the fibers, and thus $\Jn M \subset \Jn(M,p)$ is an open dense submanifold  of the (extended) submanifold jet bundle, \rf O.  However, since all our considerations are local, we can concentrate on $\Jn = \Jn(\mX,\mU)$ throughout.  For any $0 \leq k < n$, we have the jet projection
\Eq{pink}
$$\pi^n_k\colon \Jn \to \Jk,$$
together with the base projection
\[
\pi^n\colon \Jn \to \mX\qquad\text{given by}\qquad \pi^n = \pi \comp \pi^n_0.
\]

The induced coordinates on the $n$-th order jet space $\Jn \simeq \mX \times \mU\n$ are written as $z\n = (x,u\n)$ where $x\in \mX$ and $u\n \in \mU\n$.  Separating the jet coordinates by order,
\Eq{Un}
$$
\qeq{\mU\n =  \mU^0 \times \mU^1 \times \cdots \times \mU^n,\\0 \leq n < \infty ,}
$$
where
\Eq{Uk}
$$
\mU^k = \set{(\ldotsx u^\alpha_J\ldotsx)}{|J| = k, \ \alpha=1,\ldots,q},\qquad 0\leq k \leq n,
$$
denotes the space coordinatized by all $k$-th order derivatives of the dependent variables, which has dimension
\Eq{tk}
$$
\t_k = \dim\, \mU^k = q \,\binom{p+k-1}{k}.
$$
Throughout the paper we use symmetric multi-index notation for derivatives. Thus, a symmetric multi-index $J = \psubs jk$, with $1 \leq j_\nu \leq p$, corresponds to the \kth order derivative $\partial _J = \partial ^k/\partial x^{j_1} \cdots \partial x^{j_k}$, and the jet coordinate $u^ \alpha _J$ represents the $J$-th derivative of $u^\alpha (x)$ at the base point $x$.  
Thus, to each jet coordinate $u^\alpha_J$, we assign the multi-index $(\alpha;J)$.  We introduce the space of all multi-indices
\Eq{ind}
$$\ind = \set{(\alpha;J)}{1\leq \alpha \leq q \text{ and } |J| \geq 0},$$
along with the subset of indices
\Eq{indn}
$$\ind^{\geq n} = \set{(\alpha;J)}{1\leq \alpha \leq q \text{ and } |J| \geq n}$$
of order $\geq n$.   We also use the concatenation notation $J,i = (\subs jk,i)$ to denote the symmetric multi-index obtained by appending $i$ to $J$.  Inversely, we use $J\setminus k$ to denote the multi-index obtained by removing $k\in J$ from $J$.

As noted above, we can identify finite order jets of sections with Taylor polynomials.  Explicitly, for $0 \leq n < \infty$, we identify a jet $z\n = \xun \in \Jn$ with the $q$-tuple of polynomials of degrees $\leq n$ whose entries are
\Eq{Taylorpol}
$$\req{P^\alpha_n (y) = \Sumu{0 \leq |J| \leq n} \frac{\uaJ}{J\:!} \,(y - x)^J,\\\rg \alpha q.}$$
If $\xun$ is the $n$-jet of a section $u(x)$, so $\uaJ$ represents the $J$-th partial derivative of its component $u^\alpha (x)$ at $x$, then $P^\alpha_n (y)$ is the corresponding Taylor polynomial of degree $n$ based at the point $x$.

There are two inequivalent ways to define the infinite order jet bundle. The usual method is to define  $\Ji$ as the projective (or inverse) limit of the finite order jet bundles $\Jn$ under the projection maps \eq{pink}.  Thus, an infinite jet has local coordinates $x^i,\uaJ$ for all $\rg ip$, $\rg \alpha q$, and all multi-indices $|J| \geq 0$.  We identify such an infinite jet with a $q$-tuple of formal power series:
\Eq{Taylorseries}
$$\req{P^\alpha (y) = \Sumu{|J| \geq 0} \frac{\uaJ}{J\:!} \,(y - x)^J,\\\rg \alpha q.}$$
Since the coefficients $\uaJ$ are arbitrary, there is no guarantee that \eq{Taylorseries} converges.

An alternative approach is, in analogy with the finite order case, to define infinite jets as equivalence classes of sections up to infinite order contact, which is equivalent to the condition that their Taylor series \eq{Taylorseries} agree at the base point.  Since we restrict to analytic sections, the corresponding Taylor series converge and, indeed, uniquely determine the section.  Since the coefficients $\uaJ$ must now define a convergent series, with a non-zero radius of convergence, they are no longer allowed to be arbitrary.  Thus, the result of the latter construction is a subbundle $\rm A^\infty \subset \Ji$ of the preceding infinite jet bundle, which consists of infinite jets that produce convergent Taylor series, as in \eq{Taylorseries}.  We will call $\rm A^\infty$ the \is{analytic infinite jet bundle}.

Traditionally, the equivariant moving frame calculus takes place in the ordinary infinite jet bundle $\Ji$, without regard to convergence.  The goal of this paper is to provide conditions, on both the pseudo-group action and the cross-section defining the normalizations, that guarantee that the normal form determined by the moving frame normalizations belongs to the analytic infinite jet bundle $\rm A^\infty$.

%%%%%
\Subsection{sec: DE} Differential Equations.
%%%%%

A system of $n$-th order \emph{differential equations} is given by a system of equations
\begin{equation}\label{DE}
\Delta(x,u\n) = \bpa{\Delta_1(x,u\n),\ldots,\Delta_l(x,u\n)}=0
\end{equation}
involving the \nth order jet space coordinates.  To avoid singularities, the defining functions $\Delta \colon \Jn \to \mathbb{R}^l$ are assumed to be analytic, the corresponding subvariety
\Eq{den}
$$\de\n = \set{(x,u\n)}{\Delta(x,u\n)=0} \subset \Jn $$
is assumed to form an analytic fibered submanifold of the fiber bundle $\pi^n\colon \Jn \to \mX$, and the Jacobian matrix of the defining functions is of maximal rank on $\de\n$; see \rf O.

\emph{Prolongation} and \emph{projection} are two natural operations on differential equations.  The former lifts the system of differential equations to higher orders by differentiation, while the latter lowers the order by keeping only the equations (if any) of a specified lower order.  The prolongation of \eqref{DE} to order $n+k$ is the fibered submanifold $\de^{(n+k)} \subset \J{n+k}$ locally described by the system of equations
\[
\de^{(n+k)} =  \left\{
\begin{aligned}
\Delta_\nu(x,u\n) &=0,& & 1\leq \nu\leq l\\
\Dt_x^J \Delta_\nu(x,u\n) &=0,& & 1\leq |J| \leq k
\end{aligned}
\right\},
\]
where
\Eq{Dti}
$$\Dt_{x^i} = \pp{}{x^i} + \sum_{\alpha=1}^q \sum_{|J| \geq 0} u^\alpha_{J,i} \pp{}{u^\alpha_J},\qquad i=1,\ldots,p,$$
are the usual total derivative operators, which mutually commute, and
$\Dt_x^J = \Dt_{x^{j_1}}\cdots \Dt_{x^{j_k}}$, for $J=(j_1,\ldots, j_k)$ a symmetric multi-index, are their higher order iterations.  On the other hand, the projection of the \nth order differential equation $\de\n$ to a differential equation of order $n-k$, with $0< k \leq n$, which encodes the relations (if any) among derivatives of order $\leq n-k$, is given by
\[
\pi^n_{n-k}(\de\n)\subseteq \J{n-k}.
\]
To construct a local representation of $\pi^n_{n-k}(\de\n)$ one starts with \eqref{DE} and eliminates, using only algebraic operations, all derivatives of order greater than $n-k$ in as many equations as possible.  If there are no equations of order $\leq n-k$, then, at least locally, $\pi^n_{n-k}(\de\n) = \J{n-k}$.  As in \cite{S}, we assume that the systems of differential equations are \emph{regular} so that, to avoid dealing with singular points and subsets, all projections and prolongations are assumed to be fibered submanifolds.

The $k$-th prolongation and projection of a system of differential equations $\de\n$ is given by
\[
\pi_n^{n+k}(\de^{(n+k)}) \subseteq \de\n.
\]
This process may not return the original system $\de\n$ due to the existence of integrability conditions.  A system of differential equations $\de\n$ is said to be \emph{formally integrable} if for all $k\geq 0$, the equality
\begin{equation}\label{formal integrability}
\pi^{n+k+1}_{n+k}(\de^{(n+k+1)}) = \de^{(n+k)}
\end{equation}
holds.  In other words, a system of differential equations is formally integrable if, no matter the order at which the system is prolonged, no additional integrability conditions arise.

%%%%%
\Section{sec: involutivity} Involutivity.
%%%%%

Formal integrability does not in itself suffice to guarantee the existence of solutions to a system of differential equations, and, for this purpose, we need to introduce the notion of involutivity. To this end, we summarize the theory of involutive systems of partial differential equations, in the form presented by Seiler in his book \cite{S}; see also \rf{FesSei}.  In particular, we will \iz{not} use the exterior differential systems formulation of involutivity, \rf{BCGGG91,O95}.

We begin with the linearization of a system of partial differential equations \eq{den}.
Consider the tangent bundle $T\Jn \to \Jn$ parametrized by $(x,u\n,\xi,\psi\n)$, where $\xi^i,\psi^\alpha _J$ are the fiber coordinates.  Any vector field, \ie section of $T\Jn$,
is locally represented by
\[
\vv = \sum_{i=1}^p \>\xi^i\pp{}{x^i} + \sum_{0\leq |J| \leq n} \sum_{\alpha=1}^q \>\psi^\alpha_J \pp{}{u^\alpha_J},
\]
whose coefficients $\xi^i,\psi^\alpha_J$ depend on\fnote{We will often suppress the dependence on $\zn$ to avoid cluttering formulas.} $\zn = (x,u\n)$.
In view of \eq{Un}, we introduce the vertical (fiber) projection
$\pi_V \colon T\Jn\at\zn \to T\:\mU\n\at\zn$
given by removing the horizontal component:
\[
\pi_V(\vv) = \sum_{0\leq |J| \leq n} \sum_{\alpha=1}^q \>\psi^\alpha_J \pp{}{u^\alpha_J}.
\]

The (\emph{vertical}) \emph{linearization} $\LRn \at\zn\subset T\:\mU\n\at\zn$ of the system of differential equations $\de\n$ given by \eqref{DE}  at a point $\zn \in \de\n$ consists of the system of linear equations
\begin{equation}\label{eq: linearization}
\LRn= \pi_V(\vv)\,\Delta = \bigg\{
\displaystyle  \sum_{0\leq |J| \leq n} \sum_{\alpha=1}^q  \>\pp{\Delta_\nu }{u^\alpha_J}\psi^\alpha_J = 0 ,\quad \nu=1,\ldots,l\
\bigg\}.
\end{equation}
We further introduce the \emph{highest order term map} $\bH\colon T\:\mU\n\at\zn  \to T\:\mU^n\at\zn $ which only retains the terms $\psi^\alpha_J$ of order $|J| = n$ in \eqref{eq: linearization}. The resulting system of linear equations
\[
\symRn = \bH(\LRn)=
\bigg\{
\displaystyle  \sum_{|J| = n}\sum_{\alpha=1}^q\> \pp{\Delta_\nu}{u^\alpha_J}\, \psi^\alpha_J = 0,\quad \nu=1,\ldots,l\ \bigg\}
\]
is called the \emph{symbol} of the differential equation $\de\n$.  Its $\dsty l\times q\,\binom{p+n-1}{p-1}$ coefficient matrix
\[
\symMRn = \begin{pmatrix}
\displaystyle \pp{\Delta_\nu}{u^\alpha_J}
\end{pmatrix}
\]
is called the \emph{\nth order symbol matrix}.  In line with the standard regularity assumption, we suppose in the following that all intrinsic algebraic properties of the symbol, \eg its rank, are independent of the point $\zn \in \de\n$ under consideration.

The columns of the symbol matrix $\symMRn$ correspond to the unknowns $\psi^\alpha_J$ of order $|J| = n$.  In order to formulate the involutivity and solvability of the system of partial differential equations $\de\n$, we need to order the columns in an intelligent manner; our preferred ordering will be prescribed by the notion of the class of a multi-index, which relies on  a choice of ordering of the independent variables.  For general arguments, we use the natural ordering $x^1 \prec x^2 \prec \cdots \prec x^p$ throughout.  With this choice, the definition of class is as follows.

%\pjoc{I changed $k$ to $n$ and $i$ to $k$ to be more in line with later notations.}

\Df{class}
The \emph{class} of a multi-index $J=(j_1,\ldots,j_k)$ is the smallest index that appears in $J$, so
\[
\cls J = \min \{j_1,\ldots,j_k\}.
\]

Note that, in the set $\mU^k$ of jet coordinates $u^\alpha _J$ of order $|J| = k$, there are
\Eq{tki}
$$
\total{k}{i} = q\,\binom{p+k-i-1}{k-1}$$
jet coordinates whose multi-index $J$ is of class $i$.  Thus,
\Eq{tksum}
$$\t_k = \total{k}{1} + \cdotsx + \total{k}{p}$$
is the total number of jet coordinates of order exactly $k$, as in \eq{Uk}.

We sort the columns of the symbol matrix $\symMRn$ using a class-respecting term ordering so that if $\cls J > \cls K$, then the column corresponding to the unknown $\psi^\alpha_J$ must be to the \is{left} of the column corresponding to the unknown $\psi^\beta_K$.  Within a fixed class, one is free to choose any convenient ordering of the columns.  For example, if $p=3$ and we order $x \prec y \prec z$, then one possible ordering of the order $n=2$ columns of a symbol matrix is $\psi _{zz}, \psi _{yz}, \psi _{yy}, \psi _{xz}, \psi _{xy}, \psi _{xx}$, so the first column has class 3, the next two, which can be switched, have class 2, and the final three, again in any order, are of class 1.

With this column ordering, let $\symMRnREF$ be the row reduction of $\symMRn$ to its row-echelon form, \crf{OS}. Abn unknown $\psi^\alpha_J$ that corresponds to the first non-vanishing entry of a row in $\symMRnREF$, \ie the row's pivot entry, is called the \emph{leader} of the row.  We will use $\rk_n$ to denote the rank of the symbol matrix $\symMRn$, \ie the number of leaders/pivots.

The jet coordinates $u^ \alpha _J$ of order $|J|=n$ that correspond to the leader columns of the symbol matrix $\symMRn$ are known as \is{principal derivatives}. It follows that the number of principal derivatives of order $n$ is
$$\rk_n = \rank \symMRn,$$
 which also equals the number of independent differential equations of order $n$ in the system.
The other jet coordinates of order $n$ corresponding to the non-pivot columns are known as \is{parametric derivatives}.  The number of parametric derivatives of order $n$ is given by
\Eq{dn}
$$\d_n = \t_n - \rk_n.$$
We let
\Eq{rn}
$$\rk\n = \sum_{k=0}^n \> \rk_k$$
denote the total number of principal derivatives of order $\leq n$, and
\Eq{dntot}
$$\d\n = q\,\binom{p+n}{n} - \rk\n = \sum_{k=0}^n \> \d_k$$
 the total number of parametric derivatives of order $\leq n$.  By the Implicit Function Theorem and our regularity assumptions, $\d\n$ equals the fiber dimension of the \nth order system \eq{den}.

An \nth order system of partial differential equations is said to be in \is{Cartan normal form}  if all its symbol matrices of order $0 \leq k \leq n$ are either empty or in reduced row-echelon form with respect to the above class-respecting ordering of the columns.  We further say that it is in \is{reduced Cartan normal form} if, in addition, the entire symbol matrix
\[
\symMR{(n)} = \begin{pmatrix}\symMR{0} & \symMR{1} & \cdots & \symMR{n}\end{pmatrix}^T
\]
is in reduced row-echelon form, \rf{FesSei}, meaning that the entries in the column above the pivots are all $0$.  Thus, the differential equations are in reduced Cartan normal form when they take the form
\Eq{Cartannf}
$$u^\alpha _J = \Delta ^\alpha _J(x,\ldots,u^\beta _K,\ldots),$$
where $u^\alpha _J$ are the principal derivatives, and all the jet coordinates $u^\beta _K$ appearing on the right hand side are parametric and are indexed by the columns that have \is{nonzero} entries in the corresponding row of the reduced row echelon form of the entire symbol matrix.  At order $|K|=|J|$, these are all parametric derivatives that appear after $u^\alpha _J$ in the class-respecting term ordering, that is $\cls K \leq \cls J$.  Thanks to the Implicit Function Theorem, any regular system of differential equations of order $n$ can be placed in reduced Cartan normal form.

\Df{indices}
The number of leaders of class $1 \leq k \leq p$ in the row-echelon symbol matrix $\symMRnREF$ is denoted by $\index{n}{k}$.  The resulting nonnegative integers $\index{n}{1},\ldots,\index{n}{p}$ are called the \emph{indices} of the \nth order symbol $\symRn$.

We are now able to state the key definition of an involutive symbol.

\Df{involutivesymbol}
The symbol $\symRn$ with indices $\index{n}{k}$ is said to be \is{involutive} if the symbol matrix $\symMR{n+1}$ of the prolonged symbol $\symR{n+1}$ satisfies\fnote{The first equation is automatically satisfied as a consequence of the definition of the indices $\index{n}{k}$, and is included for later referencing.}
\Eq{involutivesymbol}
$$\req{\sum_{k=1}^p\, \index{n}{k} = \rk_n,\\\sum_{k=1}^p\, k\:\index{n}{k} = \rk_{n+1} = \rank \symMR{n+1}.}$$

\Rmk{2r}
We observe that the class of a derivative is not necessarily preserved under coordinate transformations. The notion of a \is{$\delta$-regular coordinate chart} is characterized by the fact that the sum on the right hand side of \eq{involutivesymbol} takes its maximal value under all possible (linear) changes of coordinates.  In particular, a necessary condition for $\delta$-regularity is that the highest index $\index{n}{p}$ takes its maximal value.  For a first order system of differential equations, this means that a maximal number of equations must be solvable for an $x^p$-derivative, and hence the surface $x^p=0$ cannot be characteristic. 
\pari
Clearly, the involutivity condition \eq{involutivesymbol} requires that we work in a $\delta$-regular coordinate system.  Indeed, we will assume throughout that we are always working in $\delta$-regular coordinates, noting that generic coordinate systems are $\delta$-regular, \rf{HSS,S}. However, in \exs{12}{CM-exm}, the most natural coordinate system is not $\delta$-regular, and so the involutivity criterion \eq{involutivesymbol} is not satisfied unless we impose a suitable change of variables before conducting the analysis.

\Df{involutivedef}
A system of differential equations $\de\n$ is \emph{involutive} if it is formally integrable and its symbol $\symRn$ is involutive.

Recall that formal integrability requires verifying \eqref{formal integrability} for \iz{all} $k\geq 0$.  But  when the symbol is involutive, it suffices to check integrability only for $k=0$.  A proof of this result, as stated below, can be found in \rf{S}.

\Th{involutive}
A system of differential equations $\de\n$ is involutive if and only if its symbol $\symRn$ is involutive and $\pi^{n+1}_n(\de^{(n+1)}) = \de\n$.

Thus, to check involutivity at order $n$, one needs to make sure that the coordinate chart is $\delta$-regular, then verify the algebraic involutivity condition \eq{involutivesymbol} for the indices of order $n$, and finally check that there are no integrability conditions at order $n+1$.

The indices $\index{n}{k}$ determine the number of principal derivatives of order $n$ and of class $k$ in the system of differential equations $\de\n$.  On the other hand, the number of parametric derivatives of order $n$ and class $k$ is given by the \emph{Cartan character} 
%$\cartan{n}{k}$, which is related to the corresponding index via the equation
\Eq{inCc}
$$\cartan{n}{k} = \total{n}{k} - \index{n}{k},\qquad 1\leq k\leq p.$$
The involutivity condition \eq{involutivesymbol} can be restated in terms of the Cartan characters as follows:
\Eq{involutivecc}
$$\req{\sum_{k=1}^p\> \cartan{n}{k} = \d_n, \\ \sum_{k=1}^p\> k\:\cartan{n}{k} = \d_{n+1}.}$$
We also note that, according to \rf{S; Proposition 8.2.2}, involutivity implies that the Cartan characters are non-increasing:
\Eq{Cartancineq}
$$\req{\cartan n1 \geq \cartan n2 \geq \cdotsx \geq \cartan np \geq 0 .}$$

\Rmk{4r} Owing to their direct relationship \eq{inCc}, when formulating results or illustrative examples, one can work either just with the indices or just with the Cartan characters, depending upon one's preference.  We have chosen to display both in order to suit readers of either persuasion.

\Rmk{3r} If $\cartan{n}{k} > \cartan{n}{k+1} = 0$
 is the last nonzero Cartan character of an involutive system of differential equations, then the general solution to the system depends on $\cartan{n}{k}$ arbitrary functions of $k$ variables, which can be identified with the initial conditions of order $k$.  On the other hand, the number of arbitrary functions of less than $k$ variables required to express a general solution is not well-defined; see also \rf{BCGGG91,CartanEinstein,O95,S}.

Any \nth order system of differential equations \eqref{DE} can be written as a first-order system of differential equations by setting the jet coordinates $u^\alpha_J$ of order $|J| \leq n-1$ to be new dependent variables.  To write down this new system, we introduce the differentiation notation
\[
\partial_i u^\alpha_J = \pp{u^\alpha_J}{x^i}.
\]
%to denote differentiation. 
Then a first order representation $\widetilde{\de}^{(1)}$ of the \nth order system $\de\n$ is given by
%\Eq{R1}$$
\begin{equation}\label{R1}
\widetilde{\de}^{(1)}=
\left\{\begin{aligned}
&\widetilde{\Delta}_\nu(x,(u^{(n-1)})^{(1)})=0,\qquad& & 1\leq \nu \leq l\\
&\partial_i u^\alpha_J = u^\alpha_{J,i},& &|J| < n-1,\quad  1 \leq i \leq p\\
\ &\partial_iu^\alpha_J = \partial_k u^\alpha_{J,i\setminus k},& &|J| = n-1,\quad  k=\cls J < i \leq p\
\end{aligned}\right\}.%$$
\end{equation}
The function $\widetilde{\Delta}_\nu$ is not uniquely defined, as there are in general several possibilities to express a higher-order derivative $u^\alpha_J$ in terms of the new coordinates.  To easily compute the indices of the symbol $\sym_{\widetilde{\de}}^1$ of the first order system \eqref{R1}, we use the mapping
\begin{equation}\label{ujet sub}
 u^\alpha_J = \begin{cases}
\ u^\alpha_J,& |J| \leq n-1,\\
\ \partial_ku^\alpha_{J\setminus k}, \quad & |J| = n,\quad \cls J = k.
\end{cases}
\end{equation}

\Pr{order 1 reduction}
Let $\tcartan{1}{1},\ldots,\tcartan{1}{p}$ be the Cartan characters of the first order representation $\widetilde{\de}^{(1)}$ and $\cartan{n}{1},\ldots,\cartan{n}{p}$ those of the original system of differential equations $\de\n$.  Then
\[
\tcartan{1}{k} = \cartan{n}{k},\qquad 1\leq  k \leq p.
\]
Moreover, the \nth order system $\de\n$ is involutive if and only if its first order representation $\widetilde{\de}^{(1)}$ is involutive.

The proof of Proposition \ref{order 1 reduction} may be found in \cite[Appendix A.3]{S}. For a first-order system of involutive differential equations $\de^{(1)}$, the reduced Cartan normal form is
\begin{equation}\label{eq: Cartan normal form}
\begin{aligned}
u^\alpha_p &= \Delta^\alpha_p(x^1,\ldots,x^p,\ldots,u^\beta_k,\ldots),&\hspace{1cm} & 1 \leq \alpha \leq \index{1}{p},\\
u^\alpha_{p-1} &= \Delta^\alpha_{p-1}(x^1,\ldots,x^p,\ldots,u^\beta_k,\ldots),& & 1 \leq \alpha \leq \index{1}{p-1}, \\
&\hspace{0.25cm}\vdots & & \\
u^\alpha_1 &= \Delta^\alpha_1(x^1,\ldots,x^p,\ldots,u^\beta_k,\ldots),& & 1 \leq \alpha \leq \index{1}{1},\\
u^\alpha &= \Delta^\alpha(x^1,\ldots,x^p,u^\delta),& & 1\leq \alpha \leq \indexzero,
\end{aligned}
\end{equation}
with, by virtue of involutivity \rf{S;Corollary 7.1.28}, the indices satisfy
$$0\leq \indexzero \leq \index{1}{1} \leq \cdots \leq \index{1}{p-1} \leq \index{1}{p} \leq q.$$
Moreover, all the derivatives appearing on the right hand side of each equation are parametric of class smaller than or equal to the class of the principal derivative occurring on the left hand side of the equation. If $\indexzero = 0$, the system does not contain algebraic equations relating the dependent variables. On the other hand, if $\indexzero > 0$, since the equations are in reduced Cartan normal form, no derivatives of order $0$ or $1$ of the principal zero-th order derivatives $u^\alpha$ can appear on the right hand side of any of the equations.   

Formally well-posed initial value conditions for the first-order system in reduced Cartan normal form \eqref{eq: Cartan normal form}
are prescribed by
\begin{equation}\label{eq: initial conditions}
\eeq{
u^\beta(0,\ldots,0) = f^\beta,&  \indexzero < \beta \leq \index{1}{1},\\
u^\beta(x^1,0,\ldots,0) = f^\beta(x^1),&   \index{1}{1} < \beta \leq \index{1}{2},\cr
&\ \vdots & \\
u^\beta(x^1,\ldots,x^{p-1},0) = f^\beta(x^1,\ldots,x^{p-1}),& \index{1}{p-1} < \beta \leq \index{1}{p},
\\
u^\beta(x^1,\ldots,x^p) = f^\beta(x^1,\ldots,x^p),&  \index{1}{p} < \beta \leq q.}
\end{equation}
%
%\begin{equation}\label{eq: initial conditions}
%\begin{aligned}
%u^\beta(x^1,\ldots,x^p) &= f^\beta(x^1,\ldots,x^p),&  \index{1}{p} &< \beta \leq q,\\
%u^\beta(x^1,\ldots,x^{p-1},0) &= f^\beta(x^1,\ldots,x^{p-1}),&\qquad  \index{1}{p-1} &< \beta \leq \index{1}{p},\\
%&\hspace{0.2cm}\vdots & & \\
%u^\beta(x^1,0,\ldots,0) &= f^\beta(x^1),&   \index{1}{1} &< \beta \leq \index{1}{2},\\
%u^\beta(0,\ldots,0) &= f^\beta,&  \indexzero &< \beta \leq \index{1}{1}.
%\end{aligned}
%\end{equation}

\Rmk{5r}
In \eqref{eq: initial conditions}, we use the convention that if, for example, $\indexzero =\index{1}{1}$,
%$\index{1}{p} = q$, 
then the first set of equations in the initial conditions \eqref{eq: initial conditions} are vacuous, and similarly for the other sets.

As they should, the initial conditions \eqref{eq: initial conditions} specify the parametric derivatives occurring on the right hand side of the system of differential equations \eqref{eq: Cartan normal form}.  For example, the parametric derivatives of class 1 are determined by differentiating the equations $u^\beta(x^1,0,\ldots,0) = f^\beta(x^1)$ for $\index{1}{1} < \beta \leq \index{1}{2}$ on the line $\{(x^1,0,0,\ldots,0)\}$.  The parametric derivatives of class 2 are obtained from the initial conditions on the plane $\{(x^1,x^2,0,\ldots,0)\}$, and so on.

Recalling our notation \eq{indn}, the reduced Cartan normal form equations \eqref{eq: Cartan normal form} and their infinite prolongation split the set of multi-indices
\Eq{indg1}
$$\ind^{\geq 1} = \ind^{\geq 1}_{\de,\prin}\, \biguplus \,\ind^{\geq 1}_{\de,\para}$$
into the disjoint subsets containing, respectively, the principal and the parametric multi-indices:
\Eq{indg1pp}
$$
\eeq{\ind^{\geq 1}_{\de,\prin} = \set{(\alpha;J)}{u^\alpha_J \text{ is a principal derivative}}
\\\ind^{\geq 1}_{\de,\para} = \{\,(\beta;K)\;|\;u^\beta_K \text{ is a parametric derivative}\,\}.}$$

The \is{Pommaret division} assigns to the multi-index $(\alpha;J)$ of class $\cls J = k$ the \is{multiplicative indices} $\{1,\ldots,k\}$, which serve to define the \is{involutive} (\is{Pommaret\/}) cone
\begin{equation}\label{involutive cone}
\cone^\alpha(J) = \set{ (\alpha; J,k^1,\ldots,k^n)}{1\leq k^j \leq \cls(J)\; \text{and}\; n\geq 0}.
\end{equation}
The set of principal indices $\ind^{\geq 1}_{\de,\prin}$ forms an ideal in $\ind$.  Involutivity of the Cartan normal form equations \eqref{eq: Cartan normal form} implies that this ideal can be decomposed into a union of non-intersecting involutive cones
\begin{equation}\label{principal jet decomposition}
\ind^{\geq 1}_{\de,\prin} = \biguplus_{i=1}^p\; \biguplus_{\alpha=1}^{\sindex{1}{i}} \; \cone^\alpha(i).
\end{equation}
The decomposition \eqref{principal jet decomposition} reflects the fact that the prolongation of the system of equations \eqref{eq: Cartan normal form} can be obtained by differentiating each equation solely with respect to the multiplicative indices of the principal derivative.  The indices $(\alpha;i)$ associated to the principal derivatives in \eqref{eq: Cartan normal form} are said to form a \is{Pommaret basis} of $\ind^{\geq 1}_{\de,\prin}$.

On the other hand, the set of parametric indices $\ind^{\geq 1}_{\de,\para}$ does not form an ideal.  That said it still admits a disjoint decomposition into involutive cones
\Eq{cones}
$$\ind^{\geq 1}_{\de,\para} = \biguplus_{\substack{\\[0.07cm] i=1}}^p\;\biguplus_{\beta=\sindex{1}{i}+1}^q\; \cone^{\alpha}(i)$$
known as its \is{Rees decomposition}.  Paraphrasing \rf{S;Proposition 5.1.6}, we have the following result, which will come into play in \sect{imf}.

\Pr{Rees}
Let $\de\n$ be a formally integrable \nth order system of differential equations.  The prolongation of $\de\n$ combined with our chosen multi-index ordering induces the splitting of multi-indices \eq{indg1}.   The ideal $\ind^{\geq n}_\prin$ has a Pommaret basis, and therefore $\de\n$ is involutive, if and only if $\ind^{\geq n}_{\de,\para}$ admits a Rees decomposition.

We end the section with the Cartan--K\"ahler existence theorem, stated, for simplicity, for first order involutive systems of differential equations in reduced Cartan normal form \eqref{eq: Cartan normal form}.  This fundamental theorem is a generalization and consequence of the basic Cauchy--Kovalevskaya  existence theorem for analytic systems of partial differential equations, \rf{O95,S}.  In essence, the Cartan--K\"ahler theorem is established by successive application of the Cauchy--Kovalevskaya theorem to the initial value problems corresponding to each line in the initial conditions \eqref{eq: initial conditions}.

\Th{thm: CK}
Let the functions $\Delta^\alpha_k$ and $f^\beta$ in \eqref{eq: Cartan normal form} and \eqref{eq: initial conditions} be real-analytic at the origin.  If the system \eqref{eq: Cartan normal form} is involutive, then it possesses one and only one solution that is analytic at the origin and satisfies the initial conditions \eqref{eq: initial conditions}.

%%%%%
\Section{pg} Lie Pseudo-Groups.
%%%%%

In this section we apply the preceding constructions to the differential equations defining Lie pseudo-group actions, referring to \rf{IOV,OP05} for details.  Let $\D = \D(M)$ denote the Lie pseudo-group of all local analytic diffeomorphisms\fnote{In general, the notation allows $\varphi$ to only be defined on an open subset of $M$.} $\varphi \colon M \to M$. We will employ Cartan's convenient notational convention and use lower case letters to denote source coordinates and the corresponding capital letters to denote target coordinates.  Thus, given a local diffeomorphism $\varphi \in \D$, its local coordinate formula will be written $Z = \varphi (z)$, so that the target coordinates $Z = \psups Zm$ are functions of the source coordinates $z = \psups zm$.

Given $0 \leq n <\infty $, let $\Dn \subset \Jn (M,M)$ be the subbundle consisting of all \nth order jets of local diffeomorphisms of $M$.  We remark that $\Dn$ forms a groupoid, \crf{Mackenzie}, under composition. We also let $\Di \subset \Ji (M,M)$ denote the corresponding space of infinite order jets of diffeomorphisms, and  $\Ai \subset \Di$ the subspace of analytic diffeomorphism jets, \ie those that define convergent Taylor series.

Given a regular analytic Lie pseudo-group $\G\subset \D$, let $\Gn \subset\Dn$ denote the subbundle (subgroupoid) consisting of \nth order jets of pseudo-group diffeomorphisms, which we can identify with the \nth order determining equations of $\G$, whose solutions are the pseudo-group transformations. Note that, by  analyticity, $\Gi \subset \Ai$.  According to Theorem 1 of \rf{IOV}, there exists an order $\ns \in \mathbb{N}$, called the \is{order of involutivity}, such that, for all finite $n\geq \ns$, the determining equations
\begin{equation}\label{eq: det eq}
\Gn = \big\{\Delta_\nu(z,Z\n) = 0,\quad \nu=1,\ldots,l_n\big\}
\end{equation}
are involutive.  Separating the pseudo-group jet coordinates by order, let
\begin{align*}
\Dn &\simeq M \times D\n = M \times D^0 \times D^1 \times \cdots \times D^n,\\
\Gn &\simeq M \times G\n = M \times G^0 \times G^1\times \cdots \times G^n,
\end{align*}
where
\[
D^k = \set{(\ldotsx Z^a_B \ldotsx)}{|B| = k,\quad a=1,\ldots,m}
\]
denotes the space of $k$-th order derivatives of the local diffeomorphism $Z=\varphi(z)\in \D$, and similarly for $G^k$. In view of \eqs{tk}{tki}, with $p=q=m$, we then have
\Eq{tkD}
$$\t_k = \dim\, D^k = m\, \binom{m+k-1}{k},$$
while the number of derivatives of order $k\geq 1$ and of class $a$ is
\Eq{tka}
$$\total{k}{a} = m\,\binom{m+k-a-1}{k-1},\qquad 1\leq a \leq m.$$
For $n\geq 1$, we have the relations
\Eq{tkall}
$$\req{\sum_{k=0}^n\> \t_k = \t\n = \dim D\n,\\\sum_{a=1}^m \>\total{n}{a} = \t_n,\\ \sum_{a=1}^m \>a\: \total{n}{a} = \t_{n+1}.}$$

For the Lie pseudo-group $\G$, and each $0 \leq n < \infty $, let $\d\n = \dim G\n$ denote the fiber dimension of the projection $\pi^n\colon \Gn \to M$.  For $0\leq k \leq n$, let $\d_k = \dim G^k$ denote the number of parametric pseudo-group parameters of order $k$, so that 
$$\d\n = \d_0+\d_1+\cdots + \d_n.$$  The number of principal pseudo-group parameters of order $k$ is then given by $\rk_k = \t_k - \d_k$.

Let $\zeta \n = (\ldotsx \zeta^a_B \ldotsx)$, for $\rg a m$ and $|B| \leq n$, be fiber coordinates on the tangent bundle  $T\:\Dn$, and let
\[
\jn\vV = \sum_{a=1}^m \sum_{0\leq|B|\leq n} \zeta^a_B \pp{}{Z^a_B}
\]
denote a vertical vector field on $\Dn$, whose coefficients are functions of $z\n$.  The linearization of the pseudo-group determining equations \eqref{eq: det eq} at the identity jet $\dsty\mathds{1}\n_z$ are the \is{linearized determining equations}
\begin{equation}\label{linearized det eq}
\LGn = \bigg\{L_\nu (z,\zeta\n)= \sum_{a=1}^m \sum_{0\leq |B|\leq n}\>\left .\pp{\Delta_\nu}{Z^a_B}\;\right |_{\dsty\mathds{1}\n_z}\> \zeta^a_B = 0,\quad \nu=1,\ldots,l_n\bigg\},
\end{equation}
which serve to define  the Lie algebroid associated with the Lie pseudo-group groupoid $\Gn$, \cite{OP05}.  As before, we introduce the highest order term map $\bH\colon T\:\Dn\at{\mathds{1}\n_z} \to T D^n\at{\mathds{1}\n_z}$, which only keeps the linear terms of order $n$ in \eqref{linearized det eq}, to obtain the \nth order pseudo-group \is{symbol}
$$\symGn = \bH(\LGn) = \bigg\{\sum_{a=1}^m \sum_{|B| = n} \>\left .\pp{\Delta_\nu}{Z^a_B}\;\right |_{\dsty \mathds{1}\n_z}\> \zeta^a_B = 0,\qquad \nu=1,\ldots,l_n\bigg\}.$$
Our regularity assumption on $\G$ requires that the intrinsic algebraic properties of the symbol are independent of the point $(z,Z\n) \in \Gn$.

Referring to \eqs{involutivesymbol}{involutivecc}, for $n\geq \ns$, the order of involutivity, the indices and Cartan characters of the determining equations $\Gn$ satisfy
\begin{equation}\label{sums}
\sum_{a=1}^m \>\index{n}{a} = \rk_n \qquad\quad \sum_{a=1}^m \>\cartan{n}{a} = \d_n,
\end{equation}
and, since the equations are involutive,
\Eq{involutivepsg}
$$\sum_{a=1}^m \>a\:\index{n}{a} = \rk_{n+1}\qquad \quad
\sum_{a=1}^m \>a\: \cartan{n}{a} = \d_{n+1}.$$

\Ex{1}
The following well-studied Lie pseudo-group, \crf{OP05,OP08},
\begin{equation}\label{eq: pg1}
X = f(x),\qquad Y = f_x(x)\, y + g(x),\qquad U = u + \frac{f_{xx}(x)\, y + g_x(x)}{f_x(x)},
\end{equation}
where $f,g$ are analytic scalar functions with $f \in \D(\mathbb{R})$, so $f_x(x) \ne 0$, will serve as our running example illustrating the constructions. The determining equations $\G^{(2)}$ of order two (in reduced Cartan normal form) are
\begin{equation}\label{eq: det eq ex}
\seq{
X_y = X_u = 0,\quad \ Y_x = (U-u)X_x,\qquad Y_y = X_x,\qquad Y_u=0,\qquad U_u = 1,\\
X_{xx} = U_y X_x,\quad X_{xy} =X_{xu} = X_{yy} = X_{yu} = X_{uu} = 0,\quad
Y_{xx} = \paz{U_x + (U-u)U_y}X_x,\\
Y_{xy} = U_y X_x,\quad \ Y_{xu} = Y_{yy}= Y_{yu} = Y_{uu} = 0,\qquad
U_{xu} = U_{yy} = U_{yu}=U_{uu} = 0.
}
\end{equation}
Thus, the parametric jet variables that serve to parametrize the fibers of $\G^{(2)}$ are
\begin{equation}\label{eq: parametric jet ex1}
\xeq{X,\\ Y,\\ U,\\ X_x,\\ U_x,\\ U_y,\\ U_{xx},\\ U_{xy};}
\end{equation}
all the other second order jet coordinates, \ie those appearing on the left hand side of the determining equations \eqref{eq: det eq ex}, are principal.  We observe that 
$$\qeq{\d_0 = \d_1 = 3,\\ \d_2 = 2, \roq{and so}\d^{(0)} = 3, \\ \d^{(1)} = 6, \\ \d^{(2)} = 8.}$$  
It is not hard to see that, in general, the order $n\geq 2$ parametric variables are $U_{x^n}, U_{x^{n-1}\:y}$, hence $\d_n = 2$ and $\d\n = 2\:n+4$.  Using the notation
\begin{equation}\label{jivV}
\ji\vV = \sum_{|B| \geq 0} \xi_B \pp{}{X_B} + \eta_B\pp{}{Y_B} + \phi_B \pp{}{U_B}
\end{equation}
to denote a vertical vector field, the corresponding linearized determining equations $\LG{2}$ of order two are obtained by applying \eqref{jivV} to the determining equations \eqref{eq: det eq ex} and then evaluating the result at the identity jet, by setting 
$$\qeq{X=x,\\ Y=y,\\ U=u, \\ X_x = Y_y = U_u = 1,}$$
 and all other jet coordinates to $0$.  The result is the linearized system
\begin{gather*}
\xi_y = \xi_u=0,\qquad \eta_x=\phi,\qquad \eta_y = \xi_x,\qquad \eta_u=0,\qquad \phi_u=0,\\
\xi_{xx} = \phi_y,\qquad \xi_{xy} = \xi_{xu} = \xi_{yy} = \xi_{yu} =\xi_{uu}=0,\\
\eta_{xx} = \phi_x,\qquad \eta_{xy} = \phi_y,\qquad \eta_{xu} = \eta_{yy}=\eta_{yu}=\eta_{uu}=\phi_{xu} = \phi_{yy} = \phi_{yu}=\phi_{uu}=0.
\end{gather*}
The order two symbol $\symG{2}$ is thus given by the equations
\begin{equation}\label{symbol ex}
\begin{gathered}
\xi_{xx} = \xi_{xy} = \xi_{yy} = \xi_{xu}=\xi_{yu} = \xi_{uu} = 0,\qquad
\eta_{xx} = \eta_{xy}  = \eta_{yy} = \eta_{xu} = \eta_{yu}=\eta_{uu} = 0, \\
\phi_{xu} = \phi_{yy} =  \phi_{yu} = \phi_{uu}=0.
\end{gathered}
\end{equation}
%\pjo{Similarly, at order $3$, all symbol variables vanish except $\phi_{xxx}$ and $\phi_{xxy}$. Observe that we can, alternatively, compute the indices \eqref{indices ex1}
%and Cartan characters \eqref{Cartan char ex1}, and thus establish involutivity, directly from the linearized symbol.}
Using the term ordering $x \prec y \prec u$, the indices of the symbol \eqref{symbol ex} are
\begin{equation}\label{indices ex1}
\index{2}{1} = 7,\qquad \index{2}{2} = 6,\qquad \index{2}{3}=3,
\end{equation}
while the Cartan characters are
\begin{equation}\label{Cartan char ex1}
\cartan{2}{1} = 2,\qquad \cartan{2}{2} = \cartan{2}{3} = 0.
\end{equation}
On the other hand, the determining equations of order three are obtained by differentiating those of order two  in \eqref{eq: det eq ex} and then replacing any principal derivatives using the preceding equations, thereby producing
\Eq{deteq3ex}
$$\hskip-.25in\eeq{X_{xxx} = (U_{xy} + U_y^2) X_x,\qquad X_{xxy} = X_{xxu} = X_{xyy} = X_{xyu} = X_{xuu} = X_{yyy} = 0,\\
X_{yyu} = X_{yuu} = X_{uuu} = 0,\qquad
Y_{xxx} = \paz{U_{xx} + (U-u)(U_{xy}+U_y^2) + 2\:U_x U_y} X_x,\\
Y_{xxy} = (U_{xy}+U_y^2)X_x,\qquad Y_{xxu} = Y_{xyy} = Y_{xyu} = Y_{xuu} = Y_{yyy} = Y_{yyu} = Y_{yuu} = Y_{uuu} = 0,\hskip-1in\\
U_{xxu} = U_{xyy} = U_{xyu} = U_{xuu} = U_{yyy} = U_{yyu} = U_{yuu} = U_{uuu} = 0,}$$
where $U_{xxx},U_{xxy}$ are the only parametric third order derivatives and the other $28$ third order derivatives are all principal.  We thus see that the algebraic involutivity constraint
\[
\index{2}{1} + 2\:\index{2}{2} + 3\:\index{2}{3} = \rk_3 = 28
\]
is satisfied. Alternatively, in terms of the Cartan characters,
\[
\cartan{2}{1} + 2\: \cartan{2}{2} + 3\:\cartan{2}{3} = \d_3 = 2.
\]
Since $\pi^3_2(\G^{(3)}) = \G^{(2)}$, there are no integrability conditions at order three, and the determining equations \eqref{eq: det eq ex} are involutive.  Based on the Cartan characters \eqref{Cartan char ex1}, the solution depends on two functions of one variable, as was already clear from the original formula \eqref{eq: pg1} for the pseudo-group transformations.

%%%%%
\Section{rpg} Reduction of Lie Pseudo-Group Actions.
%%%%%

We are now interested in the action of a Lie pseudo-group on $p$-dimensional submanifolds of the total space $M$.  To work in local coordinates, we assume that the submanifolds are transverse to the fibers, and thus form local sections of $M \to \mX$.  In this section, we formulate the reduced determining equations for the action of pseudo-group elements on sections, and prove that they form an involutive system of differential equations.  This construction is a key intermediate step towards our formulation of the system of differential equations satisfied by the normal forms of submanifolds.

We introduce the local coordinates $z = (x,u) = (\sups xp,\sups uq)$ on the total space $M$, where $p + q = m  = \dim M$, so that submanifolds are locally given as the graphs of functions $u = u(x)$.  In accordance with Cartan's notation introduced in \sect{pg}, the corresponding target coordinates are given by $Z = (X,U) = (\sups Xp,\sups Uq)$.  Let $\Jn$ denote the corresponding submanifold jet space, with coordinates $z\n = (x,u\n) = (\ldotsx x^i \ldotsx \uaJ\ldotsx)$ for $\rg ip$, \ $\rg \alpha q$, and $|J| \leq n$.  

As in \rf{OP08,OP09}, let $\E\n\to \Jn$ denote the \is{lifted bundle} obtained by pulling back the diffeomorphism jet bundle $\Dn\to M$ to the submanifold jet  space via the standard projection $\pi^n_0 \colon \Jn \to M$.  Local coordinates on $\E\n$ are given by $(z\n,Z\n) = (x,u\n,X\n,U\n)$, where $(x,u\n)$ are the preceding submanifold jet coordinates, while $Z\n = (X\n,U\n) = (\ldotsx X^i_A \ldotsx U^\alpha  _B\ldotsx)$ for $\rg ip$, \ $\rg \alpha q$, and $|A|, |B| \leq n$ are the fiber coordinates of the diffeomorphism jet bundle $\Dn$.  The lifted bundle has the structure of a groupoid using the double fibration with source map $\bs\n(z\n,Z\n)=z\n$ and target map $\bt\n(z\n,Z\n) = Z\n\cdot z\n$ prescribed by the prolonged action of the diffeomorphisms on submanifold jets.

When writing out the action of a pseudo-group transformation on a submanifold, we will continue to use, in accordance with Cartan's convention, lower case letters for the source submanifold $u = u(x)$ and its jet coordinates $\uaJ$.  However, to avoid notational confusion, especially when distinguishing submanifold jets from diffeomorphism jets, we will use hats on the dependent variable  and its derivatives to denote the target submanifold, which we thus write as $\tU = \tU(X)$ with the order zero jet being simply $\tU = U$, while the higher order jet coordinates are denoted $\tU^ \alpha _J$. Later,  once the reader becomes used to which symbol denotes which type of jet coordinate, the hats can be dropped to clean up the formulas, and, indeed, we shall do so in the examples treated in \sect{xx}.

\Ex{Ecoord}
Let $M = \Rx2$ and $\mX = \R$, which is the setting for plane curves $s = \{(x,u(x))\}$.  Given the action of a diffeomorphism of $\Rx2$ on curves, the source curve is the graph of a scalar function $u = u(x)$ for $x,u \in \R$, with jet coordinates $u,u_x,u_{xx},\ldots$, while the target is also the graph of a scalar function, which, in accordance with the above-stated convention, is written as $\tU = \tU(X)$ for $X,\tU \in \R$. Its jet coordinates are then given by $\tU,\tU_X,\tU_{XX},\ldots$.
The coordinates on the lifted bundle $\E\n$ are thus given by
\begin{align*}
(z\n,Z\n) &= (x,u\n,X\n,U\n) \\
&=(x,u,u_x,u_{xx},\ldots,X,U,X_x,X_u,U_x,U_u,X_{xx},X_{xu},X_{uu},U_{xx},U_{xu},U_{uu},\ldots),
\end{align*}
where $u$, $u_x$, $u_{xx}$, $\ldots$ are the source curve jet coordinates, while $X,U,X_x,X_u,U_x,U_u,\ldots$ (which do not have hats) are the diffeomorphism jet coordinates. The source and target maps on the lifted bundle $\E\n\to \Jn$ are
$$\eeq{\bs\n(z\n,Z\n)=(x,u,u_x,u_{xx},\ldots),\\
\bt\n(z\n,Z\n) =(X,\tU,\tU_X,\tU_{XX},\ldots)\cthn{-60}= \left(\!X,U,\frac{U_x+u_xU_u}{X_x+u_xX_u},\frac{{\hskip-.75in\dsty\big[(X_x+u_xX_u)(U_{xx}+ 2\:u_xU_{xu}+u_x^2U_{uu}+u_{xx} U_u) \atop \dsty\hskip.5in{}- (U_x+u_xU_u)(X_{xx}+2\:u_xX_{xu}+u_x^2X_{uu}+u_{xx} X_u)\big]}}{(X_x+u_xX_u)^3},\ldots \right) ,}$$
where the higher order target jets are obtained by repeatedly applying the  operator of \is{implicit differentiation}
$$\Dt_X = \frac{1}{\Dt_xX}\,\Dt_x = \frac{1}{X_x+u_xX_u}\,\Dt_x$$
to $\tU$; see also \eq{DtX} below.
 
The \is{horizontal total derivative operators} on the infinite order lifted bundles $\E\ii$ are
\begin{equation}\label{Dt}
\Dt_{x^i} = \mD_{x^i} + \sum_{\alpha = 1}^q\bigg(u^\alpha_i\, \mD_{u^\alpha} + \sum_{|J|\geq 1} u^\alpha_{J,i}\pp{}{u^\alpha_J}\bigg),\qquad i=1,\ldots,p,
\end{equation}
where\fnote{Here $z^a$ can be either $x^i$ or $u^\alpha $.}
\[
\mD_{z^a} = \pp{}{z^a} + \sum_{b=1}^m\sum_{|A| \geq 0} Z_{A,a}^b\pp{}{Z^b_A},\qquad a=1,\ldots,m,
\]
are the total derivative operators on the diffeomorphism jet bundle $\D\ii$.  We use the same notation \eq{Dti} and \eqref{Dt} for the total derivative operators on $\Ji$ and $\E\ii$, respectively, since they coincide when applied to a function $F(z\n) = F(x,u\n)$ that does not depend on the diffeomorphism jet coordinates.

Given a local section $f \colon \mX\to M$, whose graph defines a $p$-dimensional submanifold $s  = f(\mX)$, and a local diffeomorphism $\varphi\in \D(M)$, with $s \subset \dom \varphi$, we call the composition $\rvarphi = \varphi \comp f $ the \is{reduction} of $\varphi $ to the submanifold $s$.  The reduced map $\rvarphi \colon \mX \to M$ is in general not a section of $M$ since $\varphi \comp f(x)$ does not necessarily belong to the fiber of $M$ over $x \in \mX$. On the other hand, its image, namely $S = \varphi\bk{f(\mX)}= \rvarphi(s)$ is an equivalent submanifold.  If we assume that the image submanifold  $S$ is transversal to the fibers of $M$, we can locally identify it with the graph of a local section $F \colon \mX \to M$, so $S = F(\mX)$.

\Rmk{bar} We will use overbars to denote reduced maps and jet coordinates. As with the hats, these can also be dropped once the reader becomes used to which symbol denotes which jet coordinate, and, indeed, we shall do so in \sect{xx}.

For $0 \leq n < \infty$, the reduced action of local diffeomorphisms on submanifolds is encoded by  the \emph{reduction map} $\rmapn \colon \E\n \to \Jn(\mX, \mU \times M)$ given by
\Eq{rmap}
$$\rmapn(x,u\n,X\n,U\n)
= \rmapn(\zn,Z\n) = (\zn,\rZ\n)
= (x,u\n,\rX\n,\rU\n),$$
where $\rZ\n= (\rX\n,\rU\n)$ has components
$$\rZ^a_J = \Dt_x^J Z^a \forq \rg am,\quad 0 \leq |J| \leq n,$$
which are obtained by successively applying the total derivative operators \eqref{Dt} to the diffeomorphism target coordinates $Z = (X,U)$.  We call $\rZ^a_J$ the \is{reduced jet coordinates}.  The reduction map is compatible with the reduction of diffeomorphisms to submanifolds.  Namely, given  a diffeomorphism $\varphi $ and a section $s = f(x) = (x,u(x))$ contained in its domain, let $(x,u\n,X\n,U\n) \in \E\n$ be given by their combined jets, so that $(x,u\n) = \jn f\at x$ and $(x,u,X\n,U\n) = \jn \varphi \at{(x,u)}$. Then $\jn(\varphi\comp f) = \rmapn(x,u\n,X\n,U\n)$.

We will regard $\Jn(\mX, \mU \times M) \to \Jn(\mX, \mU) = \Jn$ as a fiber bundle over the submanifold jet bundle, so that the reduced jet coordinates $\rZ\n = (\ldotsx \rZ^a_J \ldotsx)$ are its fiber coordinates.

\Ex{2dx} Let $M = \Rx2$ and $\mX = \R$, as in \ex{Ecoord}.  The reduction map \eq{rmap} is computed by successively applying the total derivative operator
\Eq{Dtx1}
$$\Dt_x = \mD_x + u_x \:\mD_u + u_{xx} \pdo{u_x} + u_{xxx} \pdo{u_{xx}} + \cdotsx, $$
with
\Eq{mD1}
$$\eeq{\mD_x = \pdo x + X_x \pdo X  + U_x \pdo U  + X_{xx} \pdo{X_x}  + X_{xu} \pdo{X_u} + U_{xx} \pdo{U_x}  + U_{xu} \pdo{U_u} + \cdotsx,\\
\mD_u = \pdo u + X_u \pdo X  + U_u \pdo U  + X_{xu} \pdo{X_x}  + X_{uu} \pdo{X_u} + U_{xu} \pdo{U_x}  + U_{uu} \pdo{U_u} + \cdotsx,}$$
to $X,U$.  This produces,  at order $n = 2$,
\[
\iqeq{\rmap^{(2)}(x,u,u_x,u_{xx}, X,U,X_x,X_u,U_x,U_u,X_{xx},X_{xu},X_{uu},U_{xx},U_{xu},U_{uu}) \\= (x,u,u_x,u_{xx},  \rX,\rU,\rX_x,\rU_x,\rX_{xx},\rU_{xx} ) \\
= (x,u,u_x,u_{xx}, X,U,\Dt_xX,\Dt_xU,\Dt_x^2X,\Dt_x^2U )\\
= (x,u,u_x,u_{xx}, X,U,X_x + u_xX_u ,U_x + u_xU_u, \\
\hskip2.4cm X_{xx} + 2\: u_x X_{xu} + u_x^2 X_{uu} + u_{xx} X_u, U_{xx} + 2\:u_x X_{xu} + u_x^2 U_{uu} + u_{xx}U_u).}
\]
Observe that the expressions for the reduced jet coordinates are obtained by total differentiation  of $\mzeq{X = X(x,u),\\U= U(x,u)}$, treating $u$ as a function of $x$.

%%%%%
\Subsection{rde} The Reduced Determining Equations.
%%%%%

Just as the original pseudo-group jets satisfy a system of differential equations, 
%$\Gn \subset \Dn$, 
so do the reduced pseudo-group jets.
To construct this ``reduced'' system, first define the \is{lifted subgroupoid} $\Hn \subset \En$ to be the pullback of $\Gn$ to the submanifold jet bundle $\Jn \to M$.  We then define the \nth order \is{reduced pseudo-group jet bundle} by applying the reduction map \eq{rmap}:
\Eq{rde}
$$\rde\n =\rmapn(\Hn) \subset  \Jn(\mX, \mU \times M).$$
This can be written as a system of equations of the form
\Eq{redde}
$$\rde\n =\big\{\,\rDelta_\nu(\zn,\rZ\n) = 0,\quad \nu=1,\ldots,\rl_n\big\}.$$
If we fix a section $s = \{(x,u(x))\}$ with jet $\zn = \jn s\at x = (x,u\n(x))$, then \eq{redde} can be viewed as an \nth order system of differential equations
for the reduced diffeomorphism $\rZ = \rvarphi(x)$, that we call the \emph{reduced determining equations}, whose properties will be investigated shortly.

In local coordinates, the reduced determining equations encode all the algebraic relations that exist among the reduced jets $\rZ{}\n$.  They are obtained by writing out the formulas for the reduced jet coordinates in terms of the parametric pseudo-group derivatives, and then eliminating the latter from the resulting algebraic expressions, \ie implicitizing the resulting parametric formulae, thereby producing the identities involving only the submanifold jet coordinates and the reduced  jet coordinates.

\Ex{rdex1}
Recalling the determining equations \eqref{eq: det eq ex} of the Lie pseudo-group \eqref{eq: pg1}, we now compute the reduced determining equations, assuming that $u=u(x,y)$. The pseudo-group jet coordinates parametrizing $\G^{(2)}$ are given in \eqref{eq: parametric jet ex1}.  At order zero, we trivially have
$$\qeq{\rX = X,\\\rY = Y,\\\rU = U.}$$
Next, at order one, in view of the first order determining equations in \eqref{eq: det eq ex}, we find
\begin{align*}
& \rX_x = X_x + X_u u_x = X_x,& & \rX_y = X_y + X_u u_y= 0,\\
& \rY_x = Y_x + Y_u u_x = Y_x= (U-u)X_x,& & \rY_y = Y_y + Y_u u_y = X_x,\\
& \rU_x = U_x + U_u u_x = U_x + u_x,& & \rU_y = U_y + U_u u_y = U_y + u_y.
\end{align*}
Differentiating again, and skipping computational details, at order two we obtain
\begin{align*}
& \rX_{xx} = U_y X_x,& & \rX_{xy} = 0,& & \rX_{yy} = 0,\\
&\rY_{xx} = \paz{U_x + (U-u)U_y }X_x,& & \rY_{xy} = U_y X_x,& & \rY_{yy} = 0,\\
& \rU_{xx} = U_{xx} + u_{xx},& & \rU_{xy} = U_{xy} + u_{xy},& & \rU_{yy} = u_{yy}.
\end{align*}
Implicitization, \ie eliminating the parametric variables $X, Y, U, X_x, U_x, U_y, U_{xx}, U_{xy}$, we find that, up to order two, the relations among the reduced pseudo-group jet coordinates are
\begin{equation}\label{eq: red det eq ex1}
\begin{gathered}
\rX_y = 0,\qquad \rY_x = (\rU-u)\rX_x,\qquad \rY_y = \rX_x,\qquad
\rX_{xx} = (\rU_y-u_y)\rX_x,\\ \rX_{xy} = \rX_{yy} = 0,\qquad
\rY_{xx} = \pab{\rU_x-u_x + (\rU-u)(\rU_y-u_y)}\rX_x,\\
\rY_{xy} = (\rU_y-u_y)\rX_x,\qquad \rY_{yy} = 0,\qquad \rU_{yy} = u_{yy},
\end{gathered}
\end{equation}
which thus form the second order reduced determining equations.  We note that the parametric variables are $\rX, \rY, \rU, \rX_x, \rU_x, \rU_y, \rU_{xx}, \rU_{xy}$.

A key observation that we will need in \sect{nfde} is that the reduced determining equations must become identities when the pseudo-group element is the identity map, and hence the two sections coincide.  Algebraically, this specialization amounts to equating
\Eq{ridXU}
$$\ceq{\qeq{\rX^i = x^i, \\\rX^i_{i} = 1, \\ \rX^i_{J} = 0,\\\rg ip,\\ J \ne i, \\ |J| \geq 1,}\\\qeq{\rU^ \alpha _K = u^ \alpha _K, \\ \rg \alpha q,\\ |K| \geq  0.}}$$
The equations defining $\r{\G}\n$ must vanish identically on the affine subvariety defined by \eq{ridXU}. For example, in the case of the pseudo-group in \ex{rdex1}, every reduced determining equation in \eqref{eq: red det eq ex1} vanishes identically when
\Eq{x1id}
$$\ceq{\qeq{\rX_x = \rY_y = 1,\\\rX_y = \rY_x = \rX_{xx} = \rX_{xy} = \rX_{yy} = \rY_{xx} = \rY_{xy} = \rY_{yy} = 0,}\\\qeq{\rU=u,\\\rU_x=u_x,\\\rU_y=u_y,\\ \rU_{yy}=u_{yy}.}}$$
According to \rf{O;Proposition 2.10}, this implies that the equations \eq{redde} can be expressed as a linear combination
\Eq{rde0}
$$\rDelta _\nu = \Sum ip \Bk{A_\nu^i (\rX^i -  x^i)+ A_\nu^{i,i}(\rX^i_{i} - 1) + \Sumu{\scriptstyle J \ne i \atop \scriptstyle 1 \leq |J| \leq n} A_\nu^{i,J}\, \rX^i_{J}} + \Sum \alpha q \Sumu{0 \leq |K| \leq  n} B_\nu^{\alpha ,K} (\rU^ \alpha _K - u^ \alpha _K),$$
where the coefficient functions $A_\nu^i, A_\nu^{i,i}, A_\nu^{i,J}, B_\nu^{\alpha ,K}$ are analytic.

We now state the key condition to be imposed on the pseudo-group actions to be considered in this paper.

\Df{reducible} The pseudo-group $\G$ is order $n$ \is{reducible} on the local section $s \colon \mX \to M$  if, for all $x \in \dom s$ with $\zn = \jn s\at x$, the reduction map $\rmapn \colon \Hn\at{\zn} \to \rG\n\at{\zn}$ is one-to-one on the indicated fibers. The pseudo-group $\G$ is \is{reducible} on $s$ if it is reducible for all sufficiently large $n \geq \nEQ$.  The integer $\nEQ$ is called the \emph{order of reducibility}.

As we will see in \th{freered} below, all pseudo-groups for which the moving frame calculus is applicable automatically satisfy this condition on generic sections. In particular, this implies that any finite-dimensional Lie group action is reducible.

\Df{regularsection} A section $s \colon \mX \to M$ is called \is{regular}  if $\G$ is reducible on it.

In what follows, we will only deal with regular sections. In particular, the reduced determining equations are to be evaluated only on regular sections.   Assuming analyticity, if the pseudo-group is regular on one section, regularity holds on generic sections.

Let $\rd\n$ denote the fiber dimension of the  reduced determining equations \eq{redde}, which can be identified as the number of parametric reduced pseudo-group parameters of order $\leq n$.
A basic property of reducible Lie pseudo-groups is given in the following result.

\Lm{rlpsgparam}Let $\G$ be a reducible Lie pseudo-group with order of reducibility $\nEQ$.  Then for all $n\geq \nEQ$, the number of independent reduced pseudo-group parameters equals the number of pseudo-group parameters. That is,
\begin{equation}\label{R constraint}
\d\n = \rd\n.
\end{equation}

In other words, reducibility requires that the reduction map does not change the fiber dimensions at sufficiently high orders.  Since
\begin{equation}\label{dnineq}
0 \leq \d\n \leq (p+q)\,\binom{p+q+n}{n}\qquad \text{and}\qquad 0 \leq \rd\n \leq (p+q) \,\binom{p+n}{n},
\end{equation}
we see that reducibility imposes constraints on the size of the pseudo-group $\G$, in that it cannot be too large; see \lm{l1} below.  For example, $\G$ cannot be the full diffeomorphism pseudo-group $\D $, which maximizes the inequality \eqref{dnineq} for $\d\n$.

\Ex{rdex1a}
Returning to \ex{rdex1}, in view of \eqref{eq: red det eq ex1} and its prolongations, it follows that the parametric reduced pseudo-group jet coordinates are
\begin{equation}\label{parametric rpg}
\xeq{\rX,\\ \rY,\\ \rU,\\ \rX_x, \\\rU_{x^k},\\ \rU_{x^{k-1}y},\\ k\geq 1.}
\end{equation}
Thus, the reduced dimensions satisfy
$$\qeq{\rd^{(1)} = 6 = \d^{(1)},\\\rd^{(2)} = 8 = \d^{(2)},\roq{and, in general,} \rd\n = 2\:n+4 = \d\n,}$$
thus proving that this pseudo-group is reducible.

\Ex{5}
An example where $\nEQ > 1$ in \df{reducible} is provided by the 5-dimensional Lie group action
\[
X = a\,x+b,\qquad
U = c\,u+d\,x+e,
\]
where $a,c \neq 0$ and $b,d,e\in \mathbb{R}$.  Up to order two, the determining equations are
\[
\req{X_u = X_{xx} = X_{xu} = X_{uu} = 0,\\ U_{xx} = U_{xu} = U_{uu}=0.}
\]
Prolonging, we deduce that, as expected,
\[
\d\n  = 5\forallq n \in \mathbb{N}.
\]
On the other hand, assuming the regularity condition $u_{xx} \neq 0$, the reduced determining equations, up to order three, are
\[
\rX_{xx} = \rX_{xxx} = 0,\qquad \rU_{xxx} = \frac{u_{xxx}}{u_{xx}}\, \rU_{xx},
\]
and $\rd^{(1)}=4$, while $\rd\n = 5$ for $n\geq 2$.
Thus, $\d\n = \rd\n$ for all $n \geq \nEQ=2$.

\Ex{xfxu}
Consider the Lie pseudo-group
\[
X = x,\qquad U=f(x,u).
\]
In this case,
\[
\d\n = \binom{n+2}{2} \qquad\text{while}\qquad \rd\n = n,
\]
and hence the pseudo-group is not reducible, basically because it has a one-dimensional base but the transformations depend upon an arbitrary function of two variables.  
%It is also easily seen that the prolonged action on $\Jn(\R,\R)$ is never free.

The last example can be easily generalized, proving that a reducible pseudo-group cannot depend on functions of $\geq p+1$ variables.  We state this fact in terms of its Cartan characters.

\Lm{l1}
Let $\G$ be a reducible Lie pseudo-group whose determining equations become involutive at order $\ns$.  Then $\cartan {\ns}{p+\alpha} = 0$ for $\alpha=1,\ldots,q$.

\begin{proof}
For the purpose of contradiction, assume there is a Cartan character $\cartan {\ns}{p+\alpha} \neq 0$ for some $\alpha = 1,\ldots,q$.  The pseudo-group thus admits at least one arbitrary function depending on at least $p+1$ variables, and hence
\[
\d\n \geq \a_n = \binom{p+n+1}{n} = \frac{(p+2)\cdots(p+n+1)}{n!},
\]
where $\a_n$ is the number of jet components of order $0\leq |J|\leq n$ associated with a function $f(z^1,\ldots,z^{p+1})$ of $p+1$ variables.
On the other hand, according to \eqref{dnineq},
\[
\rd\n \leq (p+q)\, \binom{p+n}{n} = \frac{(p+q)\:(p+1)\:(p+2)\cdots(p+n)}{n!} = \frac{(p+1)(p+q)}{p+n+1} \, \a_n < \a_n \leq \d\n,
\]
whenever $n \geq \max\{\ns,\nEQ, p^2 + p\:q + q\}$.  Hence the reducibility condition \eqref{R constraint} cannot hold when $n$ is sufficiently large.
\end{proof}

%%%%%
\Subsection{lde} The Linearized Reduced Determining Equations.
%%%%%

Linearizing the reduced determining equations \eq{redde} at the  reduced identity pseudo-group jet \eq{ridXU} yields the \emph{linearized reduced determining equations}
\begin{equation}\label{linearized reduced det eq}
\LrGn = \big\{\rL_\nu(z\n,\rzeta\n)=0,\quad \nu = 1,\ldots,\rl_n\big\}.
\end{equation}
Keeping only the highest order terms, we obtain the \is{reduced symbol}
\begin{equation}\label{reduced symbol}
\symrGn = \bH(\LrGn),
\end{equation}
where, again, $\bH$ is the highest order term map which only keeps the order $n$ terms in the linearized reduced determining equations \eqref{linearized reduced det eq}.
The coefficient matrix of the reduced symbol \eqref{reduced symbol} yields the $n$-th order reduced symbol matrix $\symMrGn$, from which we can compute the reduced indices $\rindex{n}{i}$ and reduced Cartan characters $\rcartan{n}{i}$ for $i=1,\ldots,p$.

As in the previous section, we separate the reduced pseudo-group jet coordinates by order and let
\begin{align*}
\rDiff\n &\simeq \Jn \times \rD\n = \Jn \times \rD^0 \times \rD^1 \times \cdots \times \rD^n,\\
\rG\n &\simeq \Jn \times \r{G}\n = \Jn \times \r{G}^0 \times \r{G}^1\times \cdots \times \r{G}^n,
\end{align*}
where
\[
\rD^k = \set{(\ldotsx \rZ^a_B \ldotsx)}{|B| = k,\quad a=1,\ldots,m}
\]
denotes the space of $k$-th order derivatives of reduced local diffeomorphisms and similarly for $\r{G}^k$, the latter subject to the reduced determining equations.  The number of derivatives of order $k$ is
$$\rt_k = \dim\, \rD^k = m \,\binom{p+k-1}{k}.$$
Of those, the number of derivatives of class $1\leq i \leq p$ is
$$\rtotal{k}{i} = m\,\binom{p+k-i-1}{k-1},$$
so that
\[
\sum_{i=1}^p\> \rtotal{k}{i} = \rt_k,\qquad \sum_{i=1}^p\> i\,\rtotal{k}{i} = \rt_{k+1},\qquad \sum_{k=0}^n\> \rt_k = \rt\n = \dim \r{D}\n.
\]

For the reduced Lie pseudo-group $\rG$, we let $\rd_k = \dim \r{G}^k$ denote the number of parametric reduced pseudo-group parameters of order $k$, so that 
$$\rd_0 + \cdotsx + \rd_n = \rd\n = \dim \r{G}\n,$$
which is the fiber dimension of the reduced determining equations of order $n$.  The number of principal reduced pseudo-group parameters of order $k$ is then given by
\[
\rr_k =\rt_k - \rd_k.
\]
Finally, the indices and Cartan characters of the reduced determining equations \eq{redde} satisfy
\Eq{hicce}
$$\qeq{\rindex ni + \rcartan ni = \rtotal ni ,\\ i=1,\ldots,p,}$$
with
\Eq{hiccesum}
$$\req{\sum_{i=1}^p\> \rindex{n}{i} = \rr_n = \rank \symMrGn,\\
\sum_{i=1}^p\> \rcartan{n}{i} = \rd_n = \dim \symrGn.}$$

%%%%%
\Subsection{irds} Involutivity of the Reduced Determining System.
%%%%%

The aim of this section is to prove that, under the assumption that the submanifold is reducible, the reduced determining system \eq{redde} is involutive.  Moreover, the first $p$ Cartan characters of the determining  system and its reduction coincide.

\Th{reducibility thm}
Let $\G$ be a reducible Lie pseudo-group with order of reduciblity $\nEQ$ and such that $\G^{(\ns)}$ is involutive.  Then there exists $\rns\in \mathbb{N}$ such that for all $n\geq \rns\geq \max\{\ns,\nEQ\}$,
\Eq{redcartan}
$$\qeq{\cartan ni = \rcartan ni,\\ \rg ip,\\\\\cartan n{p+\alpha} = 0,\\ \rg \alpha q.}$$
In particular, the involutivity of the \nth order determining equations $\Gn$ implies the involutivity of the $n$-th order reduced determining equations $\rde\n$.

\begin{proof}
First of all, the second set of equalities in \eq{redcartan}
follows from \lm{l1} with $n\geq \ns$.  Since $\G$ is reducible, consider the reduced determining equations $\r{\G}^{(\nEQ)}$, where $\nEQ$ is the order of reducibility.  By the Cartan--Kuranishi Theorem \cite[Theorem 7.4.1]{S}, after prolongation and projection, there exists $\rns \geq \text{max}\{\ns,\nEQ\}$ such that the reduced determining equations $\r{\G}^{(\rns)}$ are involutive.

%Let $n\geq \rns \geq \text{max}\{\ns,\nEQ\}$ and $k\geq 1$.  
Let $n\geq \rns$ and $k\geq 1$.  By the definition \eq{inCc} of the Cartan characters
\[
\d^{(n+k)} = \d^{(n+k-1)} + \sum_{i=1}^p \>\cartan {n+k}i\qquad\text{and}\qquad
\rd^{(n+k)} = \rd^{(n+k-1)} + \sum_{i=1}^p\> \rcartan {n+k}i,
\]
where we used the fact that $\cartan{n+k}{p+\alpha} = 0$, $\alpha=1,\ldots,q$.
Reducibility of the pseudo-group implies that   $\d^{(n+k)} = \rd^{(n+k)}$ and $\d^{(n+k-1)} = \rd^{(n+k-1)}$, which requires
\begin{equation}\label{cartan characters constraint}
\sum_{i=1}^p \big(\cartan {n+k}i - \rcartan {n+k}i\big)=0.
\end{equation}
Using \cite[Eq.\ (8.8a)]{S}, the higher order Cartan characters are related via the equation
\begin{equation}\label{cartan relations}
\cartan {n+k}i = \sum_{j=i}^p \binom{k+j-i-1}{k-1} \cartan nj,\qquad i=1,\ldots,p.
\end{equation}
Thus,
\begin{align*}
\sum_{i=1}^p \cartan {n+k}i &= \sum_{i=1}^p \sum_{j=i}^p \binom{k+j-i-1}{k-1}\cartan nj = \sum_{j=1}^p \sum_{i=1}^j \binom{k+j-i-1}{k-1} \cartan nj \\
&= \sum_{j=1}^p \binom{k-1+j}{k}\cartan nj
= \cartan n1 + (k+1) \cartan n2 + \cdots + \frac{(k+1)\cdots (k+p-1)}{(p-1)!} \cartan np.
\end{align*}
Substituting the last expression and its reduced version into \eqref{cartan characters constraint}, we obtain
\[
\big(\cartan n1 - \rcartan n1\big) + (k+1)\big(\cartan n2 - \rcartan n2\big) + \cdots + \frac{(k+1)\cdots (k+p-1)}{(p-1)!} \big(\cartan np-\rcartan np\big)=0.
\]
Viewing this expression as a degree $p-1$ polynomial in the variable $k$ which vanishes for all $k \in \N$, we conclude that $\cartan ni - \rcartan ni = 0$ for $ i=1,\ldots,p$.
\end{proof}

\Rmk{6r} \th{reducibility thm} implies that if $\G$ is reducible, then, at a sufficiently high order, the determining equations and the reduced determining equations of the pseudo-group contain the same number of parametric pseudo-group jet coordinates, and, furthermore, their first $p$ Cartan characters are the same.

\Ex{2}
Continuing Example \ref{rdex1}, we linearize the second order reduced determining equations \eqref{eq: red det eq ex1} at the reduced identity jet and obtain
\Eq{2l}
$$\ceq{\qeq{\rxi_y = 0,&\reta_x=\rphi,&\reta_y = \rxi_x,}\\\qeq{ \rxi_{xx} = \rphi_y,&
\rxi_{xy} = \rxi_{yy}=0,& \reta_{xx}=\rphi_x,&
\reta_{xy} = \rphi_y,& \reta_{yy} = \rphi_{yy} = 0.}}$$
The order two reduced symbol is then given by the equations
\[
\rxi_{xx} = \rxi_{xy} = \rxi_{yy} = \reta_{xx} = \reta_{xy} = \reta_{yy} = \rphi_{yy} = 0
\]
so that the reduced indices and Cartan characters are
\[
\rindex{2}{1} = 4,\qquad
\rindex{2}{2} = 3,\qquad
\rcartan{2}{1} = 2,\qquad
\rcartan{2}{2} = 0.
\]
On the other hand, the order three reduced determining equations are
\begin{gather*}
\rX_{xxx} = \pab{(\rU_y-u_y)^2 + (\rU_{xy} - u_{xy})}\rX_x,\qquad
\rX_{xxy} = \rX_{xyy} = \rX_{yyy} = 0,\\
\rY_{xxx} = \pab{\rU_{xx}-u_{xx} + (\rU-u)\pab{\rU_{xy}-u_{xy} + (\rU_y-u_u)^2} + 2\:(\rU_y-u_y)(\rU_x-u_x)}\rX_x,\\
\rY_{xxy} = \pab{\rU_{xy}-u_{xy} + (\rU_y-u_y)^2}\rX_x,\quad
\rY_{xyy}=\rY_{yyy}=0,\quad
\rU_{xyy} = u_{xyy},\quad \rU_{yyy} = u_{yyy},
\end{gather*}
from which we see that $\myeq{\rr_3=10,\\\rd_3=2}$, and $\pi^3_2(\rde^{(3)}) = \rde^{(2)}$. Since
\[
\rindex{2}{1} + 2\:\rindex{2}{2} = \rr_3\qquad\text{or, equivalently,}\qquad
\rcartan{2}{1} + 2\:\rcartan{2}{2} = \rd_3,
\]
the reduced determining equations \eqref{eq: red det eq ex1} of order $\rns = 2$ are involutive.

\Rmk{7r} In the previous example, the order at which the reduced determining equations became involutive,  was the same as the order of the original determining equations (recall \ex{1}) , i.e.\ $\rns = \ns =2$.  The next example shows that this does not always hold, and that, in general, $\rns \geq \ns$.

\Ex{4}
To illustrate the second half of the preceding remark, consider the Lie pseudo-group
\[
X = x+a,\qquad Y = y+b,\qquad U=f(x)\, u + g(x)\, y + h(x),
\]
where $f,g,h$ are analytic scalar functions with $f\neq 0$,
%\fv{$f,g,h \in C^\infty(\mathbb{R})$, where $f\neq 0$}, 
%$0\neq f \in \D(\mathbb{R})$, $g,h \in C^\infty(\mathbb{R})$, 
and $a,b\in \mathbb{R}$.  The determining equations, up to order two, are
\[
\ceq{\qeq{X_x =Y_y = 1,\\ X_y = X_u = Y_x = Y_u = 0,}\\
X_{xx} = X_{xy} = X_{yy} = X_{xu} = X_{yu} = X_{uu} = 0,\\
\qeq{Y_{xx} = Y_{xy} = Y_{yy} = Y_{xu} = Y_{yu} = Y_{uu} =0,\\
U_{yy} = U_{yu} = U_{uu} = 0.}}
\]
The corresponding indices and Cartan characters are
\[
\index{2}{1} = \index{2}{2} = 6,\quad \index{2}{3} = 3,\qquad
\cartan{2}{1} = 3,\quad \cartan{2}{2} = \cartan{2}{3} = 0.
\]
Computing the order three determining equations, we obtain
\begin{gather*}
X_{xxx} = X_{xxy} = X_{xxu} = X_{xyy} = X_{xyu} = X_{xuu} = X_{yyy} = X_{yyu} = X_{yuu} = X_{uuu} = 0,\\
Y_{xxx} = Y_{xxy} = Y_{xxu} = Y_{xyy} = Y_{xyu} = Y_{xuu} = Y_{yyy} = Y_{yyu} = Y_{yuu} = Y_{uuu} = 0,\\
U_{xyy} = U_{yyy} = U_{yyu} = U_{xyu} = U_{yuu} = U_{xuu} = U_{uuu} = 0.
\end{gather*}
Since 
\[
\index{2}{1} + 2\:\index{2}{2}+3\:\index{2}{3} = 27 = \rk_3,\qquad
\cartan{2}{1} + 2\:\cartan{2}{2} + 3\:\cartan{2}{3} = 3 = \d_3,
\]
and there are no integrability conditions, this proves involutivity at order $\ns=2$.
On the other hand, the reduced determining equations, up to order two, are
\begin{equation}\label{non-involutive eq}
\qeq{\rX_x = \rY_y = 1,\\ \rX_y = \rY_x = 0,\\
\rX_{xx} = \rX_{xy} = \rX_{yy} = \rY_{xx} = \rY_{xy} = \rY_{yy} = 0.}
\end{equation}
The reduced indices are $\myeq{\rindex{2}{1} = 4,\\ \rindex{2}{2} = 2}$, while the reduced Cartan characters are $\myeq{\rcartan{2}{1}=2,\\ \rcartan{2}{2} = 1}$.  Furthermore, provided the regularity condition $u_{yy}\neq 0$ holds, the order three reduced determining equations are
$$\qeq{\rX_{xxx} = \rX_{xxy} = \rX_{xyy} = \rX_{yyy} = 0,\\
\rY_{xxx} = \rY_{xxy} = \rY_{xyy} = \rY_{yyy} = 0,\\
\rU_{yyy} = \frac{u_{yyy}}{u_{yy}}\, \rU_{yy},}$$
and the involutivity test $\rindex{2}{1} + 2\:\rindex{2}{2} = 8 \neq \rr_3 = 9$ fails, as does $\rcartan{2}{1} + 2\:\rcartan{2}{2} = 4 \neq \rd_3 = 3$.
On the other hand, omitting the computational details, the reduced determining equations become involutive at order $\rns = 3$ with
\[
\weq{\rindex{3}{1} = 6,\\ \rindex{3}{2} = 3,\\ \rcartan{3}{1} = 3,\\ \rcartan{3}{2} = 0,\\
\rindex{3}{1} + 2\:\rindex{3}{2} = 12 = \rr_4,\\
\rcartan{3}{1} + 2\:\rcartan{3}{2} = 3 = \rd_4.}
\]

\Rmk{8r} According to \th{reducibility thm}, the conditions \eq{redcartan} on the Cartan characters eventually hold whenever the Lie pseudo-group is reducible.  We note that \eq{redcartan} may also hold for some non-reducible pseudo-groups, and that these equalities imply the involutivity of the associated determining equations.  Indeed, assume \eq{redcartan} holds for all $n\geq n_\diamond\geq \ns$, for some natural number $n_\diamond$.  First, \eqref{sums} and \eq{hicce}, together with \eq{redcartan}, imply $\d_n = \rd_n$.  Similarly, at  order $n+1$ we have  $\d_{n+1}= \rd_{n+1}$.  Combining the last equality with \eq{redcartan}, we conclude that
\[
\sum_{i=1}^p\, i\,\rcartan{n}{i} = \sum_{a=1}^m\, a\,\cartan{n}{a} = \d_{n+1} = \rd_{n+1}.
\]
Thus, the reduced determining equations $\rde\n$ satisfy the algebraic involutivity test.  Moreover, since $\Gn$ is involutive, $\pi^{n+1}_n(\G^{(n+1)}) = \Gn$, which implies $\pi^{n+1}_n(\cH^{(n+1)}) = \Hn$.  Then, using \eq{rde},
\begin{align*}
\rpi^{n+1}_n\big(\rde^{(n+1)}\big) &= \rpi^{n+1}_n\big(\rmap^{n+1}( \cH^{(n+1)})\big) = \rmapn\big( \pi^{n+1}_n (\cH^{(n+1)}) \big) = \rmapn (\Hn) = \rde\n,
\end{align*}
which thereby proves involutivity of the reduced determining equations $\rde\n$.

 We now illustrate the remark with an example.

\Ex{99} Consider the pseudo-group action
\[
X=x+a,\qquad U = \lambda\, u + f(x),
\]
where $f$ is an analytic scalar function, while $a,\lambda \in \mathbb{R}$, with $\lambda \neq 0$.  Up to order two, the determining equations are
\[
X_x = 1,\quad X_u = 0,\qquad X_{xx} = X_{xu} = X_{uu} = U_{xu} = U_{uu} = 0.
\]
These equations are involutive with indices and Cartan characters
\begin{equation}\label{non-reducible cartan characters}
\req{\index{2}{1} = 3,\\\index{2}{2} = 2,\\\cartan{2}{1} = 1,\\ \cartan{2}{2} = 0.}
\end{equation}
The number of parametric pseudo-group jet coordinates of order $\leq k \in \mathbb{N}$ is $\d^{(k)} = k+3$.  
\pari
On the other hand, assuming $u=u(x)$, the reduced determining equations of order $\leq 2$ are
\[
\req{\rX_x=1,\\ \rX_{xx} = 0.}
\]
At order two, the reduced index and reduced Cartan character are
\begin{equation}\label{non-reducible rcartan characters}
\req{\rindex{2}{1} =1,\\\rcartan{2}{1} = 1,}
\end{equation}
while the dimension of the reduced pseudo-group jet bundles are $\rd^{(k)} = k+2$.  Since $\rd^{(k)} < \d^{(k)}$, the pseudo-group is non-reducible. But \eqref{non-reducible cartan characters} and \eqref{non-reducible rcartan characters} satisfy \eq{redcartan} when $n \geq 2$.

%%%%%
\Section{mfnf} Reduced Moving Frames and Normal Forms.
%%%%%

In this section, we review the moving frame construction for infinite-dimensional Lie pseudo-groups, as originally introduced in \rf{OP08}.  Restricting ourselves to reducible Lie pseudo-groups, we will work with the reduced pseudo-group jets rather than the original jets, keeping in mind that when the pseudo-group is reducible, they are in one-to-one correspondence.

Let $\G$ be a reducible Lie pseudo-group acting on (local) sections  $s=\{(x,u(x))\}$ of the bundle $\pi\colon M \to \mX$.  For transformations near the identity $\mathds{1}_M$, the transformed submanifold $S=\varphi(s)$ remains a section.  The prolonged action on the $n$-th order submanifold jet space $\Jn$ is obtained by applying the \is{implicit total derivative operators}
\Eq{DtX}
$$\Dt_{X^i} = \sum_{j=1}^p W^j_i\, \Dt_{x^j},\qquad i=1,\ldots,p,$$
where $(W^j_i) = (\rX^i_j)^{-1}$ denotes the entries of the inverse reduced total Jacobian matrix (which can be simplified using the determining equations), to the reduced target dependent variables $\tU^\alpha = \rU^\alpha$:
\begin{equation}\label{eq: pr action}
\tU^\alpha_J = \Dt_X^J \tU^\alpha = \Dt_{X^{j_1}}\cdots \Dt_{X^{j_k}} \tU^\alpha.
\end{equation}

If $\r{g}\n$ denotes the \is{parametric} reduced pseudo-group parameters of $\r{\G}\n$, then, as a consequence of formula \eq{DtX} for the implicit total derivative operators, the prolonged action \eqref{eq: pr action} can be written in terms of the submanifold jet coordinates $(x,u\n)$ and the parametric reduced pseudo-group parameters $\r{g}\n$:
\begin{equation}\label{explicit pr action}
(\rX,\tU\n) = P\n(x,u\n,\r{g}\n).
\end{equation}

\Ex{5a}
We compute the prolonged action for the Lie pseudo-group \eqref{eq: pg1} acting on surfaces $u=u(x,y)$.  We streamline the computations by taking  the reduced determining equations \eqref{eq: red det eq ex1} into account.  In particular, we recall that the reduced parametric pseudo-group jet coordinates are given in \eqref{parametric rpg}.
Thus, the lifted total derivative operators \eq{DtX} are
\[
\Dt_X = \frac{1}{\rX_x}\,\Dt_x - \frac{\rY_x}{\rX_x^2}\,\Dt_y = \frac{1}{\rX_x}\bbk{\Dt_x + (u-\rU)\:\Dt_y},\qquad
\Dt_Y = \frac{1}{\rX_x}\,\Dt_y,
\]
and the coordinate expressions for the prolonged action up to order two are found to be
\begin{align}
\tU_X &= \frac{\rU_x+(u-\rU)\,\rU_y}{\rX_x}, \qquad \quad
\tU_Y = \frac{\rU_y}{\rX_x},\nonumber\\
\tU_{XX} &= \! \frac{\rU_{xx}+(u_y-\rU_y)\,\rU_x + (u_x-\rU_x)\,\rU_y+(u-\rU)\paz{2\rU_{xy} + 2(u-\rU)u_{yy}+(u_y-\rU_y)\,\rU_x}}{\rX_x^2}, \nonumber\\
\tU_{XY} &= \frac{\rU_{xy} + (u_y-\rU_y)\,\rU_y + (u-\rU)u_{yy}}{\rX_x^2},\qquad \quad
\tU_{YY} = \frac{u_{yy}}{\rX_x^2}.\label{pr action ex 2}
%,\nonumber\\
%\tU_{XYY} &= \frac{u_{xyy}+2(u_y-\rU_y)u_{yy}+(u-\rU)u_{yyy}}{\rX_x^3},\qquad \quad 
%\tU_{YYY} = \frac{u_{yyy}}{\rX_x^3}.\label{pr action ex}
\end{align}
We will also use the following third order expressions
\begin{align}
\tU_{XYY} &= \frac{u_{xyy}+2(u_y-\rU_y)u_{yy}+(u-\rU)u_{yyy}}{\rX_x^3},\qquad \quad 
\tU_{YYY} = \frac{u_{yyy}}{\rX_x^3};\label{pr action ex 3}
\end{align}
the other two, \ie $\tU_{XXX}$, $\tU_{XXY}$, are more complicated and not required.  Observe that, as stated in \eqref{explicit pr action}, the resulting formulas only depend on the reduced parametric pseudo-group parameters and the submanifold jet coordinates.

We are now ready to introduce the notion of a reduced moving frame.

\Df{mvff} Let $\mrH\n \to \Jn$ denote  the  lifted subgroupoid   obtained by pulling back $\r{\G}\n \to M$ to $\Jn$.
A \emph{reduced moving frame} $\rrho\n$ of order $n$ is a $\r{\G}\n$ equivariant local section $\rrho\n \colon \Jn \to \mrH\n$.

\Rmk{20r} The moving frame introduced in \df{mvff} differs from the original definition given in \rf{OP08} since it is based on the prolonged action of the reduced pseudo-group $\r{\G}$ rather than the original pseudo-group $\G$.  For non-reducible Lie pseudo-group actions, the two notions differ, whereas, as we now explain, for reducible pseudo-groups they are equivalent.  We will discuss the explicit construction of a reduced moving frame through the choice of a cross-section to the pseudo-group orbits in \sect m below.

In the original implementation \rf{OP08}, a moving frame exists at order $n$ provided the prolonged action is regular and (locally) free, as specified in the following definition.

\Df{free} The pseudo-group $\G$ is said to act \emph{freely} at a submanifold jets $\zn \in \Jn$ if its isotropy group $\G\n_{\zn} = \set{g\n \in \Gn}{g\n \cdot \zn = \zn}$ is trivial, which means that $\G\n_{\zn}=\set{\mathds{1}\n_z}{\pi^n_0(z\n) = z}$, \ie the only pseudo-group jet fixing $\zn$ is the identity.  More generally, the pseudo-group acts \emph{locally freely} at $\zn$ if $\G\n_{\zn}$ is a discrete group.

Once the pseudo-group acts (locally) freely\footnote{In general, one expects a subvariety of singular jets in $\Jn$ where the prolonged action is not locally free.} on an open subset $V\n \subset \Jn$ for some $n$, persistence of freeness, \rf{OP08,OP12}, implies that $\G$ acts freely on the open subset $V\ps{n+k} = (\pi^{n+k})^{-1} V\n$.  We now observe that freeness of the prolonged action implies reducibility of the Lie pseudo-group action.

\Th{freered}  If $\G$ acts freely on the open subset $V\n \subset \Jn(\mX,M)$ then it is order $n$ reducible on any section whose jet lies in $V\n$.

\begin{proof}
Note that the identity reduced jet $\dsty\barr {\mathds{1}}\n_z$ fixes any jet $\zn \in \Jn$, where $z = \pi^n_0(\zn)$.  Thus, because the action of $\G$ on $\Jn$ factors through the reduced action, each element of
\Eq{rmapnid}
$$(\rmapn)^{-1}\{\barr {\mathds{1}}\n_z \}\capz \Hn$$
 fixes $\zn$.  If the action is not reducible, the subset \eq{rmapnid} will contain non-identity jets, and hence the isotropy subgroup of $\zn$ will be non-trivial.
\end{proof}

\th{freered} implies that once the prolonged action becomes free, the reduced prolonged action is also free, that is, the isotropy group $\r{\G}\n_{\zn}$ is trivial.  For a reducible Lie pseudo-group, the converse is also true.

\Th{conversethm}
Let $\G$ be reducible on $z\n$.  If the prolonged action of the reduced pseudo-group $\r{\G}$ is free at $z\n$, then $\G\n$ acts freely at $\zn$.

\begin{proof}
Since $\G$ is reducible and $\r{\G}\n_{\zn} = \{\r{\mathds{1}}_z\n\}$, the isotropy group $\G\n_{\zn}$ must also contain a single jet.  Since $\dsty\mathds{1}\n_{z} \in \G\n_{\zn}$, it follows that $\G\n_{\zn} = \{\mathds{1}\n_z\}$.
\end{proof}

Theorems \ref{freered} and \ref{conversethm} imply that for reducible Lie pseudo-groups we can go back and forth between the construction of a moving frame for the original pseudo-group $\G$ and for the reduced pseudo-group $\r{\G}$.  This allows us to state the main existence theorem for a reduced moving frame.

\Th{mf existence} Let $\G$ act freely and regularly on the open set $V\n \subset \Jn(\mX,M)$.  Then for any $z\n \in V\n$ there exists a reduced moving frame of order $n$ in a neighborhood $N\n \subset V\n$ containing $z\n$.

%%%%%
\Subsection i Isotropy.
%%%%%

According to the preceding discussion, there are two types of isotropy of a submanifold jet --- those where the reduced action fixes the jet, and, more restrictively, those with trivial reduced action.  Let us characterize them for better understanding of the underlying geometry.  Note that the observations in this subsection are not used in the subsequent developments, and can thus be skipped without loss of continuity.

Given the submanifold jet $\zn\in \Jn$, let $\Dn_{\zn}\subset \Dn$ denote its isotropy subgroup of order $n$, \ie the set of $n$-jets of local diffeomorphisms which fix $\zn$. Let $\mTn_{\zn} \subset \Dn_{\zn}$ be those isotropy elements which have trivial reduction. We can thus identify $\dsty\mTn_{\zn} \simeq (\rmapn)^{-1}\{\barr {\mathds{1}}\n_z\}$ where we are now applying the reduction map $\rmapn$ --- see \eq{rmap} --- to an arbitrary diffeomorphism jet.  Let $\cQn_{\zn} = \Dn_{\zn}/\mTn_{\zn}$ denote the quotient space.

We now investigate $\Dn_{\zn}$, $\mTn_{\zn}$, and $\cQn_{\zn}$.  By applying a suitable diffeomorphism, we can, without loss of generality, assume that our section $s$ is, locally, the trivial zero section, $u(x) \equiv 0$, with zero $n$ jet, so $\zn = 0^{(n)}$. In this setting, a diffeomorphism $1$-jet $Z\ps 1= (X\ps 1,U\ps 1)$ belongs to $\mT\ps{1}_{0^{(1)}}$ if and only if
\[
X=U=0,\roq{and}
\begin{aligned}
\delta^i_j &= \rX^i_j = X^i_{x^j} + \sum_{\beta = 1}^q u^\beta_j X^i_{u^\beta} = X^i_{x^j},\\
0 &= u^\alpha_j = \rU^\alpha_j = U^\alpha_{x^j} + \sum_{\beta=1}^q u^\beta_j U^\alpha_{u^\beta} = U^\alpha_{x^j},
\end{aligned}\qquad
\begin{aligned}
i,j &=1,\ldots, p,\\
 \alpha &= 1,\ldots,q,
 \end{aligned}
\]
where $\delta^i_j$ is the Kronecker delta.  On the other hand, $Z\ps 1 \in \D\ps{1}_{0^{(1)}}$ if and only if $X=U = \tU^\alpha_{X^j} =\Dt_{X^j}(U^\alpha)= 0$, with $j=1,\ldots,p$, and $\alpha=1,\ldots,q$. Since the matrix $(W^i_j)$ in the definition of the total derivative operators \eq{DtX} is invertible, the constraints for $Z\ps 1$ to be in $\D\ps{1}_{0\ps{1}}$ are
\[
X=U=0,\qquad 0 = u^\alpha_j = \rU^\alpha_j = U^\alpha_{x^j} + \sum_{\beta=1}^q u^\beta_j U^\alpha_{u^\beta} = U^\alpha_{x^j},\qquad
\begin{aligned}
j &=1,\ldots,p,\\
\alpha &=1,\ldots,q.
\end{aligned}
\]
By similar computations, $Z\n \in \mTn_{0\ps{n}}$ if and only if
\[
X=U=0,\quad X^i_{x^j} = \delta^i_j,\qquad X^i_J =0,\qquad |J|\geq 2,\qquad U^\alpha_K = 0,\qquad |K|\geq 1,
\]
while $Z\n \in \Dn_{0\ps{n}}$ if and only if
\[
X=U=0,\qquad U^\alpha_K =0,\qquad |K|\geq 1.
\]
In other words, at $x=0$, $\Dn_{0\ps{n}}$ consists of $n$-jets of diffeomorphisms of the form
$$\qeq{X = f(x,u),\\U =  u \: g(x,u),\\f(0,0) = 0,\\\det(f^i_{x^j})\at{(0,0)} \ne 0,\\\prod_{\alpha=1}^q g^\alpha (0,0) \ne 0,}$$
while $\mTn_{0\ps{n}}$ consists of $n$ jets of diffeomorphisms of the form
$$\qeq{X = x + u \: h(x,u),\\U =  u \: g(x,u),\\ \prod_{\alpha=1}^q g^\alpha(0,0) \ne 0.}$$
In particular, on the zero section, we have $X = x$ and hence $\mTn_{0\ps{n}}$ consists of $n$-jets of diffeomorphisms which fix every single point of $s$, \ie the jets of the global isotropy subgroup of $s$.  On the other hand, the quotient space $\cQn_{0\ps{n}} = \Dn_{0\ps{n}}/\mTn_{0\ps{n}}$ can be identified with the space of local diffeomorphisms of the form
$$\qeq{X = a(x),\\U = u\\ \text{with}\\a(0) = 0,\\ \det(a^i_j)(0) \ne 0.}$$
These are just the reparametrizations of the zero section, which are extended to be diffeomorphisms with identical reparametrizations of the parallel sections, although the method of extension is unimportant and just selects a particular representative of the quotient space.

Thus, pseudo-groups whose reduced action is free differ from freely acting pseudo-groups only by the inclusion of some additional transformations that belong to the global isotropy subgroup of the section and/or perform reparametrizations. These all preserve the section, and thus do not affect the moving frame calculation nor the computations of differential invariants.

\Ex{redfree}  Suppose $p=q=1$, and consider the Lie pseudo-group action
\Eq{pg1}
$$\qeq{X=x+a,\\ U = f(x,u),}$$
where $f_u \ne 0$.  Since the reduced parametric pseudo-group jet coordinates are $\rX$, $\rU_{x^n}$, $n\geq 0$, and the prolonged action is $\tU_{X^n} = \rU_{x^n}$, this pseudo-group admits a free reduced action.  On the other hand, the pseudo-group \eq{pg1} does not act freely anywhere on the jet space $\Ji$.  When $\meq{p=1, \\q=2}$, the extended pseudo-group
\Eq{pg2}
$$\qeq{X=x+a,\\ U = f(x,u),\\ V = v+b,}$$
is of the same form, and furthermore is intransitive and so has nontrivial differential invariants, namely $v_{x^n}$ for all $n \geq 1$, despite the fact that it does not act freely. On the other hand, when $\meq{p=2,\\ q=1}$, the same pseudo-group
\Eq{pg3}
$$\qeq{X=x+a,\\Y = y+b,\\ U = f(x,u),}$$
acts freely and transitively on the subset  of jet space where $u_y \ne 0$ at all orders $\geq 1$.  We note that the pseudo-groups \eq{pg1} and \eq{pg2} are not reducible, while \eq{pg3} is reducible by virtue of \th{freered}.

%%%%%
\Subsection{m} The Reduced Moving Frame Construction.
%%%%%

Coming back to the construction of a reduced moving frame, this is accomplished by selecting a cross-section $\K\n \subset \Jn$ that is transversal to the orbits of the prolonged group action \eqref{eq: pr action}.  As in most applications, we will always assume that $\K\n$ is a coordinate cross-section defined by fixing $\d\n$ values of the individual jet coordinates $z\n = (x,u\n)$ to suitable constants.  Let
\Eq{indKn}
$$\indKn \subset \set{i, \ (\alpha;J)}{\xeq{\rg ip, \\ \rg \alpha q, \\ |J| \leq n}}$$
denote the set of indices of jet coordinates  of order $\leq n$ that determine the cross-section, which is thus  prescribed by $\d\n = \#\: \indKn$  equations, of the form
\begin{equation}\label{Kn}
\K\n = \set{x^i = c^i,\; u^\alpha_J = c^\alpha_J}{i,\, (\alpha;J) \in \indKn},
\end{equation}
for suitable constants $c^i,c^\alpha_J$.  %At times we will make the abuse of notation $u^\alpha_J \in \indKn$ when referring to a coordinate jet whose indices $(\alpha;J)$ are in $\indKn$.

Given a cross-section \eqref{Kn}, the \is{reduced \ro(right\/\ro) moving frame}\footnote{By an abuse of notation, we use the same symbol to denote the pseudo-group normalization function and the corresponding moving frame section in \df{mvff}. Also, in \eqref{mf formula} the moving frame only specifies the parametric reduced pseudo-group jets, the principal pseudo-group parameters being determined by the reduced determining equations \eq{redde}.} 
\begin{equation}\label{mf formula}
\r{g}\n = \rrho\n(x,u\n)
\end{equation}
gives the reduced pseudo-group element that  maps a submanifold jet $(x,u\n)$ belonging to a suitable neighborhood of the cross-section to the cross-section jet $(\rX,\tU\n)\in \K\n$ that lies in the same pseudo-group orbit.  The freeness assumption in \th{mf existence} guarantees that the reduced pseudo-group element \eqref{mf formula} is uniquely determined.  %Conversely, the pseudo-group inverse $(\r{g}\n)^{-1} = (\rrho\n(X,\tU\n))^{-1}$ defines the reduced left moving frame that sends the cross-section jet $(x,u\n)$ to the submanifold jet $(X,\tU\n)$.

To explicitly determine the moving frame, we apply the cross-section normalizations, in the form $\myeq{\rX^i = c^i,\\\tU^\alpha_J = c^\alpha_J}$, to the corresponding components of the formulas \eqref{explicit pr action} for the prolonged pseudo-group action, and solve the resulting algebraic equations
\begin{equation}\label{norm eq}
P^i(x,u\n,\r{g}\n) = c^i,\qquad
P^\alpha_J(x,u\n,\r{g}\n) = c^\alpha_J,\qquad\text{with}\qquad i,\,(\alpha;J) \in \indKn,
\end{equation}
 for the reduced pseudo-group parameters.  Transversality of the cross-section and freeness of the reduced action guarantee, via the Implicit Function Theorem, that the normalization equations \eqref{norm eq} can be locally solved for $\r{g}\n$ near the cross-section, thereby producing the reduced moving frame \eqref{mf formula}.
%
%\begin{equation}\label{mf formula}
%\r{g}\n = \rrho\n(x,u\n).
%\end{equation}
%
Furthermore, substituting the moving frame expressions \eqref{mf formula} into the formulas \eqref{explicit pr action} for the prolonged action produces the \is{normalized differential invariants}.  Those corresponding to the cross-section coordinates, namely
\[
\qeq{c^i = P^i(x,u\n,\rrho\n(x,u\n)),\\ c^\alpha_J = P^\alpha_J(x,u\n,\rrho\n(x,u\n)) ,\\  i,\; (\alpha;J) \in \indKn,}
\]
reduce, by construction, to the normalization constants, and are known as the \is{phantom differential invariants}, whereas the remaining functions
\begin{equation}\label{norm inv}
\eeq{H^j(x,u\n) = P^j(x,u\n,\rrho\n(x,u\n)),\\ I^\beta_K(x,u\n) = P^\beta_K(x,u\n,\rrho\n(x,u\n)),}\qquad \qquad  j,\; (\beta;K) \notin \indKn,
\end{equation}
form a complete system of functionally independent differential invariants of order $\leq n$, known as the \is{basic normalized differential invariants}, although in what follows ``basic'' will often be dropped. 
%In light of equality \eqref{explicit pr action left}, w
%We extend this terminology to the jet coordinates $u^\alpha_J$.

%\Df{jet df}
%A jet coordinate $u^\alpha_J \in \indK\ii$ is called a \emph{phantom coordinate} while the remaining coordinates $u^\beta_K \notin \indK\ii$ are called \emph{(basic) normalized coordinates}.

\Ex{mf ex}
Returning to our running example, under the assumption that $u_{yy} >0$, a possible cross-section to the second order prolonged action \eqref{pr action ex 2} is given by
\begin{equation}\label{K2ex}
\K^{(2)}=\{\,x=0,\, y=0,\, u=c_0,\, u_x = c_1,\, u_y = d_0,\, u_{xx} = c_2,\, u_{xy} = d_1,\, u_{yy} = 1\,\},
\end{equation}
where $c_0,c_1,c_2,d_0,d_1$ are arbitrary constants.  More generally,
\begin{equation}\label{genKex}
\K\ii = \{\,x=0,\, y=0,\, u_{yy}=1,\, u_{x^k} = c_k,\, u_{x^ky} = d_k,\quad \text{for all} \quad k\geq 0\,\}.
\end{equation}
Following the original papers \rf{OP08,OP09}, and to simplify the computations, we set the arbitrary constants to zero, \ie $c_k=d_k=0$, when computing the moving frame. 
\pari
Referring to the formulas \eqref{pr action ex 2} for  the prolonged action, the normalization equations, up to order two, are obtained by substituting the cross-section determining equations \eqref{K2ex} into the prolonged action:
\Eq{mvfeqsex}
$$\seq{\qeq{0 = \rX,\\ 0 = \rY,\\ 0 = \tU = \rU,\\ 
0 = \tU_X =\frac{\rU_x + (u - \rU) \rU_y}{\rX_x},\\
0 = \tU_Y = \frac{\rU_y}{\rX_x},} \\
0 = \tU_{XX} = \frac{\dsty\ibeq{97.5}{\rU_{xx} - (u_y - \rU_y)\,\rU_x + (u_x-\rU_x)\,\rU_y \\{}+
(u-\rU)(2\: \rU_{xy}+2(u-\rU)u_{yy} + (u_y-\rU_y)\,\rU_x)}}{\rX_x^2},\\
0 = \tU_{XY} = \frac{\rU_{xy} + (u_y - \rU_y)\,\rU_y + (u-\rU)u_{yy}}{\rX_x^2},\qquad
1 = \tU_{YY} = \frac{u_{yy}}{\rX_x^2}.}$$
%
%where the reduced pseudo-group jet coordinates $(\rX,\rY,\rU,\rX_x,\rU_x,\rU_y,\rU_{xx},\rU_{xy})$ are evaluated at the source $(X,Y,\tU(X,Y))$.   
Solving these equations for the reduced pseudo-group parameters yields the reduced second order moving frame
\begin{equation}\label{right mf ex}
\begin{gathered}
\rX = 0,\qquad \rY=0,\qquad \rU = 0,\qquad \rX_x = \sqrt{u_{yy}},\qquad \rU_x = 0,\qquad \rU_y=0,\\
\rU_{xx} = 0,\qquad \rU_{xy} = -\:u u_{yy}.
\end{gathered}
\end{equation}
Substituting the pseudo-group normalizations \eqref{right mf ex} into the right hand side of the formulas \eqref{pr action ex 3} produces the third order normalized differential invariants 
\begin{equation}\label{norm inv ex}
I_{1,2} = \frac{u_{xyy} + uu_{yyy} + 2\:u_yu_{yy}}{u_{yy}^{3/2}},\qquad
I_{0,3} = \frac{u_{yyy}}{u_{yy}^{3/2}}.
\end{equation}
%%
%The left moving frame is obtained by inverting the pseudo-group normalizations \eqref{right mf ex}.  The result is
%%
%\begin{equation}\label{left mf ex}
%\begin{gathered}
%\rX(0) = X,\qquad \rY(0) = Y,\qquad \rU(0) = \tU,\qquad
%\rX_x(0) = \frac{1}{\sqrt{\tU_{YY}}},\\
%\rU_x(0) = \frac{\tU_X + \tU \tU_Y}{\sqrt{\tU_{YY}}},\qquad
%\rU_y(0) = \frac{\tU_Y}{\sqrt{\tU_{YY}}}, \qquad
%\rU_{xy}(0) = \frac{\tU_{XY} + \tU_Y^2 + \tU \tU_{YY}}{\tU_{YY}},\\
%\rU_{xx}(0) = \frac{\tU_{XX} + 2\:\tU\tU_{XY} + 2\:\tU_Y \tU_X + 4\tU \tU_Y^2 -  \tU \tU_Y\tU_X - \tU^2\tU_Y^2}{\tU_{YY}},
%\end{gathered}
%\end{equation}
%where the pseudo-group parameters are now evaluated at the origin $0=(0,0,0)$.  One can verify the validity of \eqref{left mf ex} by substituting the cross-section \eqref{K2ex}, with $c_0=c_1=d_0=d_1=0$, and the pseudo-group normalizations \eqref{left mf ex} into the prolonged action \eqref{pr action ex} to obtain identities.

\Rmk{99r} Since the prolonged pseudo-group transformations \eqref{explicit pr action} only depend on the reduced pseudo-group jets, the moving frame method applies equally well to non-free actions whose reduced action is eventually free.  However, we have, as yet, been unable to come up with any truly interesting examples, beyond the rather trivial ones that are based on \ex{redfree}.  Therefore, as in almost all other treatments of moving frames, we have restricted our attention to pseudo-groups which act freely on an open subset of jet space of suitably high order.

%%%%%
\Section{nf} Normal Forms.
%%%%%

As shown in \rf{OP08; Section 8} (see also \rf{O2018}), the method of (reduced) moving frames can naturally be formulated in terms of power series.  As explained in \sect{sec: jet}, we can identify a submanifold jet $(x,u\ii) \in \Ji$ with a formal power series
\begin{equation}\label{u series}
u^\alpha(y) = \sum_J\> \frac{u^\alpha_J}{J!}\,(y-x)^J,\qquad \alpha=1,\ldots,q,
\end{equation}
centered at the point $x \in \mX$.  By definition, the power series converges to an analytic function in a neighborhood of $x$ if and only if $(x,u\ii) \in A^\infty$.

\Df{normalform} Given an infinite order coordinate cross-section $\K = \K\ii\subset \Ji$, a power series \eqref{u series} is said to be in \is{normal form} if the corresponding submanifold jet lies in the cross-section:  $(x,u\ii) \in \K$.

Thus, the normal form will depend upon the choice of cross-section.
In particular, the normal form power series converges to an analytic function if and only if the corresponding jet $(x,u\ii) \in \K \capz A^\infty$ lies in the analytic part of the infinite order cross-section.  In general, the moving frame method does not make any guarantees that this occurs and so such normal forms are merely formal power series.  The main result of this paper is to establish convergence of normal form power series under suitable assumptions, which include most examples that arise in applications.

In a little more detail, as in \eq{indKn}, let $\indK$ denote the set of indices $i$, $(\alpha;J)$ of jet coordinates that prescribe the coordinate cross-section \eqref{Kn} at order $n=\infty$.  
Thus, the coefficients $u^\alpha_J$ with $(\alpha; J) \in \indK$ represent the normalization constants prescribed by the cross-section, \ie the phantom invariants, which serve to fix the  normal form power series.  The remaining coefficients $u^\beta_K$ with $(\beta; K) \not\in \indK$ will represent the corresponding complete set of basic normalized differential invariants, as described below. We further set
\Eq{indKa}
$$\indKa = \set J{(\alpha; J) \in \indK}\qquad \text{where}\qquad \alpha=1,\ldots,q.$$
We can then extract from \eqref{u series} the \emph{cross-section power series} 
%From the normal form power series \eqref{u series}, we can extract the \emph{cross-section power series}
%
\begin{equation}\label{csSeries}
C^\alpha(y) = \sum_{J\,\in\, \indKa}\> \frac{c^\alpha_J}{J!}\,(y-x)^J,\qquad \alpha=1,\ldots,q,
\end{equation}
whose indicated Taylor coefficients are the normalization constants, \ie the phantom invariants.  If $\indKa$ is a finite set, then $C^\alpha(y)$ is a polynomial, while if $\indKa = \emptyset$, our convention is that  $C^\alpha(y)$ does not exist.  

Consider two sections $s,S \subset M$ of the fibered manifold $\pi\colon M \to \mX$.  In local coordinates, the ``source section'' has the form $s=\{(x,u(x))\}$, while the ``target section'' is given by $S=\{(X,\tU(X))\}$.  
We will assume that the source section represents the normal form, meaning that its jet $(x,u\ii) \in \K$ lies in the cross-section. On the other hand, the target section will be a prescribed analytic section  that we seek to normalize via a suitable pseudo-group diffeomorphism.  In other words, we seek a diffeomorphism  $\varphi \in \G$ such that, locally, $S = \varphi (s)$.  In terms of the reduced pseudo-group, this requires
\begin{equation}\label{nf eq}
\rU = \tU(\rX)\qquad \text{or, more explicitly,}\qquad U(x,u(x)) = \tU(X(x,u(x))).
\end{equation}

\Warnt This is the opposite point of view from that was used  in \sect m to construct the moving frame, where the target submanifold belonged to the cross-section.  Thus, to be in alignment with our current point of view, we should switch the source and target coordinates in the constructions.  While the calculations could clearly be implemented  in this manner from the outset, using a suitable change in notation, in our view the resulting notations are confusing and are at odds with the traditional way moving frames are constructed for both finite-dimensional Lie group actions and infinite-dimensional pseudo-groups.  We will explicitly note when the required switch is necessary. 
In addition, it is preferable, due to other notational considerations, to make the normal form the source submanifold.  

Thus, given a prescribed section with analytic power series
\begin{equation}\label{U series}
\tU^\alpha(Y) = \sum_{J}\> \frac{\tU^\alpha_J}{J!}\, (Y - X)^J,\qquad \alpha=1,\ldots,q,
\end{equation}
centered at the point $X \in \mX$,
the moving frame will map it to a normal form power series \eqref{u series} whose phantom coefficients are constants and whose remaining coefficients are the basic differential invariants, expressed  in terms of the jet coordinates $X^i,\tU^\alpha_J$.  In other words, if $I^\alpha_J(x,u\n)$ is a normalized differential invariant, then the corresponding coefficient in the normal form power series \eqref{u series} is $u^\alpha_J = I^\alpha_J(X,\tU\n)$. In view of \eqref{csSeries}, the normal form power series thus takes the form
\Eq{nfps}
$$u^\alpha(y) = C^\alpha(y) + \sum_{J\not\in \indKa}\> \frac{I^\alpha_J}{J!}\,(y-x)^J = \sum_{J\,\in\, \indKa}\> \frac{c^\alpha_J}{J!}\,(y-x)^J + \sum_{J\not\in \indKa}\> \frac{I^\alpha_J}{J!}\,(y-x)^J,\quad \alpha=1,\ldots,q.$$
%where $I^\alpha_J$ are the basic normalized differential invariants \eqref{norm inv}.}
The key issue to be addressed in this paper is whether the resulting normal form power series \eq{nfps} converges.  In general this is not the case --- for instance, it is trivially not convergent if the cross-section power series $C^\alpha(y)$ do not converge --- and an additional requirement must be imposed. Namely, 
the coordinate cross-section must be ``well-posed'', as formulated in \sect{well-posed}, and the corresponding cross-section power series \eqref{csSeries} must converge.  Fortunately, choosing a well-posed cross-section is not difficult, and leads to a practical algorithm for constructing convergent normal forms for reducible submanifolds. 

\Ex{9}
For our running example, the normal form corresponding to the cross-section \eqref{genKex} is the (formal) Taylor series at the origin corresponding to the function
\begin{equation}\label{nf_ex}
u(x,y) = c(x) + y\: d(x) + \f2\:{y^2}\:w(x,y),
\end{equation}
where $c_k,d_k$ are the Taylor coefficients, at $x=0$, for the scalar functions $c(x),d(x)$, respectively, and where $w(0,0) = 1$. In this example, the cross-section power series is
\Eq{ex9csps}
$$C(x,y) = c(x) + y\: d(x) + \f2\:{y^2}.$$
Except for the constant term, the Taylor coefficients of $w(x,y)$ are the normalized differential invariants.  In the calculations of \ex{mf ex}, we took $c(x) \equiv d(x) \equiv 0$, in which case, the third order invariants \eqref{norm inv ex} are (up to multiple) the coefficients of the linear terms in $w(x,y)$.  In this case, the first few terms of the normal form power series are
$$u(x,y) = \frac{y^2}{2} + \frac{I_{1,2}}{2}\,x\:y^2 + \frac{I_{0,3}}{6}\,y^3 + \frac{I_{2,2}}{4}\,x^2\:y^2 + \frac{I_{1,3}}{6}\,x\:y^3 + \frac{I_{0,4}}{24}\,y^4 + \cdotsx,$$
where 
\begin{equation}\label{norm inv ex U}
I_{1,2} = \frac{\tU_{XYY} + \tU \tU_{YYY} + 2\:\tU_Y \tU_{YY}}{\tU_{YY}^{3/2}},\qquad
I_{0,3} = \frac{\tU_{YYY}}{\tU_{YY}^{3/2}},
\end{equation}
are the third order differential invariants \eqref{norm inv ex} evaluated on the target section $U = \tU(X,Y)$, while $I_{2,2},I_{1,3},I_{0,4}$ are the normalized fourth order differential invariants, again evaluated on the target section, whose explicit formulae can be deduced from \rf{OP08; Example 8.6}.  And similarly at higher order.  
%The key question, to be addressed by our main theorem, is whether the normal form series corresponding to \eqref{nf_ex} converges.

%%%%%
\Subsection{nfde} The Normal Form Determining Equations.
%%%%%

We now formulate a system of differential equations that a normal form must satisfy. These equations are obtained by suitably manipulating the reduced determining equations for the pseudo-group. As noted above, the normal form is denoted by source coordinates, \ie $u = u(x)$, while the prescribed submanifold is written in target coordinates as in \eqref{nf eq}.

Consider the reduced determining equations
\begin{equation}\label{psi red det eq}
\rde\n = \big\{\,\rDelta\n \big(x,u\n,\rX\n,\rU\n\big) = 0\,\big\}
\end{equation}
for the reduced pseudo-group diffeomorphism $\rvarphi(x)=\pab{\rX(x,u(x)),\rU(x,u(x))}$ evaluated on a section $u(x)$. Recall that $u\n, \rX\n,\rU\n$ denote derivatives with respect to the source variables $x$ up to order $n$.  Applying the chain rule to differentiate the first equation in \eqref{nf eq}  yields formulae for the $x$ derivatives of $\rU$ in terms of the $x$ derivatives of $\rX$ and the $X$ derivatives of $\tU$:
\begin{equation}\label{chain rule}
\rU\n = \rH\n(\rX\n,\tU\n),
\end{equation}
where $\tU\n$ denotes the derivatives of $\tU$ with respect to the target independent variables $X$ up to order $n$. These formulae can be explicitly computed by successively applying the \is{chain rule total derivative operators}
\Eq{chaindops}
$$\req{\Dt_{x^i} = \Sum jp \rX^j_{x^i} \, \Dt_{X^j},\\\rg ip,}$$
to each $\tU^ \alpha $.
% for $\rg \alpha q$.
For example, when $p=q=1$, we have  $\Dt_x = \rX_x \, \Dt_X$, and hence, up to order two,
\[
\rU_x = \tU_X \rX_x,\qquad \rU_{xx} = \tU_{XX}\rX_x^2 + \tU_X \rX_{xx}.
\]

Substituting the expressions \eqref{chain rule} into the reduced determining equations
 \eqref{psi red det eq} produces the \emph{normal form determining equations}
\begin{equation}\label{nf det eq}
\nde\n = \big\{\widetilde{\Delta}\n\big(x,u\n,\rX\n,\tU\n\big) = 0\big\}.
\end{equation}
Given a prescribed function $\tU = \tU(\rX)$ defining a submanifold (section), whose derivatives $\tU\n$ are known, we view \eqref{nf det eq} as an \nth order system of differential equations for the unknown functions $\rX(x),u(x)$, the latter, when subject to the appropriate initial conditions, prescribing the normal form of the given submanifold, and our goal is to establish their involutivity.

\Ex{7}
To illustrate the construction, let us compute the normal form determining equations for the Lie pseudo-group \eqref{eq: pg1}. We begin by applying the chain rule total differential operators
\Eq{ex7crop}
$$\qeq{\tD_x = \rX_x \tD_X + \rY_x \tD_Y,\\\tD_y = \rX_y \tD_X + \rY_y \tD_Y,}$$
once and twice to the equation $\rU = \tU$ to produce the first and second order chain rule formulas 
\Eq{21cr}
$$\eeq{\rU_x = \tU_X \rX_x + \tU_Y \rY_x,\qquad \rU_y = \tU_X \rX_y + \tU_Y \rY_y,\\
\rU_{xx} = \tU_{XX} \rX_x^2 + 2\:\tU_{XY} \rX_x\rY_x + \tU_{YY} \rY_x^2 + \tU_X \rX_{xx} + \tU_Y \rY_{xx},\\
\rU_{xy} = \tU_{XX} \rX_x\rX_y + \tU_{XY} (\rX_x\rY_y +\rX_y\rY_x) + \tU_{YY} \rY_x\rY_y + \tU_X \rX_{xy} + \tU_Y \rY_{xy},\\
\rU_{yy} = \tU_{XX} \rX_y^2 + 2\:\tU_{XY} \rX_y\rY_y + \tU_{YY} \rY_y^2 + \tU_X \rX_{yy} + \tU_Y \rY_{yy}.}$$
We substitute these into the reduced determining equations \eqref{eq: red det eq ex1}.  The resulting equations, once simplified, are the normal form determining equations
\begin{equation}\label{nf det eq pg1}
\req{\eeq{
\rX_y = 0,\\ \rY_x = (\tU-u)\rX_x,\\\rY_y = \rX_x,}\\
\eeq{\rX_{xx} = \tU_Y \rX_x^2-u_y\rX_x,\qquad \rX_{xy} = \rX_{yy}=0,\\
\rY_{xx} = \paz{\tU_X + 2\:(\tU-u)\tU_Y}\rX_x^2 - \paz{u_x + (\tU-u)u_y} \rX_x ,\\
\rY_{xy} = \tU_Y\rX_x^2-u_y \rX_x,\qquad
\rY_{yy} = 0,\qquad u_{yy} = \tU_{YY} \rX_x^2.}}
\end{equation}
Observe that the parametric derivatives are $\rX,\rY,u,\rX_x,u_x,u_y,u_{xx},u_{xy}$, while the left hand sides are the principal derivatives.
\pari
For later use in \ex{ex: I,J pg1}, we further compute the normal form determining equations of order three.  The most direct way to perform this computation is to apply the chain rule operators \eq{ex7crop} to directly differentiate the second order equations \eqref{nf det eq pg1}. The result is
\begin{equation}\label{order 3 nf det eq pg1}
\hskip10pt\eeq{\rX_{xxx} =  (u_y^2 - u_{xy})\:\rX_x - 3\:u_y \tU_Y \rX_x^2+ (\tU_{XY} + 2\:\tU_Y^2+(\tU-u)\tU_{YY}) \:\rX_x^3,\\
\rX_{xxy} = \rX_{xyy} = \rX_{yyy} = \rY_{xyy}=\rY_{yyy}=0,\\
\rY_{xxx} =  (2u_x u_y - u_{xx}  + (\tU-u)(u_y^2-u_{xy}))\rX_x - 3\:( u_x \tU_Y + u_y \tU_X + 2\:(\tU-u)u_y\tU_Y)\rX_x^2 \hskip-1in\cthn{30}
+ (\tU_{XX}+4\:\tU_X \tU_Y + 3\:(\tU-u)(\tU_{XY}+2\:\tU_Y^2) + 2(\tU-u)^2\tU_{YY})\rX_x^3,\\
\rY_{xxy} = (u_y^2-u_{xy})\rX_x - 3\:u_y \tU_Y \rX_x^2 + (\tU_{XY} + 2\:\tU_Y^2 + (\tU-u) \tU_{YY})\rX_x^3,\\
u_{xyy} = -2\:u_y \tU_{YY}\rX_x^2 + (\tU_{XYY} + 2\:\tU_Y \tU_{YY} - u \:\tU_{YYY} + \tU \tU_{YYY})\rX_x^3,\\ u_{yyy} = \tU_{YYY} \rX_x^3,}
\end{equation}
with parametric derivatives $\rX,\rY,u,\rX_x,u_x,u_y,u_{xx},u_{xy},u_{xxx},u_{xxy}$.

To investigate involutivity of the normal form determining equations, we linearize at the identity jet, keeping in mind that $\rX\n$ and $u\n$ vary, while $\tU\n$ is fixed.  The vector field used for linearization is
\[
\sum_{0\leq|J|\leq n} \bigg(\sum_{i=1}^p \>\rxi^i_J \,\pp{}{\rX^i_J}+ \sum_{\alpha=1}^q \>\psi^\alpha_J \,\pp{}{u^\alpha_J}\bigg).
\]
We begin by linearizing the chain rule formula \eqref{chain rule},  writing out its individual components.

\Lm{lincov}  For any $\rg \alpha q$ and multi-index $J = \psubs jn$, the linearization of the chain rule equation
\Eq{rUaJ}
$$\rU^\alpha _J = \rH^\alpha _J(\rX\n,\tU\n)$$
at the identity jet is
\Eq{lincov}
$$\rphi ^\alpha _J = \Dt_{x}^J \Pa{\Sum ip u^\alpha_i \:\rxi^i} - \Sum ip u^\alpha_{J,i} \:\rxi^i .$$

\Proof
Linearization at the identity amounts to computing the infinitesimal generator of a one parameter group.  In the case of \eqref{nf eq}, the group can be identified with the induced action of the inverse of the change of independent variables prescribed by $X = \rX(x) = X(x,u(x))$ on the dependent variables $u$; for details, see the discussion on pages 105--106 of \rf O.  Because we are dealing with the inverse, the infinitesimal generator is
$$ - \, \Sum ip \rxi^i \pp{}{x^i},$$
which only acts on the independent variables.
Linearizing the induced action on the derivatives \eq{rUaJ} is the same as computing the prolongation of this vector field, which, according to \rf{O; Theorem 2.36} is exactly given by the prolongation formula \eq{lincov}, the quantity in parentheses being its characteristic.
 \qed
 
\Ex{lincovex}
Linearizing the particular chain rule formulas \eq{21cr} at the identity, where \eq{x1id} holds, produces
\Eq{21s}
$$\hskip-30pt\seq{\rphi_x = u_x\: \rxi_x + u_y \:\reta_x = \Dt_x(u_x\: \rxi + u_y \:\reta) - (u_{xx}\: \rxi + u_{xy} \:\reta),\\ \rphi_y = u_x \:\rxi_y + u_y \:\reta_y= \Dt_y(u_x\: \rxi + u_y \:\reta) - (u_{xy}\: \rxi + u_{yy} \:\reta),\\
\rphi_{xx} = u_x \:\rxi_{xx} + u_y \:\reta_{xx} + 2\:u_{xx} \:\rxi_x + 2\:u_{xy} \:\reta_x = \Dt_x^2(u_x\: \rxi + u_y \:\reta) - (u_{xxx}\: \rxi + u_{xxy} \:\reta),\\
\rphi_{xy} = u_x \:\rxi_{xy} + u_y \:\reta_{xy} + u_{xx} \:\rxi_x + u_{xy} (\rxi_y+ \reta_x) + u_{yy}\: \reta_x = \Dt_x\Dt_y(u_x\: \rxi + u_y \:\reta) \!-\! (u_{xxy}\: \rxi + u_{xyy} \:\reta),\hskip-1in\\
\rphi_{yy} = u_x \:\rxi_{yy} + u_y \:\reta_{yy} + 2\:u_{xy} \:\rxi_y + 2 \:u_{yy}\:\reta_y = \Dt_y^2(u_x\: \rxi + u_y \:\reta) - (u_{xyy}\: \rxi + u_{yyy} \:\reta),}$$
which are in accordance with the general \eqf{lincov}.

\Th{lnfde} The linearization of the normal form determining equations \eqref{nf det eq} at the identity, where $(\rX,u\n) = (x,\tU\n)$, coincides with the linearization of the reduced determining equations \eqref{psi red det eq} at the identity $(\rX,\rU\n) = (x,u\n)$ after the substitutions
\Eq{lnfdesubs}
$$\qeq{\rphi^ \alpha _J \longmapstox \Dt_x^J \Pa{\Sum ip u^\alpha_i \rxi^i} - \Sum ip u^\alpha_{J,i} \rxi^i - \psi^ \alpha _J,\\\rg \alpha q, \\ J = \psubs jk.}$$

The proof of \th{lnfde} appears after the following illustrative example.

\Rmk{11r}
The linearization of the normal form determining equations \eqref{nf det eq} in \th{lnfde} occurs at the point $(x,\tU\n)$.  But since  $u\n = \tU\n$ at the identity, we may substitute $u\n$ for $\tU\n$ in the linearization, which is implicitly done in \th{lnfde}.

\Ex{7l}
Returning to \ex7, let us linearize the normal form determining equations \eqref{nf det eq pg1} and \eqref{order 3 nf det eq pg1} at the identity transformation. To do so, we apply the vector field
\[
\sum_{i,j=0}^\infty\> \bigg(\rxi_{ij}\pp{}{\rX_{ij}} + \reta_{ij}\pp{}{\rY_{ij}} + \psi_{ij}\pp{}{u_{ij}}\bigg)
\]
to the equations and then set $\meq{\rX = x,\\ \rY = y,\\ \tU_{X^i Y^j}=u_{x^iy^j}}$ for all $ i,j \geq 0$. At order $2$, this yields the linear system
\begin{equation}\label{order 2 lin nf det eq pg1}
\eeq{
\rxi_y = 0,\qquad \reta_x = -\psi,\qquad \reta_y = \rxi_x,\qquad
\rxi_{xx} = u_y \:\rxi_x - \psi_y,\qquad \rxi_{xy} = \rxi_{yy} = 0,\\
\reta_{xx} = u_x \:\rxi_x - u_y \:\psi - \psi_x,\qquad
\reta_{xy} = u_y \:\rxi_x - \psi_y,\qquad \reta_{yy} =0,\qquad
\psi_{yy} = 2\:u_{yy}\:\rxi_x,}
\end{equation}
while at order $3$ we append the equations
\begin{equation}\label{order 3 lin nf det eq pg1}
\begin{aligned}
\rxi_{xxx} &= (2\:u_{xy}+ u_y^2)\:\rxi_x - \psi_{xy} - u_y \psi_y - u_{yy} \psi,& &\hspace{-1cm}\rxi_{xxy} = \rxi_{xyy} = \rxi_{yyy} = 0,\\
\reta_{xxx}&= (2\:u_{xy} + 2\: u_x u_y)\:\rxi_x - \psi_{xx} - u_y \psi_x - u_x \psi_y -(2\:u_{xy} + u_y^2)\:\psi,& &\\
\reta_{xxy} &= (2\:u_{xy}+u_y^2)\rxi_x - \psi_{xy} - u_y\psi_y - u_{yy}\psi,& &\hspace{-1cm}\reta_{xyy} = \reta_{yyy} =0,\\
\psi_{xyy} &= (3\:u_{xyy} + 2\:u_y u_{yy})\:\rxi_x - 2\:u_{yy}\psi_y - u_{yyy}\psi,& &\hspace{-1cm}
\psi_{yyy} = 3\:u_{yyy}\rxi_x.
\end{aligned}
\end{equation}

\Proofth{lnfde}
In view of \eq{rde0}, the linearized reduced determining equations have the form
\Eq{lrde0}
$$\rL_\nu = \Sum ip \Sumu{0 \leq |J| \leq  n} A_{\nu,\mathds{1}}^{i,J}\, \rxi^i_{J} + \Sum \alpha q \Sumu{0 \leq |K| \leq  n} B_{\nu,\mathds{1}}^{\alpha ,K}\, \rphi^ \alpha _K,$$
where the additional $\mathds{1}$ subscript means that we evaluate the indicated coefficients at the identity.  On the other hand, substituting \eq{rUaJ} into \eq{rde0}, we deduce that the normal form determining equations take the form
$$\ibeq{200}{\widetilde\Delta _\nu = \Sum ip \Bk{\widetilde A_\nu^i (\rX^i -  x^i)+ \widetilde A_\nu^{i,i}(\rX^i_{i} - 1) + \Sumu{\scriptstyle J \ne i \atop \scriptstyle 1 \leq |J| \leq n} \widetilde A_\nu^{i,J} \rX^i_{J}} \\{}+ \Sum \alpha q \Sumu{0 \leq |K| \leq  n} \widetilde B_\nu^{\alpha ,K} \bbk{\rH^ \alpha _K(\rX\k,\tU\k) - u^ \alpha _K},}$$
whose coefficients are obtained from those of \eq{rde0} by using the chain rule substitution  \eqref{chain rule}.
Linearizing the latter expressions at the identity, using \eq{lincov}, and noting that at the identity  \eqref{chain rule} reduces to $\rU\n = \tU\n = u\n$, produces
\Eq{lnde0}
$$\widetilde L_\nu = \Sum ip \Sumu{0 \leq |J| \leq  n} A_{\nu,\mathds{1}}^{i,J}\, \rxi^i_{J} + \Sum \alpha q \Sumu{0 \leq |K| \leq  n} B_{\nu,\mathds{1}}^{\alpha ,K}\,\Bk{\Dt_x^K \Pa{\Sum ip u^\alpha_i \rxi^i} - \Sum ip u^\alpha_{K,i} \rxi^i - \psi^ \alpha _K}.$$
Comparing \eqas{lrde0}{lnde0} completes the proof. \qed

\Rmk{12r}
Inverting the substitutions \eq{lnfdesubs} for $\psi^\alpha_J$, we recover the usual formula for the prolongation of the vector field
\begin{equation}\label{restricted v}
- \,\overline{\vv} = - \Pa{ \Sum ip \rxi^i(x) \pp{}{x^i} +  \Sum \alpha q \rphi^\alpha(x) \pp{}{u^\alpha}}
\end{equation}
to jet space.  More explicitly, recall from \rf{O} that the \nth order prolongation of $\overline{\vv}$ is the vector field
\begin{equation}\label{prv}
\overline{\vv}\n = \Sum ip \rxi^i \pp{}{x^i} + \sum_{\alpha=1}^q \sum_{0\leq |J| \leq n} \widehat{\phi}^\alpha_J \pp{}{u^\alpha_J},
\end{equation}
where the prolonged vector field coefficients are given by the formula
\begin{equation}\label{prCoeff}
\widehat{\phi}^\alpha_J = \rphi^\alpha_J - \Dt_x^J \Pa{\sum_{i=1}^p \rxi^i u^\alpha_i} + \sum_{i=1}^p \rxi^i\: u^\alpha_{J,i}.
\end{equation}
Then, under the substitution \eq{lnfdesubs}, the prolonged vector field $- \,\overline{\vv}\n$ given in \eqref{prv} is mapped to the vector field
\begin{equation}\label{v equality}
\widetilde{\vv}\n = -\,\Sum ip \rxi^i \pp{}{x^i} + \sum_{\alpha=1}^q \sum_{0\leq |J|\leq n} \psi^\alpha_J\,\pp{}{u^\alpha_J}.
\end{equation}

As an immediate corollary, we are able to characterize the involutivity of the normal form determining equations.

\Th{infdeq} For any order $n$, if the reduced determining equations are involutive, then so are the normal form determining equations.

\Proof
Keeping only the highest order terms in the substitution \eq{lnfdesubs} we have, at the level of the symbol, that
\begin{equation}\label{symbol sub}
\rphi^\alpha_J \longmapstox \Sum ip u_i \,\rxi^i_J - \psi^\alpha_J.
\end{equation}
Using the freedom that still remains within a given class, we order the columns of the symbol matrix $M_{\rG}^n$ of the reduced determining equations so that the columns associated to $\rphi^\alpha_J$ are to the left of the columns corresponding to $\rxi^i_K$ when $\cls J = \cls K$.  Now consider the row-echelon form $M_{\rG,\text{REF}}^n$ of the symbol matrix.
If $\rphi^\alpha_J$ is a pivot of $M_{\rG,\text{REF}}^n$, then, under the substitution \eqref{symbol sub}, $\psi^\alpha_J$ is a pivot of the row-echelon  symbol matrix $M_{\nde,\text{REF}}^n$ of the normal form determining equations.  If $\rxi^i_J$ is a pivot of $M_{\rG,\text{REF}}^n$, then all the matrix components to the left of $\rxi^i_J$ in that row are zero.  Based on our ordering of the columns of $M_{\rG}^n$, the substitution \eqref{symbol sub} does not alter the fact that $\rxi^i_J$ is a pivot of $M_{\nde,\text{REF}}^n$.  Therefore, the symbol matrices for the reduced determining equations  and the normal form determining equations have the same indices and ranks and involutivity of $M_{\rG}^n$ implies involutivity of $M_{\nde}^n$.  

Next, the normal form determining equations are integrable as any integrability condition would map back to an integrability condition of the reduced determining equations.  Therefore, the normal form determining equations are involutive. \qed
%{\color{red}Question: What about integrability conditions?} \pjoc{I think it suffices to say that any integrability condition in the new variables would map back to an integrability condition in the original variables which, by assumption, do not exist.} \qed
%The substitution \eq{lnfdesubs} is an invertible linear map, which does not alter the algebraic properties of the symbol, and hence does not affect its involutivity. \qed

%%%%%
\Section{imf} Involutivity and Reduced Moving Frames.
%%%%%

We now have reached the heart of the paper where we complete the proof of our general convergence result for normal forms of submanifolds.  The key remaining step is to establish compatibility of the cross-section normalizations producing the moving frame  with the involutivity of the normal form determining system.  The main complication is that they are not necessarily compatible at low order. However, as we will demonstrate, once we are beyond the order of freeness of the prolonged pseudo-group action and the order of involutivity of the normal form determining equations, this identification can be made.  Indeed, this is to be suspected, since the order of freeness is also where the algebraic moving frame constructions used in \rf{OP09} apply.  As noted in \rf{OP09}, the finite number of normalizations imposed at or below the order of freeness are not, in general, compatible with the algebraic framework used to establish generating sets of differential invariants and syzygies, and so must be appended to the former to obtain a complete system of differential invariants. Here we will see a similar behavior within the involutivity framework. Before establishing this connection, we illustrate the incompatibility at low order by revisiting our running example.

The involutivity of the normal form determining equations \eqref{nf det eq} relies on the class-based ordering of multi-indices, which imposes some restrictions on which jet coordinates are parametric and principal.  For example, in the normal form determining equations \eqref{nf det eq pg1}, the equation
\[
u_{yy} = \tU_{YY} \rX_x^2
\]
is solved for $u_{yy}$ since it is a principal derivative according to the involutivity framework.  On the other hand, recalling the moving frame computations in \ex{mf ex} and the fact that we switch the source and target so that the source jet coordinates are in the cross-section, the normalization equation is written as
\[
\tU_{YY} = \frac{u_{yy}}{\rX_x^2} = \frac{1}{\rX_x^2}
\roq{and is to be solved for}
\rX_x = \sqrt{\frac{u_{yy}}{\tU_{YY}}}.
%= \frac{1}{\sqrt{\tU_{YY}}}.
\]
Thus, the same equation may be solved for different jet coordinates depending on whether we implement the involutivity formalism or the moving frame construction.
At the level of the linearized equations, the equation in question is
\[
2\: u_{yy}\:\rxi_x - \psi_{yy} = 0.
\]
Since the symbol of the equation is $\psi_{yy}=0$, involutivity involves solving for $\psi_{yy}$, while the moving frame construction requires solving for $\rxi_x$.

The aim of this section is to show that, while they may differ at low order, if the normal form determining equations are prolonged beyond the order of freeness of the prolonged pseudo-group action, then the determination of the parametric derivatives via the symbol of the normal form determining equations is compatible with the moving frame construction provided we use an appropriately well-posed cross-section.

%%%%%
\Subsection{prolonged N} Beyond the Order of Freeness.
%%%%%

Let $\nf\geq 1$ be the order of freeness of the prolonged Lie pseudo-group action.  In this section we describe the structure of the order $\nf+1$ normal form determining equations.

To simplify the exposition, we assume that the pseudo-group acts transitively on the space of independent variables $\mX$.  This implies that the order zero reduced pseudo-group jet coordinates $\rX$ are parametric parameters in the normal form determining equations. Freeness implies that at order $\nf$ all the reduced horizontal pseudo-group jet coordinates $\rX^i_J$ of orders $1 \leq |J| \leq \nf$ can be normalized by a suitable choice of cross-section.  On the other hand, this implies that the same jet coordinates can be solved for in the normal form determining system $\nde\ps{\nf}$:
%
%\begin{subequations}\label{nde nf}
\begin{equation}\label{XiJ norm eq}
\rX^i_J = \Xi^i_J(x,\tU^{(\nf)},\rX,\ldots,u^\beta_K,\ldots),\qquad i=1,\ldots,p,\quad 1\leq |J|\leq \nf,
\end{equation}
where $u^\beta_K$ are parametric normal form jet coordinates of order $|K|\leq \nf$. In particular, no derivatives $\rX^j_K$ of order $|K| \geq 1$ appear on the right hand side of these equations.   The remaining equations in $\nde\ps{\nf}$ will specify the principal normal form jet coordinates
\begin{equation}\label{ualphaI}
u^\alpha_J = \Delta^\alpha_J(x,\tU^{(\nf)},\rX,\ldots,u^\beta_K,\ldots),\qquad\text{where}\qquad  1\leq \alpha \leq q,\quad  |J|\leq \nf,
\end{equation}
%\end{subequations}
%
and $u^\beta_K$ are again parametric derivatives of order $|K|\leq \nf$.  Since equations \eqref{XiJ norm eq}, \eqref{ualphaI} are obtained by implementing the reduced moving construction, these equations are not necessarily class-respecting at order $\nf$. This means that the class of the parametric normal form jet coordinates of order $\nf$ on the right hand side of an equation may be greater than the class of the order $\nf$ derivative occurring on the left hand side of the same equation.  

To obtain class-respecting equations for the reduced horizontal pseudo-group jet coordinates, we differentiate the equations  in \eqref{XiJ norm eq} for the reduced pseudo-group parameters $\rX^i_J$ of order $|J|=\nf$ with respect to the multiplicative variables $j \leq \cls(J)$, thereby obtaining the following subset of normal form determining equations:
%
%\begin{subequations}\label{nf+1 nfeq}
\begin{equation}\label{XiJj-tmp}
\rX^i_{J,j} = \Xi^i_{J,j}(x,\tU^{(\nf+1)},\rX,\ldots,u^\beta_K,\ldots,u^\beta_{K,j},\ldots),\qquad i=1,\ldots,p,\quad |J|=\nf,
\end{equation}
of order $\nf+1$.
Note that when we differentiate, the resulting expressions include the first order derivatives $\rX^i_j$, but these can be replaced by their expressions in  \eqref{XiJ norm eq} and hence the right hand sides of the resulting equations continue to be independent of the derivatives of the $\rX^i$.  We also note that all reduced horizontal pseudo-group jet coordinates of order $\nf+1$ appear on the left hand sides of \eqref{XiJj-tmp}.  In other words, all reduced horizontal pseudo-group parameters $\rX^i_J$ of order $|J|=\nf+1$ are principal.  Also, the class of $\rX^i_{J,j}$ in \eqref{XiJj-tmp} is now
$\cls(J,j) = j$,
and the normal form jet coordinates $u^\beta_{K,j}$ on the right hand side of \eqref{XiJj-tmp} satisfy the class requirement
\[
\cls(K,j) = \min\{\cls(K),j\} \leq j.
\]
Therefore, the equations \eqref{XiJj-tmp} are class-respecting.  The remaining order $\nf+1$ normal form determining equations are equations specifying the order $\nf+1$ principal normal form jet coordinates
\begin{equation}\label{u normal form eq}
u^\alpha_J =  \Delta^\alpha_J(x,\tU^{(\nf)},\rX,\ldots,u^\beta_K,\ldots),\qquad 1\leq \alpha \leq q,\quad |J| =  \nf+1,
\end{equation}
%\end{subequations}
%
where we used \eqref{XiJ norm eq}, \eqref{XiJj-tmp} to remove the reduced horizontal pseudo-group parameters $\rX^i_J$ of orders $1\leq |J|\leq \nf+1$.  Without loss of generality, we can assume that the equations \eqref{u normal form eq} are class-respecting.  As we will see in the next section, this can be achieved by considering the vertical symbol of the normal form determining equations, which we now introduce.

%%%%%
\Subsection{symbols} Vertical and Prolonged Annihilator Symbols.
%%%%%

Let $\LNn$ denote the linearization of the $n$-th order normal form determining equations \eqref{nf det eq} at the identity, and let
\[
\symNn = \bH(\LNn)
\]
be its symbol.  Also, let $\symMNn$ be the corresponding symbol matrix.  We first fix some of the freedom that exists when ordering the columns of $\symMNn$ within a fixed class.  To be compatible with the moving frame construction, we require the columns associated to the reduced pseudo-group parameters $\rX^i_J$ to appear to the left of the columns corresponding to the normal form jet coordinates $u^\beta_K$ when $\cls K = \cls J$.  This ordering stems from the fact that, in the moving frame method, we prioritize solving for the reduced horizontal pseudo-group parameters $\rX^i_J$ over the normal form jet coordinates $u^\beta_K$ within a fixed class. We note that this convention is the opposite of that used in the proof of \th{infdeq}.  But as we show in \sect{Consistency}, this discrepancy becomes immaterial once we pass beyond the order of freeness of the prolonged pseudo-group action.

Since all the reduced horizontal pseudo-group parameters $\rX^i_J$ of order $|J|=\nf+1$ are principal variables in the order $\nf+1$ normal form equations \eqref{XiJj-tmp}, \eqref{u normal form eq}, the involutivity of  $\nde\ps{\nf+1}$ is solely dependent on the equations \eqref{u normal form eq}, which relate the normal form jets.  This observation leads us to introduce the \is{$n$-th order vertical symbol}
\Eq{vsnfn}
$$\vsnf^n = \symNn\capz \spn \{\psi\n\} ,$$
consisting of all the equations in the $n$-th order symbol that only depend on the coefficients $\psi^\alpha_J$ of order $|J|=n$.  Combining these spaces, we define the \emph{vertical symbol}
\begin{equation}\label{vsnf}
\vsnf = \bigcup_{n=0}^\infty \>\vsnf^n.
\end{equation}

\Rmk{} Coming back to the system \eqref{u normal form eq}, the principal normal form jet coordinates are indexed by the pivots of the row reduced order $\nf+1$ vertical symbol $\vsnf^{\nf+1}_{\REF}$.

\Ex{8}
In our running example, keeping only the highest order terms in the linearized equations \eqref{order 2 lin nf det eq pg1}, we obtain the symbol equations
\begin{gather*}
\rxi_y=0,\qquad \reta_x = 0,\qquad \reta_y = \rxi_x,\\
\rxi_{xx} = \rxi_{xy} = \rxi_{yy}=0,\qquad
\reta_{xx} = \reta_{xy} = \reta_{yy} = 0,\qquad \psi_{yy}=0.
\end{gather*}
Therefore, the vertical symbols of order $\leq 2$ are
\[
\vsnf^0 = \vsnf^1 = \emptyset, \roq{and} \vsnf^2 = \{\psi_{yy} = 0\}.
\]
Similarly, from the order three linearized equations \eqref{order 3 lin nf det eq pg1}, we find that the order three vertical symbol is $\vsnf^3 = \{\psi_{xyy} =  \psi_{yyy} = 0\}$, and more generally,
\[
\vsnf^n = \{\,\psi_{x^j y^{n-j}} = 0\;|\; 0\leq j \leq n-2\,\} \forq n\geq 2.
\]

Upon row reducing the vertical symbol $\vsnf$, the pivots of $\vsnf\REF$ identify principal normal form jet coordinates in accordance with the theory of involutivity.
Now the question is whether this identification of principal normal form jet coordinates is compatible with the  moving frame construction.  To answer this question, we introduce the $n$-th order \emph{prolonged annihilator subbundle}
\begin{equation}\label{Zn}
\mZ\n =  \LNn \capz \spn \bbc{\rxi,\psi\n}
\end{equation}
containing the linearized normal form equations that only depend on $\rxi$ and $\psi\n$.  The name for \eqref{Zn} originates from the observation that $\{\rxi,\psi\n\}$ are the coordinates of the prolonged vector field \eqref{v equality}, and that equations in $\mZ\n$ are linear combinations of $\{\rxi,\psi\n\}$ that vanish. To better understand the origin of \eqref{Zn}, we recall that a function $I(x,u\n)$ is a differential invariant of $\r{\G}$ if and only if it is annihilated by all prolonged infinitesimal generators \eqref{prv} of the reduced pseudo-group action. In view of the alternative form \eqref{v equality}, this is equivalent to the infinitesimal constraint
\begin{equation}\label{infinv}
\vvn\n(I) = - \sum_{i=1}^p \>\rxi^i\pp{I}{x^i} + \sum_{\alpha=1}^q \sum_{0\leq |J| \leq n} \psi^\alpha_J \pp{I}{u^\alpha_J} = 0,
\end{equation}
from which we deduce the following result.

\Th{invariant}
If $I(x,u\n)$ is a differential invariant, then the infinitesimal invariance equation $\vvn\n(I)=0$ belongs to the \nth order prolonged annihilator subbundle $\mZ\n$.

\begin{proof}
By definition, $\mZ\n$ contains all the linear combinations of $\rxi$ and $\psi\n$ that vanish.  Since the infinitesimal invariance criterion \eqref{infinv} is of this form, it must belong to $\mZ\n$.
\end{proof}

Applying \th{invariant} to the basic normalized differential invariants \eqref{norm inv}, evaluated at the source variables $(x,u\n)$ rather than the target variables $(X,\tU\n)$, we conclude that the infinitesimal invariance conditions
\Eq{vvnI}
$$\vvn\n(H^j) = \vvn\n(I^\beta_K)=0,\qquad\text{with}\qquad j,\; (\beta;K)\notin \indKn,$$
are equations in $\mZ\n$.  Since the basic normalized differential invariants form a complete set of functionally independent differential invariants  of order $\leq n$, it follows that, at each regular jet,
\Eq{mZnI}
$$
\mZ\n\at\zn = \set{\vvn\n(H^j)\at\zn=\vvn\n(I^\beta_K)\at\zn = 0}{j,\; (\beta;K) \notin \indKn}.$$

\Rmk Z One needs to be a little careful here.  Not every equation defining $\mZ\n$ is necessarily of the form \eq{vvnI} as its coefficients need not be partial derivatives of some function.  On the other hand, \eq{mZnI} says that, at a fixed regular jet, the linear subvariety defined by the differential invariant conditions \eq{vvnI} coincides with the $n$-th order prolonged annihilator subbundle $\mZ\n$.

Keeping only the highest order terms in \eqref{Zn}, we introduce the \is{$n$-th order prolonged annihilator symbol}
$$\disym^n = \bH(\mZ\n).$$
Since $\rxi$ has order zero, it follows that for $n\geq 1$, the $n$-th order prolonged annihilator symbol $\disym^n$ only involves linear equations in $\psi^\alpha_J$ of order $|J|=n$.

\Ex{ex: I,J pg1}
\label{examp-psi=gamma}
Recalling the linearized normal form determining equations \eqref{order 2 lin nf det eq pg1}, \eqref{order 3 lin nf det eq pg1}, we conclude that  when $u_{yy}\neq 0$,
\begin{equation}\label{mZ3 ex}
\mZ^{(3)} = \bigg\{\psi_{yyy} = \frac{3\: u_{yyy}}{2\:u_{yy}}\psi_{yy},\quad
\psi_{xyy} = \bigg(\frac{3}{2}\frac{u_{xyy}}{u_{yy}}+u_y\bigg)\psi_{yy} - 2\:u_{yy} \psi_y - u_{yyy} \psi\bigg\}.
\end{equation}
We observe that, in accordance with the preceding remarks, the equations in \eqref{mZ3 ex} can also found by imposing the infinitesimal invariance conditions \eq{vvnI} for the normalized invariants \eqref{norm inv ex}, re-expressed in terms of the source variables $(x,u\n)$.
Keeping only the highest order terms,
\begin{equation}\label{low order vert symbols pg1}
\disym^0 = \disym^1=\disym^2=\emptyset, \roq{while}\disym^3 = \{\psi_{xyy} = \psi_{yyy} = 0\} = \vsnf^3.
\end{equation}
More generally, $\disym^n = \vsnf^n$ for all $n\geq 3$.
\pari
On the other hand, when $u_{yy} = 0$, we have
\[
\mZ^{(2)} =\{ \psi_{yy} = 0\} \qquad\text{and}\qquad
\mZ^{(3)} = \{ \psi_{yy} = \psi_{xyy} = \psi_{yyy} = 0\},
\]
so that
\[
\disym^2 = \{\psi_{yy} =0\}\qquad\text{and}\qquad
\disym^3 = \{\psi_{xyy} = \psi_{yyy} = 0\}.
\]
In this case, the equality $\disym^n = \vsnf^n$ holds for all $n\geq 2$.

\Rmk{13r}
It is worth reiterating that all the symbol computations are done at a fixed jet, whose dependence has been omitted throughout the paper.  The last example reminds us that we need to pay attention to the base jet when performing computations, which can vary from point to point.  This is important when, for example, analyzing singular normal forms, \rf{OSV}.

\Ex{n2} As a second example, consider the Lie pseudo-group
\begin{equation}\label{n2pg}
X = f(x),\qquad Y = \lambda\, y,\qquad U = u+b,\qquad V=v+c,
\end{equation}
where $f \in \D(\R)$, $\lambda >0$, and $b,c \in \mathbb{R}$. Here we assume that $p=q=2$ with $u=u(x,y)$, $v=v(x,y)$.  Working under the assumption that $y\neq 0$, the normal form determining equations $\nde^{(2)}$ of order two are
\begin{gather*}
\rX_x = \frac{u_x}{\tU_X},\quad \rX_y=0,\quad
\rY_x = 0,\quad \rY_y = \frac{\rY}{y},\quad
u_y = \frac{\rY\:\tU_Y}{y},\quad v_x = \frac{u_x\: \tV_X}{\tU_X},\quad
v_y = \frac{\rY \:\tV_Y}{y},\\
\rX_{xx} = \frac{u_{xx}}{\tU_X} - \frac{u_x^2\: \tU_{XX}}{\tU_X^3},\quad \rX_{xy} = \rX_{yy} = \rY_{xx} = \rY_{xy} = \rY_{yy} = 0,\quad u_{xy} = \frac{u_x \rY \:\tU_{XY}}{y\: \tU_X},\\
\xeq{u_{yy} = \frac{\rY^2 \:\tU_{YY}}{y^2},\\
v_{xx} = \frac{u_{xx} \:\tV_X}{\tU_X} + u_x^2\bigg(\frac{\tV_{XX}\tU_X - \tU_{XX} \tV_X}{\tU_X^3}\bigg),\\
v_{xy} = \frac{u_x \rY \:\tV_{XY}}{y\: \tU_X},\\ v_{yy} = \frac{\rY^2 \:\tV_{YY}}{y^2}.}
\end{gather*}
We remark that the equations for $u_y$, $v_x$, $v_y$, $\ldots$, can be obtained by successively applying the chain rule operators
$$\req{\tD_x = \rX_x \tD_X = \frac{u_x}{\tU_X}\,\tD_X,\\\tD_y = \rY_y \tD_Y = \frac\rY y\, \tD_Y,}$$
to the last two transformations in \eqref{n2pg}.  Linearization at the identity yields the system of linear equations $\LN{2}$ given by
\begin{gather*}
\rxi_x = \frac{\psi_x}{u_x},\quad \rxi_y=0,\quad \reta_x=0,\quad
\reta_y = \frac{\reta}{y},\quad \psi_y = \frac{u_y}{y},\quad \gamma_x = \frac{v_x}{u_x}\,\psi_x ,\quad \gamma_y = \frac{v_y}{y}\,\reta,\\
\rxi_{xx} = \frac{\psi_{xx}}{u_x}-2\frac{u_{xx}\psi_x }{u_x^2},\quad
\rxi_{xy}=\rxi_{yy}=\reta_{xx}=\reta_{xy}=\reta_{yy}=0,\quad
\psi_{xy} =u_{xy}\bigg( \frac{\psi_x}{u_x} + \frac{\reta}{y}\bigg),\\
\xeq{\psi_{yy} = 2\frac{u_{yy}}{y}\,\reta,\\
\gamma_{xx} = \frac{v_x}{u_x}\,\psi_{xx}  + 2\bigg(\frac{v_{xx}u_x-u_{xx}v_x}{u_x^2}\bigg)\psi_x,\\
\gamma_{xy} = v_{xy}\bigg(\frac{\psi_x}{u_x} + \frac{\reta}{y}\bigg),\\
\gamma_{yy} = 2\frac{v_{yy}}{y}\,\reta,}
\end{gather*}
where $\rxi$, $\reta$, $\psi$, $\gamma$ denote the linearizations of $\rX$, $\rY$, $u$, $v$, respectively.  Up to order two, the symbols are
\begin{align*}
\symN{0} &= \emptyset,\qquad
\symN{1} = \bigg\{\rxi_x = \frac{\psi_x}{u_x},\quad \rxi_y=\reta_x=\reta_y=0,\quad \psi_y=0,\quad \gamma_x=\frac{v_x}{u_x}\,\psi_x ,\quad \gamma_y=0\bigg\},\\
\symN{2} &= \bigg\{\rxi_{xx} = \frac{\psi_{xx}}{u_x},\quad \rxi_{xy} = \rxi_{yy} = \reta_{xx} = \reta_{xy} = \reta_{yy}= \psi_{xy}=\psi_{yy}=0,\\
&\hspace{9.25cm} \gamma_{xx} = \frac{v_x}{u_x}\,\psi_{xx},\quad \gamma_{xy}=\gamma_{yy}=0\bigg\}.
\end{align*}
In this example,
\begin{align*}
\mZ^{(2)} = \bigg\{ \psi_y &= \frac{ u_y}{y}\,\reta,\quad
\gamma_x = \frac{ v_x}{u_x}\,\psi_x,\quad \gamma_y = \frac{v_y}{y}\,\reta,\quad \psi_{xy} = u_{xy}\bigg(\frac{\psi_x}{u_x} + \frac{\reta}{y}\bigg),\quad \psi_{yy} = 2\frac{ u_{yy}}{y}\,\reta,\\
&\gamma_{xx} = \frac{v_x}{u_x}\,\psi_{xx}+2\bigg(\frac{v_{xx}u_x - u_{xx}v_x}{u_x^2}\bigg)\psi_x,\quad \gamma_{xy} = v_{xy}\bigg(\frac{\psi_x}{u_x}+\frac{\reta}{y}\bigg),\quad \gamma_{yy} = 2\frac{v_{yy}}{y}\,\reta\bigg\},
\end{align*}
which can also be found by applying the vector field
\[
\vvn\ii =  -\rxi \pp{}{x} - \reta \pp{}{y} + \sum_J \bigg(\psi_J \pp{}{u_J} + \gamma_J \pp{}{v_J}\bigg)
\]
to the differential invariants
\begin{gather*}
I_{0,1} = y\: u_y,\qquad I_{1,1} = \frac{y\: u_{xy}}{u_x},\qquad
I_{0,2} = y^2 u_{yy},\\
J_{1,0} = \frac{v_x}{u_x},\qquad J_{0,1} = y\, v_y,\qquad
J_{2,0} = \frac{v_{xx}u_x-v_x u_{xx}}{u_x^3},\qquad
J_{1,2} = \frac{y\: v_{xy}}{u_x},\qquad J_{0,2} = y^2 v_{yy},
\end{gather*}
and setting the result to zero.  Finally, we note that
$$\eeq{\disym^0 = \vsnf^0 = \emptyset,\qquad
\disym^1 = \vsnf^1 = \bigg\{\psi_y=0,\quad \gamma_x = \frac{v_x}{u_x}\,\psi_x ,\quad \gamma_y = 0\bigg\},\\
\disym^2 = \vsnf^2 = \bigg\{\psi_{xy}=\psi_{yy}=0,\quad \gamma_{xx}=\frac{ v_x}{u_x}\,\psi_{xx},\quad \gamma_{xy} = \gamma_{yy}=0\bigg\}.}$$

\Rmk{14r}
The linear spaces defined above are closely related to the algebraic constructions introduced in \rf{OP09}.  First, the vertical symbol \eqref{vsnf} is related to the prolonged symbol submodule defined in \rf{OP09; Definition 4.2}.  On the other hand, the prolonged annihilator subbundle $\mZ\n$ is equivalent to the bundle introduced in \rf{OP09; eq.\ (4.26)}.  As in \rf{OP09}, we note that the vertical symbol
\eqref{vsnf} forms a submodule, while, in general, this is not the case for the prolonged annihilator symbol $\disym = \bigcup_{n=0}^\infty \disym^n$.  That said, as we will show in the next section, beyond the order of freeness $\nf$ of the prolonged pseudo-group action, the equality $\disym^{> \nf} = \vsnf^{>\nf}$ holds, which implies that the truncated prolonged annihilator symbol acquires the algebraic structure of the vertical symbol.  Again, this is completely analogous to the constructions in \rf{OP09}, where the algebraic structure of the differential invariants produced by the moving frame normalizations only appears after the order of freeness.

%%%%%
\Subsection{Consistency} Compatibility.
%%%%%

As we observed in \sect{prolonged N}, beyond the order of freeness $\nf$, the reduced horizontal pseudo-group jet coordinates $\rX^i_J$ of order $|J| >\nf$ can all be chosen to be principal derivatives.  Therefore, the only parametric derivatives of order $>\nf$ are found among the normal form jet coordinates $u^\alpha_J$, where the splitting of the normal form jets into principal and parametric variables is determined by the vertical symbol $\vsnf^{>\nf}$.  Since we wish to prioritize choosing the $\rX^i_J$ as principal derivatives, in the following, we assign the index $(\alpha; J)$ for $1 \leq \alpha \leq q$ to the normal form jet $u^\alpha_J$ and $(q+i;J)$ for $1 \leq i \leq p$ to the reduced horizontal pseudo-group jet $\rX^i_J$.  Then, beyond the order of freeness, the splitting of the multi-indices
\[
\ind^{>\nf} =  \ind_{\mathcal{N},\prin}^{>\nf}\; \biguplus\; \ind_{\mathcal{N},\para}^{>\nf}
\]
of order $>\nf$ induced by the normal form determining equations has $(q+i; J) \in \ind_{\mathcal{N},\prin}^{>\nf}$ for all $i=1,\ldots,p$ and all $|J| > \nf$, while  $\ind_{\mathcal{N},\para}^{>\nf}$ only contains tuples of the form $(\alpha;J)$ with $1\leq \alpha \leq q$ and $|J|>\nf$.

As outlined in \sect{m}, the construction of a moving frame is based on the introduction of a (coordinate) cross-section \eqref{Kn}, where the phantom invariants $u^\alpha_J$ with $(\alpha; J) \in \indK$ determine the cross-section, and hence prescribe the constant Taylor coefficients in the normal form series \eq{nfps}, and thus play the role of ``parametric" jet coordinates in the moving frame framework.  For the moving frame construction to be compatible with the involutivity of the normal form determining equations we need, starting at a certain order $n$, for the parametric derivatives in the normal form determining equations to coincide with the phantom derivatives. In other words, at a certain order $n$, the equality
\begin{equation}\label{constant jet equality}
\ind_{\mathcal{N},\para}^{>n} = \ind^{>n}_\mK
\end{equation}
should hold.  To show that \eqref{constant jet equality} can be achieved, we show below that the equality
\begin{equation}\label{symbol equality}
\vsnf^{>\nf} = \disym^{>\nf}
\end{equation}
between the truncated vertical and prolonged annihilator symbols holds.  Equality \eqref{symbol equality} will then allow us to define in \sect{well-posed} the notion of a \is{well-posed cross-section} for which \eqref{constant jet equality} holds at order $n=\nf$.

%\Rmk{} When \eqref{constant jet equality} holds, we note that the principal normal form derivatives of order $>n$ determined by the normal form determining system are the same as the normalized jets $u^\beta_K \notin \indK^{>n}$ in the moving frame construction.  In fact, as explained in \sect{well-posed}, the equality \eqref{symbol equality} among the symbols will allow us to split the jets $u^\alpha_J$ of order $|J| > \nf$ into principal and parametric variables in a way that is compatible with involutivity of the normal form determining equation and the moving frame construction.  \pjoc{Is this remark now a bit repetitive of the previous discussion?}

According to \df{free}, the reduced Lie pseudo-group $\r{\G}$ acts freely at $z\n \in \Jn$ if and only if the isotropy group is trivial, meaning
\begin{equation}\label{FreenessEqs}
\rG\n_{z\n} = \bbc{(\rX\n,\rU\n) \in\rG\n \;|\;P\n(z\n,\rX\n,\rU\n) = z\n } = \bbc{\overline{\mathds{1}}\n_{z}},
\end{equation}
where $P\n(z\n,\rX\n,\rU\n)$ is the function that prescribes the prolonged action \eqref{eq: pr action} at order $n$.  At the infinitesimal level, the Lie pseudo-group acts locally freely if and only if
\begin{equation}\label{InfFreenessEqs}
\overline{\mathfrak{g}}\n_{z\n} = \bbc{(\rxi\n,\rphi\n)\in \LrGn(z\n,\rxi\n,\rphi\n)\;|\;\overline{\vv}\n = 0} = \{0\},
\end{equation}
where the prolonged vector field $\r{\vv}\n$ is defined in \eqref{prv}.

The next result shows that \is{persistence of freeness}, \cf \rf{OP08,OP12}, also holds for reduced Lie pseudo-group actions.

\Th{persist}
If the reduced pseudo-group $\rG$ acts \ro(locally\/\ro) freely at $z\n \in \Jn$, then for all $k>0$ it acts \ro(locally\/\ro) freely at $z\ps{n+k} \in \J{n+k}$ where $\pi^{n+k}_n(z\ps{n+k}) = z\n$.

\begin{proof}
The linearized equations \eqref{InfFreenessEqs} imply that the symbol of the system of equations \eqref{FreenessEqs} is trivial.  Therefore the system \eqref{FreenessEqs} is involutive with vanishing Cartan characters $\rcartan{n}{i} = 0$ for $i=1,\ldots,p$.
Since $\rG\ps{n+k}_{z\ps{n+k}}$ can be obtained by prolonging $\rG\n_{z\n}$, and involutivity is preserved under prolongation, we conclude, recalling \eqref{cartan relations}, that the Cartan characters of $\rG\ps{n+k}_{z\ps{n+k}}$ also vanish,
which means that all jets of order $n+k$ are uniquely determined.  Since $\r{\mathds{1}}^{(n+k)}_{z^{(n+k)}} \in \rG\ps{n+k}_{z\ps{n+k}}$, this is the only solution and the reduced pseudo-group remains free at order $n+k$.
\end{proof}

We now make the substitutions \eqref{chain rule} in  \eqref{FreenessEqs} to obtain\fnote{Keep in mind, that, in accordance with Remark \ref{11r},  $u\n = \tU\n$ at the identity transformation.}
\begin{equation}\label{NFree}
\bbc{(\rX\n,u\n)\in \nde\n \;|\;P\n(z\n,\rX\n,\rH\n(\rX\n,\tU\n))=z\n} = \bbc{(\mathds{1}\n_x,\tU\n)},
\end{equation}
which holds whenever the reduced pseudo-group acts freely.  At the infinitesimal level, we use equality \eqref{v equality} to conclude that $0=\r{\vv}\n = - \vvn\n$,  the latter being equivalent to 
$$\bbc{\rxi = 0,\ \ \psi\n=0}.$$
  Thus, the linearization of \eqref{NFree}, at the identity transformation, yields
\begin{equation}\label{LinNFree}
\bbc{(\rxi\n,\psi\n)\in\LNn(z\n,\rxi\n,\psi\n)\;|\;\rxi=0,\ \psi\n=0} = \{0\}.
\end{equation}

\Rmk{Rmkfree}
The local freeness condition \eqref{LinNFree} implies that the system of equations $\LNn(z\n,\rxi\n,\psi\n) \capz \bbc{\rxi=0, \ \psi\n=0}$ is equivalent to $\bbc{\rxi\n = 0, \ \psi\n=0}$.  Therefore any linear combination $Y \in \text{span}\bbc{\rxi\n,\;\psi\n}$ can be written in the form $Y= U + V$, with $U \in \text{span}\{\rxi,\psi\n\}$ and the equation $V=0$ belonging to $\LNn$.

We now establish the key moving frame/involutivity compatibility result.

\Th{Psi=Gamma}
If $\rG$ acts \ro(locally\/\ro) freely at $z\n \in \Jn$, then  the equality $\Psi^k\at{z\ps{k}} = \disym^k\at{z\ps{k}}$ holds for all $k > n$ and all  $z\ps{k} \in (\pi^k_n)^{-1}\{ z\n\}$.

\begin{proof}
By an inductive argument that relies on the persistence of freeness, it suffices to prove the equality for $k=n+1$.  Since $\Psi^n \subseteq \disym^n$ for any $n\in \mathbb{N}$, it suffices to show the reverse inclusion. In other words, if $Q=0$ is in $\Psi^{n+1}$, by which we mean that $Q=0$ is one of the defining equations of $\Psi^{n+1}$, we must show that there exists $U \in \text{span}\bbc{\rxi,\;\psi\n}$ such that
\[
Q+U = 0 \ \in\ \mZ^{(n+1)}.
\]
If this is the case, then $Q = \bH(Q+U) = 0$ is in $\disym^{n+1}$.

Now, since $Q=0$ is an equation in the symbol $\symN{n+1}$, there exists $Y \in \text{span}\bbc{\rxi\n,\;\psi\n}$ such that
\[
Q+Y = 0\  \in \ \LN{n+1}.
\]
Using \rmk{Rmkfree}, we have
%Remark \ref{Rmkfree}, we have \pjoc{The macro rmk is not working.  Reference missing.}
\[
Y = U+V,
\]
with $U \in \text{span}\bbc{\rxi,\;\psi\n}$ and $V=0$ in $\LNn$. Thus, the equation
\[
Q+U = (Q+Y) - V = 0 \ \in\ \LN{n+1}.
\]
Since $Q=0$ is in $\Psi^{n+1}$ and  $U \in \text{span}\bbc{\rxi,\;\psi\n}$, we conclude that $Q+U =0\in \mZ\ps{n+1}$.
\end{proof}

\Co{inv-free}
Let $\nf$ be the order at which the prolonged action of the Lie group $\G$ becomes free.  Then
\begin{equation}\label{psi=gamma nf}
\vsnf^{>\nf} = \disym^{>\nf}.
\end{equation}

\begin{proof}
By \th{freered}, the pseudo-group is reducible and the prolonged action of $\r{\G}$ also becomes free at order $\nf$.  By \th{Psi=Gamma}, $\vsnf^n = \disym^n$ for all $n> \nf$, which yields \eqref{psi=gamma nf}.
\end{proof}

Assuming the normal form determining equations become involutive at order $\rns$ and that the prolonged pseudo-group action becomes free at order $\nf \geq \rns$, we say, in light of the equality \eqref{psi=gamma nf} between the vertical and prolonged annihilator symbols of order $\geq \nf+1$, that the involutivity of the normal form determining equations becomes \is{compatible} with the moving frame construction at order $\nf+1$.

\Rmk{15r}
Example \ref{ex: I,J pg1} shows that freeness is not necessary to obtain the equality $\vsnf^{>n} = \disym^{>n}$ for some $n\in \N$.  Non-free actions will arise, in particular, in equivalence problems where there are non-trivial isotropy groups.  By appropriately dealing with the isotropy group, a modified version of \co{inv-free} should still hold.  The details are, however, deferred to a future study.

%\Rmk{16r} The normal form equations $\nde^{(\nf)}$ are given by the equations \eqref{XiJ norm eq} and \eqref{ualphaI}.  However, we note that the jets occurring on the left hand sides of the equations are not necessarily in accordance with the pivots of the symbol matrix $\symMN{(\nf)}$.

\Ex{7c}
Continuing \ex{7}, we saw that the order two normal form determining equations are not compatible with the moving frame construction. But since the prolonged action becomes free at order two, in view of \th{inv-free}, those of order three, as given in \eqref{order 3 nf det eq pg1}, will be compatible.  The normal form determining equations of order three remain involutive with Cartan characters
$\myeq{\rcartan{3}{1} = 2,\\ \rcartan{3}{2} = 0}$.

%%%%%
\Subsection{well-posed} Well-Posed Cross-Sections.
%%%%%

We now explain how to define a cross-section to be used in the construction of a moving frame that is compatible with the algebraic constructions and results of the previous sections.  As seen in \sect{m}, a moving frame is obtained by selecting a coordinate cross-section \eqref{Kn} transversal to the prolonged pseudo-group orbits.  To construct such a cross-section, at any order $n\geq 0$ it suffices to consider the prolonged annihilator symbol $\disym^n$ with coefficient matrix $\symM^n_\Upsilon$.  The jet coordinates $u^\alpha_J$ of order $|J| = n$ specifying the cross-section equations $u^\alpha_J = c^\alpha_J$ in \eqref{Kn} are then given by the non-pivot columns of the reduced coefficient matrix $\symM^n_{\Upsilon,\REF}$.

In light of \co{inv-free}, for all $n>\nf$, the jet coordinates that specify the coordinate cross-section can be chosen so as to coincide with the parametric derivatives in the normal form determining equations, provided the columns of $\symM^n_\Upsilon$ are sorted using the same class-respecting ordering imposed on the \nth order vertical symbol $\vsnf^n$.  Therefore, the cross-section equations 
$$\mK^{>\nf} = \set{u^\alpha_J = c^\alpha_J}{(\alpha;J) \in \indK^{>\nf}}$$
can be chosen so as to specify the parametric derivatives $u^\alpha_J$ of order $>\nf$ in the normal form determining equations, or, equivalently, the constant Taylor coefficients in the normal form power series \eq{nfps}.  On the other hand, the system of equations obtained by combining the cross-section equations
$$\mK^{(\nf)} = \set{x^i=c^i, \ u^\alpha_J=c^\alpha_J}{i,\; (\alpha;J) \in \indK^{(\nf)}}$$
with  the normal form equations \eqref{XiJ norm eq}, \eqref{ualphaI} of order $\leq \nf$ determines all the parametric normal form derivatives therein.  The equations $x^i=c^i$ in 
 $\mK^{(\nf)}$ are used to define the order zero reduced pseudo-group parameters
\begin{equation}\label{X constraints}
\rX^i(c^1,\ldots,c^p) = X^i_0,\qquad i=1,\ldots,p,
\end{equation}
where $X^i_0$ are arbitrary constants.  When combined, the entire cross-section $\mK$, together with \eqref{X constraints}, determines all the parametric derivatives in the normal form determining equations.  We call such a cross-section a \emph{well-posed cross-section}. 

This terminology stems from the fact that, as we explain in more detail in \sect{convergence}, it specifies formally well-posed initial conditions for the normal form determining equations.  A well-posed cross-section is a refinement of the notion of algebraic cross-section introduced in \cite{OP09}, which is prescribed by a Gr\"obner basis of the submodule $\vsnf$. On the other hand, implicit in our implementation of the theory of involutivity is the fact that the determination of a well-posed cross-section is prescribed by a Pommaret basis, \cite{S}. The main difference between Gr\"obner and Pommaret bases occurs in the definition of the multiplicative variables of a multi-index $J$, \rf{S}.  As seen in \sect{sec: involutivity}, for the Pommaret division the assignment of multiplicative variables depends on the class of $J$.  On the other hand, there is no constraint on the multiplicative variables for a Gr\"obner basis.   Thus, in general, Pommaret and Gr\"obner bases are not necessarily the same\fnote{Only when the ideal is stable, which we do not require, can one guarantee that its reduced Pommaret basis equals its reduced Gr\"obner basis, \cite{M1998}.}. Finally, with a Pommaret basis, an ideal can be decomposed into a finite union of disjoint involutive Pommaret cones, as in \eqref{involutive cone}, while the cones associated with a Gr\"obner basis may have non-trival intersections, and hence the connection with involutivity is not evident. 

\Rmk{29r}   A well-posed cross-section is the Lie pseudo-group analogue of a minimal order cross-section introduced in \rf{O2007} for finite-dimensional Lie group actions.  In both cases, the cross-section has the property that pseudo-group jet coordinates are normalized as soon as possible.   More precisely, a cross-section $\K \subset \Ji$ is of \emph{minimal order} if for all $n \geq 0$ its projection $\K\n = \pi^\infty_n(\K) \subset \Jn$ forms a cross-section to the orbits of $\r{\G}\n$ on $\Jn$.

%In light of \co{inv-free}, for all $n>\nf$, the pivots of $\symKnREF$ are complementary to the pivots of $\Psi^{n}\REF$.  Therefore, the cross-section equations $\mK^{>\nf} = \set{u^\alpha_J = c^\alpha_J}{(\alpha;J) \in \indK^{>\nf}}$  specify the parametric derivatives of order $>\nf$ in the normal form determining equations, or, equivalently, the constant Taylor coefficients in the normal form series.  On the other hand, the cross-section  $\mK^{(\nf)} = \set{x^i=c^i, u^\alpha_J=c^\alpha_J}{i,\; (\alpha;J) \in \indK^{(\nf)}}$ of order $\leq \nf$ determines the parametric derivatives in the normal form equations \eqref{XiJ norm eq}, \eqref{ualphaI} of order $\leq \nf$ resulting from the moving frame implementation.  When combined, the whole cross-section $\mK$ therefore determines all the parametric normal form jet coordinates in the normal form determining equations.

\Rmk{18r}
In \cite{KZ2015}, Kossovskiy and Zaitsev realized the importance of working with well-posed cross-sections in order to construct convergent normal forms. As mentioned in the first paragraph of section two of their work, they resolve the problem of divergence of Kol\'a\v{r}'s normal form for degenerate hypersurfaces in $\mathbb C^2$, \cite{Kolar}, by selecting a well-posed/minimal order cross-section.

We now introduce a simple criterion that guarantees that a prescribed cross-section is well-posed without having to compute the normal form determining equations. Assume that the determining equations of the reduced pseudo-group $\r{\G}$ become involutive at order $\rns$ and that the prolonged pseudo-group action becomes free at order $\nf\geq \rns$.  By \th{infdeq}, the normal form determining equations $\nde\ps{\nf+1}$ are involutive.   \pr{Rees} implies that the set of parametric indices $\ind^{>\nf}_{\nde,\para}$ admits a Rees decomposition.  Since a well-posed cross-section is constructed such that \eqref{constant jet equality} holds with $n=\nf$, it follows that its set of defining indices admits the same Rees decomposition, so
\begin{equation}\label{decomposition}
\indK^{>\nf} = \biguplus_{(\alpha;J)\: \in\: \cI_{\K}^{\nf+1}} \cone^\alpha(J).
\end{equation}
%
%admits the same Rees decomposition as $\ind^{>\nf}_{\nde,\para}$.  In
Conversely, \pr{Rees} states that the Rees decomposition \eqref{decomposition} is sufficient to guarantee the existence of a Pommaret basis for the ideal $\disym^{>\nf} = \vsnf^{>\nf}$.  %This allows us to introduce a simple criterion to determine that a cross-section is well-posed without having to compute the normal form determining equations.
This implies the following well-posedness criterion.

\Th{well-posedK}
Let $\G$ be a Lie pseudo-group whose prolonged action becomes free at order $\nf$.  A cross-section $\K$ is well-posed if it is of minimal order and its set of defining indices $\indK^{>\nf}$ admits a Rees decomposition \eqref{decomposition}.

\begin{proof}
We need to show that there exists a system of normal form determining equations that is involutive at order $\nf+1$ with $\K$ providing well-posed initial conditions.

Since the prolonged action becomes free, by \th{freered} the pseudo-group $\G$ is reducible with reduced determining equations \eqref{psi red det eq}.  The normal form determining equations are then obtained by substituting the chain rule formulas \eqref{chain rule} into \eqref{psi red det eq} to obtain \eqref{nf det eq}.  These equations are subsequently solved for the principal pseudo-group jets $\rX^i_J$, and the principal normal form jets $u^\alpha_J$ with $(\alpha;J) \notin \indKn$.  In order for  $\indKn$ to be as large as possible, we must require that as many reduced pseudo-group parameters $\rX^i_J$ be principal as possible.   This, in other words, is equivalent to requiring that the cross-section $\K$ be of minimal order.  Recall from \sect{prolonged N} that once $n=\nf$ all reduced pseudo-group derivatives $\rX^i_J$ are principal. 

The order $\nf+1$ normal form determining equations are given by equations of the form \eqref{XiJj-tmp}, \eqref{u normal form eq}.  The equations \eqref{XiJj-tmp} for the reduced pseudo-group jets  do not provide any obstruction to involutivity, and therefore it suffices to consider the equations for the normal form jets \eqref{u normal form eq}.  By assumption, since $\indK^{>\nf}$ admits a Rees decomposition \eqref{decomposition}, \pr{Rees} guarantees that the symbols $\disym^{>\nf} = \vsnf^{>\nf}$ admit a Pommaret basis.  The existence of the Pommaret basis implies that it is possible to express the differential equations for the normal form jets of order $\nf+1$ in such a way that \eqref{XiJj-tmp}, \eqref{u normal form eq} is involutive with the parametric normal form jets $u^\beta_K$ of order $|K| = \nf+1$ specified by the cross-section.
\end{proof}

\Ex{Well-posedKex}
For our running example --- the Lie pseudo-group \eqref{eq: pg1} --- a well-posed cross-section is given by \eqref{genKex}, which we can verify satisfies the hypotheses of \th{well-posedK}.  First, for all $n\geq 0$, $\K\n$ is transversal to the prolonged pseudo-group action and thus is of minimal order.
Next, since the prolonged action becomes free at order $\nf=2$, we must consider the cross-section determining equations of order $>2$, namely
\[
\K^{>2} = \set{ u_{x^{k+1}}=c_{k+1},\, u_{x^ky}=d_k}{k\geq 2}.
\]
The corresponding set of determining indices has the Rees decomposition
\begin{align*}
\indK^{>2} &= \set{ (k+1,0),\, (k,1)}{n\geq 2} \\
&= \set{(k+1,0)}{k\geq 2}\; \biguplus\; \set{(k,1)}{k\geq 2} = \cone(3,0) \;\biguplus\; \cone(2,1).
\end{align*}

%%%%%
\Subsection{convergence} Convergence of the Normal Form Power Series.
%%%%%

We are now ready to state and prove the main result of the paper.  As in the previous section, let $\nf\geq \rns$ denote the order of freeness, taken to be at least as large as the order of involutivity.  We follow the discussion on page \pageref{R1} to rewrite the order $\nf+1$ involutive normal form determining equations $\nde^{(\nf+1)}$ as an equivalent system of first order differential equations
\begin{equation}\label{order 1 nf det eq}
\widetilde{\nde}^{(1)}=
\left\{
\begin{aligned}
\ & \widetilde{\Delta}^{(1)}(x,\tU^{(n_f+1)},(\rX^{(n_f)})^{(1)},(u^{(n_f)})^{(1)})=0, \\
&\begin{aligned}
&\partial_i \rX^j_J = X^j_{J,i},& &
\partial_i u^\alpha_J =u^\alpha_{J,i}, & \hspace{1cm} &
|J|\leq n_f,\; 1\leq i \leq p, \\
&\partial_i\rX^j_J = \partial_k \rX^j_{J,i\setminus k},& &
\partial_i u^\alpha_J = \partial_k u^\alpha_{J,i\setminus k}, & &
|J| = n_f,\; k = \cls J < i \leq p. \
\end{aligned}
\end{aligned}
\right\}.
\end{equation}
According to Proposition \ref{order 1 reduction}, this first order system remains involutive with the same Cartan characters as the original normal form determining system $\nde^{(\nf+1)}$.  Furthermore, we write \eqref{order 1 nf det eq} in reduced Cartan normal form.  Since the second and third lines of \eqref{order 1 nf det eq} are already in Cartan normal form, we focus on the equations $\widetilde{\Delta}^{(1)}=0$.  When expressing the order $\nf+1$ jets as first order derivatives, we use
the substitutions \eqref{ujet sub} and make the blanket assumption that when writing $\partial_k u^\gamma_{K\setminus k}$, the multi-index $K$ is of order $\nf+1$ and class $k$.  Doing so, we obtain the first order system of differential equations
%
%\begin{subequations}\label{Cartan nf}
\begin{equation}\label{Cartan nf det eq1}
\begin{aligned}
\partial_j u^\alpha_{J\setminus j} &= \Delta^\alpha_J\big(x,\tU^{(n_f+1)},\rX,\ldots,u^
\beta_K,\ldots, \partial_i u^\gamma_{I\setminus i},\ldots\big), \\
\partial_\ell \rX^i_{L} &= \Xi^i_{L,\ell} \big(x,\tU^{(n_f+1)},\rX,\ldots, u^\beta_K,\ldots, \partial_n u^\kappa_{N\setminus n},\ldots \big),
\end{aligned}
\end{equation}
where all the normal form jets $u^\beta_K$,  $\partial_i u^\gamma_{I\setminus i}$, $\partial_n u^\kappa_{N\setminus n}$ appearing on the right hand side of the equations are parametric with $|K|\leq \nf$, $|L| = \nf$, $|I| = |J| = |N| =\nf+1$, while $i\leq j$ and $n\leq \ell \leq \cls(L)$.  We note that the equations \eqref{Cartan nf det eq1} are just the equations \eqref{XiJj-tmp}, \eqref{u normal form eq} written as first order differential equations. The equations \eqref{Cartan nf det eq1} are supplemented with the algebraic equations
\begin{equation}\label{Cartan nf det eq2}
u^\alpha_J = \Delta^\alpha_J(x,\tU^{(n_f)},\rX,\ldots,u^\beta_K,\ldots),\qquad
\rX^i_J = \Xi^i_J(x,\tU^{(n_f)},\rX,\ldots,u^\beta_K,\ldots),
\end{equation}
%\end{subequations}
%
given by \eqref{XiJ norm eq}, \eqref{ualphaI}, where $u^\alpha_J$ and $\rX^i_J$ are principal derivatives and $u^\beta_K$ are parametric derivatives of order $\leq \nf$.  According to Theorem \ref{thm: CK}, provided all the functions $\Delta^\alpha_J$, $\Xi^i_J$, and $\Xi^i_{L,\ell}$ in \eqref{Cartan nf det eq1}, \eqref{Cartan nf det eq2} are real-analytic at the origin, the formally well-posed initial conditions
%
%\begin{subequations}\label{complete initial cond}
\begin{equation}\label{Cartan nf initial cond}
\begin{aligned}
u^\beta_K(0,\ldots,0) &= f^\beta_K, \\
u^\gamma_{I\setminus 1}(x^1,0,\ldots,0) &= f^\gamma_{I\setminus 1}(x^1),\\
&\hspace{0.2cm}\vdots \\
u^\gamma_{I\setminus p-1}(x^1,\ldots,x^{p-1},0) &= f^\gamma_{I\setminus p-1}(x^1,\ldots,x^{p-1}),\\
u^\gamma_{I\setminus p} (x^1,\ldots,x^p) &= f^\gamma_{I\setminus p}(x^1,\ldots,x^p),
\end{aligned}
\end{equation}
specifying the parametric derivatives occurring on the right hand side of the equations \eqref{Cartan nf det eq1}, \eqref{Cartan nf det eq2} are analytic at the origin, and the algebraic equations
\begin{equation}\label{alg eq}
\begin{aligned}
u^\alpha_J(0,\ldots,0) &= \Delta^\alpha_J(x,\tU^{(n_f)},\rX,\ldots,u^\beta_K,\ldots)\at{(0,\ldots,0)},\\
\rX^i_J(0,\ldots,0) &= \Xi^i_J(x,\tU^{(n_f)},\rX,\ldots,u^\beta_K,\ldots)\at{(0,\ldots,0)},\\
\rX^i(0,\ldots,0) & =X^i_0,
\end{aligned}
\end{equation}
%\end{subequations}
%
are satisfied, then the normal form determining system admits one and only one solution that is analytic at the origin.  In particular, the normal form $u(x)$, which forms part of the solution is analytic.  In \eqref{alg eq}, the right hand side of the third equation are the components of the point $X_0=(X_0^1,\ldots,X_0^p) \in \mX$ at which the submanifold is being considered.

\Rmk{17r}
The initial conditions \eqref{Cartan nf initial cond}, \eqref{alg eq} are stated under the assumption that the pseudo-group $\G$ can map the
origin $0 \in \mX$ to the point $X_0$.  In applications, the origin can be replaced by any convenient point $\bp \in \mathcal{X}$.  For example, the points where $y=0$  are singular for the pseudo-group \eqref{n2pg}, and here the origin can be replaced by the point $\bp=(0,1)$, so that any point $(X_0,Y_0)$ with $Y_0 > 0$ lies on its group orbit.  In general, given $\bp \in \mX$, the initial conditions \eqref{Cartan nf initial cond} can be modified by considering hyperplanes passing through $\bp$.  Of course, it is also possible to make a local change of coordinates preserving $\delta$-regularity so that $\bp$ is mapped to $0$ and the initial conditions are given by \eqref{Cartan nf initial cond}, \eqref{alg eq}.

\Ex{9z}
For our running example, based on the cross-section \eqref{genKex}, the standard moving frame implementation yields the general normal form \eqref{nf_ex}.
%\begin{equation}\label{nf_ex}
%u(x,y) = c(x) + y\, d(x) + \frac{y^2}{2}\,w(x,y),
%\end{equation}
%where $w(0,0) = 1$.  
Since the prolonged action becomes free at order $\nf=2$, we must consider the order three normal form determining equations given in \eqref{order 3 nf det eq pg1} to show that 
formal power series \eqref{nf_ex} converges. We note that the last two equations of \eqref{order 3 nf det eq pg1} are solved for the principal normal form jets $u_{xyy}$ and $u_{yyy}$, in accordance with the order three vertical symbol \eqref{low order vert symbols pg1}.  As first order partial differential equations, these determine $\partial_x(u_{yy})$ and $\partial_y(u_{yy})$.  On the other hand, the order three normal form jets $u_{xxy} = \partial_x(u_{xy})$ and $u_{xxx} = \partial_x(u_{xx})$ are parametric of class one.  
\pari
In accordance with \eqref{Cartan nf initial cond}, $u_{xx}$ and $u_{xy}$ are fixed by imposing initial conditions along the $x$-axis.  Differentiating \eqref{nf_ex}, those are given by
\begin{equation}\label{ic_ex1}
u_{xx}(x,0) = c_{xx}(x),\qquad u_{xy}(x,0) = d_x(x),
\end{equation}
These are supplemented with the algebraic initial conditions
\begin{equation}\label{ic_ex2}
\begin{aligned}
& & &X(0,0) = X_0,& &Y(0,0)=Y_0, & &\\
&u(0,0) = c_0,&\quad &u_x(0,0) = c_1,&\quad &
u_y(0,0) = d_0,&\quad &u_{yy}(0,0) = 1,
\end{aligned}
\end{equation}
that come from the low order normalizations. We note that the initial conditions \eqref{ic_ex1}, \eqref{ic_ex2} can be simplified to
\[
\qeq{X(0,0) = X_0,\\ Y(0,0)=Y_0,\\ u(x,0) = c(x),\\ u_y(x,0) = d(x),\\ u_{yy}(0,0)=1.}
\]
Assuming the functions $c(x)$ and $d(x)$ are both analytic, the solution to the involutive normal form determining equations \eqref{order 3 nf det eq pg1}
includes the normal form function \eqref{nf_ex}, thereby establishing its analyticity and hence convergence of the corresponding power series.

Keeping this example in mind, we are now able to state our general convergence result.  Further illustrative examples will appear in \sect{xx}.

\Th{main}
Let $\G$ be an analytic Lie pseudo-group acting transitively on $\mathcal{X}$ with its prolonged action acting eventually freely on an analytic submanifold $\tU(X)$.
If the cross-section
\begin{equation}\label{K main}
\mathcal{K} = \set{x^i=0, \ u^\alpha_J = c^\alpha_J}{i,\;(\alpha;J) \in \indK}
\end{equation}
is well-posed, and the cross-section power series \eqref{csSeries} determined by the normalization constants $c^\alpha_J$ are convergent and so define analytic functions, then the corresponding normal form power series \eq{nfps} converges and defines an analytic function in the neighborhood of the origin.

%\Th{main}
%Let $\G$ be an analytic Lie pseudo-group acting transitively on $\mathcal{X}$ with its prolonged action acting eventually freely on an analytic submanifold $\tU(X)$.
%If the cross-section
%%
%\begin{equation}\label{K main}
%\mathcal{K} = \set{x^i=0, u^\alpha_J = c^\alpha_J}{i,\;(\alpha;J) \in \indK}
%\end{equation}
%%
%is well-posed, with the normalization constants $c^\alpha_J$ defining analytic functions
%%
%\begin{equation}\label{csSeries}
%C^\alpha(y) = \sum_{J\, \in\, \indKa} \frac{c^\alpha_J}{J!} \,y^J,\qquad \alpha = 1,\ldots,q,
%\end{equation}
%%
%for the cross-section power series, then the corresponding normal form power series
%%
%\begin{equation}\label{analytic u}
%u^\alpha(y) = \sum_{J} \frac{u^\alpha_J}{J!} y^J
%\end{equation}
%%
%defines an analytic function in the neighborhood of the origin.

\begin{proof}
Since the prolonged action of $\G$ becomes eventually free, the pseudo-group is reducible by \th{freered}, and by \th{reducibility thm} there exists $\rns \in \N$ such that reduced determining equations $\r{\G}\ps{\rns}$ are involutive.  By \th{infdeq} the normal form determining equations $\nde\ps{\rns}$ are involutive.  Let $\nf \geq \rns$ be an order at which the prolonged action is free.  Since involutivity is preserved under prolongation, the normal form determining equations $\nde\ps{\nf+1}$ are involutive.  Since $\G$ and $\tU(X)$ are analytic, the normal form determining equations $\nde\ps{\nf+1}$ are also analytic, and, when written as a system of first order differential equations, are given by \eqref{Cartan nf det eq1}, \eqref{Cartan nf det eq2}.  With $\K$ being a well-posed cross-section, the analytic cross-section power series \eqref{csSeries} provides the analytic well-posed initial conditions \eqref{Cartan nf initial cond}, \eqref{alg eq}.   That is,
\[
\qeq{C^\gamma_{I\setminus i} (x^1,\ldots,x^i,0,\ldots,0) =  f^\gamma_{I\setminus i}(x^1,\ldots,x^i),\\ i=1,\ldots,p,\\C^\beta_K(0,\ldots,0) = f^\beta_K.}
\]
Theorem \ref{thm: CK} implies that the solution to the initial value problem is unique and analytic including the part of the solution corresponding to the normal form power series \eq{nfps}.
\end{proof}

\Rmk{regularity-r}
Implicit in the statement of \th{main} is the fact that the coordinates used to express the well-posed cross-section \eqref{K main} are $\delta$-regular.  Indeed, by definition $\G$ is a Lie pseudo-group if its elements are the solutions to an involutive system of differential equations, and involutivity, within our framework, requires $\delta$-regularity of the underlying coordinate system.

%%%%%
\Section{chain} Chains.
%%%%%

In their paper \cite{CM74}, Chern and Moser introduced the concept of a chain as a tool for proving the convergence of their normal form power series for CR hypersurfaces $S\subset \mathbb{C}^m$.   A regular curve $\chain \subset S$ in the hypersurface $S$ is said to be a \is{chain} if its projection $\pi(\chain) \subset \mX$ onto the space of independent variables can be rectified by a biholomorphic transformation that also normalizes the Taylor coefficients of the hypersurface $S$ appearing in the Chern--Moser normal form.  They employ a finite sequence of transformations that successively rectify the chains and thereby place the Taylor expansion of the transformed surface in normal form.  Each transformation is analytic since it either satisfies an algebraic constraint or is the solution to an analytic system of ordinary differential equations.  Therefore, the final transformed hypersurface is analytic, and its Taylor series, which is in normal form, converges.

To make the discussion more precise, let us first review the convergence argument in \cite{CM74} when $m=2$ so that $S \subset \mathbb{C}^2$ is a three-dimensional hypersurface.  We introduce complex (source) coordinates $\mzeq{z = x + \i y,\\ w = u + \i v}$, so that $\mzeq{Z = X + \i y,\\W = U + \i V}$ are the corresponding target coordinates.  As in \cite{CM74}, we assume that the prescribed hypersurface is locally parametrized by
\[
S = \{(Z,\oZ,U,\tV(Z,\oZ,U))\},
\]
so that $(Z,\oZ,U)\in \mX$ are viewed as independent variables.
 %with $Z\in \mathbb{C}$ and $U\in \mathbb{R}$.  We refer the reader to \cite{M2020} for a detailed account of this particular case.   As in \cite{CM74}, let $W = U + \i V$.  
 After translation, we can work at the origin and consider the Taylor expansion
\begin{equation}\label{V}
\tV(Z,\oZ,U) = \sum_{j,k=0}^\infty  Z^j\oZ{}^kF_{j,k}(U),
\end{equation}
where the Taylor coefficients and powers of $U$ are contained in the functions $F_{j,k}$.  By assumption $F_{0,0}(0) = 0$, since the hypersurface has been translated to the origin. One then seeks a chain, meaning a curve
\begin{equation}\label{MC chain}
\chain = \{(\psi(u),\varphi(u))\} \subset S\qquad\text{with}\qquad \varphi_u(0)\neq 0,
\end{equation}
whose projection $\pi(\chain)$ is holomorphically rectified onto the line $\ell=\{(0,0,u)\}$ and sends the hypersurface Taylor series \eqref{V} to the Chern--Moser normal form, which is given in \eq{CMnf} below.  This is accomplished by the following sequence of analytic transformations, each of which serves to normalize some of the Taylor coefficients in the expansion \eqref{V}.

\begin{description}
\item[Step 1:] The holomorphic transformation
\[
Z = z+\psi(w),\qquad W = \varphi(w)
\]
takes $\pi(\chain)$ into $\ell$ and sends \eqref{V} to\fnote{During the course of the procedure, the expressions for the Taylor coefficient functions $F_{jk}$ will change.  We avoid introducing new notation for each version.}
\[
v = \sum_{j+k\geq 1} z^j\:\oz{}^k F_{j,k}(u).
\]
We observe that such a transformation does not impose any constraint on the chain.

\item[Step 2:] Cancel the harmonic terms $z^j F_{j,0}(u)$ and $\oz{}^k F_{0,k}(u)$ using a transformation of the form
\begin{equation}\label{step2pg}
z^* = z,\qquad w^* = w+g(z,w)\qquad \text{with}\qquad g(0,w) =0,
\end{equation}
so that the new power series is
\[
v = \sum_{j\geq 1\text{ or }k\geq 1} z^j \:\oz{}^k F_{j,k}(u).
\]
The function $g(z,w)$ is derived in the proof of \rf{CM74; Lemma 3.2} and is found by solving an algebraic equation.  We note that \eqref{step2pg} does not affect the line $\ell=\{(0,0,u)\}$, which is also the case for all upcoming transformations.

\item[Step 3:] Under the assumption that the hypersurface is Levi nondegenerate, which means that $\tV_{Z\oZ} \ne 0$, normalize $z \oz{}^k F_{1,k}(u)=0$ and $z^j \oz F_{j,1}(u)=0$ using
\[
z^* = z + f(z,w),\qquad w^*=w,
\]
with $f(0,w)=0$, $f_z(0,w)=0$, so that
\[
v = z\:\oz\: F_{1,1}(u) + \sum_{j,k\geq 2} z^j\:\oz{}^k F_{j,k}(u),
\]
where $F_{1,1}(0) \neq 0$. The function $f(z,w)$ satisfies an algebraic equation given in the proof of \rf{CM74; Lemma 3.3}.

\item[Step 4:] Normalize $F_{1,1}(u)=1$ using a transformation of the form
\begin{equation}\label{C transformation}
z^* = C(w)\:z,\qquad w^*=w
\end{equation}
so that the transformed power series is
\[
v = z\:\oz + \sum_{j,k\geq 2} z^j \:\oz{}^k F_{j,k}(u).
\]
To do so, it suffices to take
\begin{equation}\label{C}
C(u) = \sqrt{F_{1,1}(u)}
\end{equation}
and then replace $u$ by $w$ to obtain the transformation \eqref{C transformation}.

\item[Step 5:] Normalize 
$$\qeq{F_{2,2}(u)=0,\\F_{3,2}(u)=\overline{F_{3,2}(u)}=0,\\F_{3,3}(u)=0,}$$
 so that the \is{Chern--Moser normal form} is
\Eq{CMnf}
$$v = z\:\oz + z^4\:\oz{}^2 F_{4,2}(u) + z^2\:\oz{}^4F_{2,4}(u) + \sum_{\substack{j+k\geq 7\\ j,k\geq 2}}z^j\: \oz{}^k F_{j,k}(u).$$
The normalization $F_{3,2}(u)=0$ imposes a differential constraint on the first component of the chain \eqref{MC chain} given by a second order ordinary differential equation for $\rpsi(u)$:
\[
\psi_{uu}=Q(u,\psi,\rpsi,\psi_u,\rpsi_u).
\]
The explicit formula for $Q$ is not provided in \rf{CM74}.  For three-dimensional hypersurfaces, a Lie theoretic description of this equation is given in \cite{M2020}.

The normalization $F_{22}(u)=0$ is achieved using the transformation
\begin{equation}\label{lambda transformation}
z^* = \lambda(w) \:z,\qquad w^*=w,
\end{equation}
such that when $w=u$ is real, the restricted transformation satisfies
%\pjoc{I'm a litle confused here. Is $\lambda$ defined for complex $w$, but only constrained when $w=u$ is real?}  satisfying 
$\lambda(u)\,\overline{\lambda(u)}=1$, $\lambda(0)=1$, and is a solution to the first order ordinary differential equation
\[
\lambda_u = -\frac{\i}{2} F_{2,2}(u)\: \lambda.
\]
The transformations \eqref{C transformation} and \eqref{lambda transformation} are slightly different. In light of \eqref{C}, the function $C(w)$ in \eqref{C transformation} is real-valued or purely imaginary depending on whether $F_{1,1}(u) >0$ or $F_{1,1}(u)<0$, while the function $\lambda(w)$ in \eqref{lambda transformation} is complex-valued.

Finally, the normalization $F_{3,3}(u)=0$ is achieved via the transformation
\[
z^* = z \sqrt{\varphi_w(w)},\qquad w^* = \varphi(w),
\]
with $\varphi(\mathbb{R}) \subset \mathbb{R}$, $\varphi(0)=0$, $\varphi_w(0) >0$, and satisfies the third order ordinary differential equation
\[
\varphi_{uuu} = \frac{3\:\varphi^2_{uu}}{2\:\varphi_u} - 3 \:F_{3,3}(u)\:\varphi_u.
\]
This provides constraints on the second component of the chain \eqref{MC chain}.
\end{description}

We now explain how this particular Chern--Moser construction can be formulated within our general framework. To make the connection evident, let us assume for the time being that the class one Cartan character  of the involutive normal form determining equations is the only nonzero character, so
\Eq{chain1}
$$\rcartan{n}{1} \neq 0,\qquad \rcartan{n}{2} = \cdots = \rcartan{n}{p} = 0.$$
In this particular setting, the general solution depends only on functions of one variable, and the initial conditions \eqref{Cartan nf initial cond} reduce to
\begin{equation}\label{InitCond}
u^\beta_K(0,\ldots,0) = f^\beta_K,\qquad u^\gamma_{I\setminus 1}(x^1,0,\ldots,0) = f^\gamma_{I\setminus 1}(x^1).
\end{equation}
Since the Taylor coefficients of the initial conditions \eqref{InitCond} determine the cross-section $\K$, the left hand side of the equations \eqref{InitCond} can be replaced by the cross-section functions \eqref{csSeries} so that
\begin{equation}\label{Cf}
C^\beta_K(0,\ldots,0) = f^\beta_K \qquad \text{and}\qquad
C^\gamma_{I\setminus 1}(x^1,0\ldots,0) = f^\gamma_{I\setminus 1}(x^1).
\end{equation}
We observe that the equations \eqref{Cf} are defined on the line $\ell=\{(x^1,0,\ldots,0)\} \subset \mX$.  Then, a one-dimensional chain $\chain$ is a regular curve in the section $S$ with the property that there exists a pseudo-group transformation $\varphi^{-1} \in \G$ mapping $\chain$ to the curve $\varphi^{-1}(\chain) = \lc = (\ell,C(\ell))$ contained in the normal form $s$, where $C(y)$ is the cross-section function \eqref{csSeries}.   In particular, we note that the projection of the chain onto the space of independent variables $\pi(\chain)\subset \mX$ is rectified to the line $\varphi^{-1}\at{\mX}(\pi(\chain)) = \ell$.  In other words, $\pi(\chain) = \varphi\at{\mX}(\ell)$.

Thus, to find the chain $\chain = \varphi(\ell,C(\ell))$ passing through $(X_0,\tU(X_0))$, it suffices to find $\varphi \in \G$ such that
\begin{equation}\label{chain det eq}
(\varphi\at{\mX}(\ell),\tU(\varphi\at{\mX}(\ell))) = \varphi(\ell,C(\ell)).
\end{equation}
Setting $(X(x,u),U(x,u)) = \varphi(x,u)$ and $(\rX\at{\lc},\rU\at{\lc}) =  \varphi(\ell,C(\ell)) = (X(\ell,C(\ell)),U(\ell,C(\ell))$, equation \eqref{chain det eq} reduces to solving
\begin{equation}\label{chain eq}
\tU(\rX\at{\lc}) = \rU\at{\lc}.
\end{equation}
We note that \eqref{chain eq} is the same equation as \eqref{nf eq} but restricted to the curve $\lc = (\ell,C(\ell))$.  More explicitly, \eqref{chain eq} is obtained by replacing $(x,u(x))$ in the second equation of \eqref{nf eq} by $(\ell,C(\ell))$ to obtain
\begin{equation}\label{chain eq2}
 \tU(X(\ell,C(\ell)) = U(\ell,C(\ell)).
\end{equation}
With $\lc$ being one-dimensional, the equations \eqref{chain eq2} form a system of ordinary differential equations for the parametric reduced pseudo-group jets with initial value $(\rX\at{\lc}(0),\rU\at{\lc}(0)) = (X_0,\tU(X_0))$.  We now show how this works with two examples.

\Ex{10}
Consider the Lie pseudo-group
\begin{equation}\label{chain pg1}
X = f(x),\qquad Y = y+b,\qquad U=\frac{u}{f_x(x)},
\end{equation}
acting on surfaces $u(x,y)$, where $f\in \D(\R)$ and $b\in \R$. We assume $u \ne 0$ in what follows, and similarly for $\tU$.  Furthermore, we choose $\rY$,
$\rX_{x^k}$, with $k\geq 0$, as parametric reduced pseudo-group jets in the computations below.  
\pari
The normal form determining equations of order one are 
\begin{equation}\label{pg2 rdeteq}
\rX_y = \rY_x = 0,\qquad \rY_y = 1,\qquad \rX_x = \frac u\tU,\qquad u_y = \frac {\tU_Y}\tU\,u.
\end{equation}
%\begin{equation}\label{pg2 rdeteq}
%\rX_y = \rY_x = 0,\qquad \rY_y = 1,\qquad \tU\,\rX_x = u,\qquad \tU_Y\rX_x = u_y.
%\end{equation}
%
These equations are involutive with indices and Cartan characters
\[
\rindex{1}{1} = 2,\qquad \rindex{1}{2} = 3,\qquad
\rcartan{1}{1} = 1,\qquad \rcartan{1}{2} = 0.
\]
%As it can be seen from the pseudo-group \eqref{chain pg1}, t

\pari
A moving frame for the pseudo-group \eqref{chain pg1} was constructed in \rf{OP08} using the cross-section
\[
\K = \{x=y=0,\; u=1,\; u_{x^k} = 0,\; k\geq 1\},
\]
which induces the initial conditions
\[
X(0,0) = X_0,\qquad Y(0,0)=Y_0,\qquad u(x,0) = 1
\]
for the system of partial differential equations \eqref{pg2 rdeteq} and corresponds to the normal form
\Eq{ex10nf}
$$u(x,y) = 1 + y\:w(x,y),$$
where the Taylor coefficients of $w(x,y)$ give the basic differential invariants expressed in terms of the jet coordinates of $\tU$.  The  cross-section function is $u(x,0) = C(x) = 1$, and defines the line
\begin{equation}\label{rectification}
\lc = \{(x,0,1)\} \subset s
\end{equation}
contained in the graph of the normal form \eq{ex10nf}.
\pari
A chain is a regular curve  $\chain=\bbc{(\rX(x),Y_0,\tU(\rX(x),Y_0))} \subset S$ contained in the submanifold that is rectified to the line \eqref{rectification} by a pseudo-group transformation \eqref{chain pg1}.  First, for the $y$-coordinate of $\lc$ to be sent to $Y_0$ in the chain, a translation with $b=Y_0$ is performed.  On the other hand, the function $\rX(x)$ satisfies the chain determining equation \eqref{chain eq2}, which yields the differential equation
\[
\tU(\rX(x),Y_0) = \rU = \frac{1}{\rX_x(x)}.
\]
In other words,
\[
\rX_x(x) = \frac{1}{\tU(\rX(x),Y_0)}\qquad\text{with the initial condition}\qquad \rX(0) = X_0.
\]
This is an ordinary differential equation for $\rX(x)$, whose right hand side is analytic when the surface $\tU(X,Y)$ is analytic, and hence defines an analytic normalizing transformation.

\Ex{11}
Let us return to our running example, which consists of the Lie pseudo-group \eqref{eq: pg1} with normal form \eqref{nf_ex}  and cross-section power series \eq{ex9csps}.
% is
%%
%\begin{equation}\label{cs function ex2}
%C(x,y) = c(x) + y\, d(x) + \f2\:{y^2},
%\end{equation}
%%
%where $c(x)$ and $d(x)$ are specified functions, which in the simplest version can be set to zero $c(x) = d(x) = 0$, such as in \ex{mf ex}.   
%\pari
Let us determine the chain corresponding to the two initial conditions
\[
\qeq{u(x,0)=c(x),\\
u_y(x,0)=d(x).}
\]
These provide the pair of equations
\[
\req{\tU(\rX(x),\rY(x,0)) = c(x) + \frac{\rY_x(x,0)}{\rX_x(0)},\\\tU_Y(\rX(x),\rY(x,0)) \rX_x(x) = d(x) + \frac{\rX_{xx}(x)}{\rX_x(x)}}.
\]
Thus, the chain $\chain=\bbc{(\rX(x),\rY(x,0),\tU(\rX(x),\rY(x,0))}$ is obtained by solving a pair of ordinary differential equations
\begin{equation}\label{chain ODEs pg2}
\begin{aligned}
\rX_{xx}(x) &= \tU_Y(\rX(x),\rY(x,0))\rX_x^2(x)-d(x)\rX_x(x),\\
\rY_x(x,0) &= \bbk{\tU(\rX(x),\rY(x,0)) - c(x)}\rX_x(x),
\end{aligned}
\end{equation}
subject to the initial conditions
\Eq{chain ics pg2}
$$\rY(0,0) = Y_0,\qquad \rX(0,0) = X_0,\qquad \rX_x(0) = \rX_x^0.$$
Again, analyticity of the surface $\tU(X,Y)$ and of the cross-section function \eq{ex9csps} implies analyticity of the right hand sides of the differential equations \eqref{chain ODEs pg2}, and thus analyticity of the normalizing transformation. To obtain the quadratic term in $y$ in  the normal form series \eq{ex9csps}, we need to impose the algebraic constraint $\rX_x^0 = \sqrt{\tU_{YY}(X_0,Y_0)}$ on the initial conditions.  We note that the ordinary differential equations \eqref{chain ODEs pg2} for the inverse pseudo-group transformation had originally been derived in \rf{OP08; Example 32} with $c(x)=d(x)=0$.

The preceding discussion focused on  one-dimensional chains (curves), where the constraint on the Cartan characters \eq{chain1} holds.  In the more general situation, when there are one or more nonzero higher order Cartan characters, the appropriate analog of chains will include submanifolds of dimension $\geq 2$.  For example, if the largest nonzero Cartan character is $\rcartan{n}{k}$, then a $k$-dimensional chain $\chain_k$ is a submanifold in $S$ that can be mapped to
\[
\bm{\mathcal{P}}_k = (\mathcal{P}_k,C(\mathcal{P}_k)),
\]
where $C(y)$ is the cross-section function \eqref{csSeries}, and such that the projection $\pi(\chain_k)\subset \mX$ is rectified to the particular $k$-dimensional coordinate subspace $\mathcal{P}_k = \{(x^1,\ldots,x^k,0,\ldots,0)\}$.  The pseudo-group transformation rectifying the chain will satisfy a system of partial differential equations for the parametric reduced pseudo-group jets that are given by
\[
\tU(\rX\at{\bm{\mathcal{P}}_k}) = \rU\at{\bm{\mathcal{P}}_k}.
\]
Inside the $k$-dimensional chain $\chain_k$ there may be  a sequence of lower dimensional chains $\chain_1 \subset \chain_2\subset \cdots \subset \chain_{k-1} \subset \chain_k$, with each projection $\pi(\chain_j)$ mapped to the $j$-dimensional coordinate subspace $\mathcal{P}_j=\{(x^1,\ldots,x^j,0,\ldots,0)\}$ under a suitable pseudo-group transformation.  The existence of these subchains will depend on the form of the initial values \eqref{Cartan nf initial cond}, which is ultimately determined by the Cartan characters, \rf{S; Proposition 8.2.10}.  We remark that such higher dimensional chains can be found in \rf{ES1996}, which  introduces two-dimensional chains when studying normal forms for elliptic CR submanifolds in $\mathbb C^4$.

%%%%%
\Section{xx} Additional Examples.
%%%%%

In this section we provide four more relatively simple examples illustrating the results of the paper.  We conclude by showing how the convergence theorem of Chern and Moser, \rf{CM74}, can be
deduced from our general theorem; this requires finding suitable coordinates that assure involutivity of the determining equations.  In these examples, we will omit the bar notation over $X$ and $Y$ and the hat notation on $U$ and its derivatives, which will unclutter the equations while hopefully not leading to any confusion now that the procedures and meanings are clear.

\Ex{12} In our running example, the pseudo-group considered only involved functions depending on one independent variable, namely $x$.  In this example we
consider the pseudo-group
\[
X=f(x),\qquad Y=g(y),\qquad U=u+c,
\]
where $f,g\in \D(\mathbb{R})$, $c\in \mathbb{R}$, and $x$, $y$ are assumed to be independent variables so that we consider the action on surfaces $u=u(x,y)$. Thus, the first order reduced determining equations are
\begin{equation}\label{rdet ex12}
X_y = Y_x=0,\qquad U_x = u_x,\qquad U_y = u_y,
\end{equation}
while the order two equations are
\Eq{ex12rdet2}
$$X_{xy} = X_{yy} = Y_{xx}=Y_{xy} = 0,\quad U_{xx}=u_{xx},\quad
U_{xy} = u_{xy},\quad U_{yy} = u_{yy}.$$
Using the ordering $x \prec y$, the indices for the order one equations \eqref{rdet ex12} are $\index{1}{1}=2$ and $\index{1}{2}=2$ so that
\[
\index{1}{1} + 2\:\index{1}{2} = 6\neq 7 = \rk_2.
\]
Alternatively, the Cartan characters are $\cartan{1}{1} = 1$, $\cartan{1}{2} = 1$ and
\[
\cartan{1}{1} + 2\:\cartan{1}{2} = 3 \neq 2 = \d_2.
\]
It follows that the equations \eqref{rdet ex12} are not involutive.  In fact,  the reduced determining equations are not involutive at any order $n$.  To see this,  we observe that the order $n$ determining equations for $Y$ are
\[
Y_{x^n} = Y_{x^{n-1}y} = \cdots = Y_{xy^{n-1}}=0,
\]
which are all of class one with respect to our chosen ordering.  From those equations it is not possible to obtain the equation $Y_{xy^n}=0$ at order $n+1$ since $y$ is not a multiplicative variable.  Changing the ordering to $y \prec x$ would not resolve the issue because the same problem would now appear among the determining equations for $X$.
The conclusion is that the current coordinates are not $\delta$-regular.
\pari
As emphasized in \sect{sec: involutivity}, we must therefore introduce new  coordinates that are $\delta$-regular.  This can be done, for example, by setting
\[
x=t+s\qquad\text{and}\qquad y=t-s.
\]
The pseudo-group then becomes
\[
T+S = f(t+s),\qquad T-S=g(t-s),\qquad U=u+c
\]
or
\[
T = \frac{f(t+s)+g(t-s)}{2},\qquad
S = \frac{f(t+s)-g(t-s)}{2},\qquad U=u+c.
\]
\pari
Relabeling the variables and functions, we now consider the Lie pseudo-group
\Eq{pg12}
$$X = f(x+y) + g(x-y),\qquad Y=f(x+y)-g(x-y),\qquad U=u+c.$$
The normal form determining equations can be obtained by recursively applying the total derivative operators
\Eq{12Dt}
$$\qeq{\Dt_x = X_x\:\Dt_X + Y_x\: \Dt_Y,\\\Dt_y = X_y\:\Dt_X + Y_y\: \Dt_Y,}$$
 to the pseudo-group transformations \eq{pg12} and eliminating the derivatives of the functions $f,g$ from the resulting equations.  At first order, this results in
\Eq{12e1}
$$\ceq{\qeq{X_x = f_t + g_t,\\ X_y = f_t - g_t,\\Y_x = f_t - g_t,\\ Y_y = f_t + g_t, }\\\req{X_x U_X + Y_x U_Y= u_x ,\\ X_y U_X + Y_y U_Y= u_y,}}$$
where $f_t,g_t$ represent the first order derivatives of $f,g$.  Provided $U_X^2-U_Y^2\neq 0$, we can eliminate $f_t,g_t$ to produce the first order normal form determining equations:
\begin{equation}\label{nf order1 eq ex12}
X_x = Y_y = \frac{u_x U_X - u_y U_Y}{U_X^2-U_Y^2},\qquad X_y = Y_x = \frac{u_y U_X - u_x U_Y}{U_X^2-U_Y^2},
\end{equation}
where we take $u_x,u_y$ to be the parametric derivatives.  This is consistent with the moving frame construction, but not with the theory of involutivity, which would require solving for $u_y$, assuming the ordering $x\prec y$.  In accordance with the discussion in \sect{imf}, this is a second example illustrating the discrepancy between the two theories at low order.
\pari
The normal form determining equations of order two can be obtained by differentiating \eqref{nf order1 eq ex12} using the total differential operators \eq{12Dt} --- or, alternatively applying \eq{12Dt} to \eq{12e1} and eliminating the first and second derivatives of $f,g$ --- which produces
\begin{equation}\label{nf det ex12 eq1}
X_{xy} = Y_{xx},\qquad
X_{yy} = X_{xx},\qquad
Y_{xy} = X_{xx},\qquad
Y_{yy} = Y_{xx},
\end{equation}
along with
\begin{equation}\label{nf det ex12 eq2}
\eeq{X_{xx} = \frac{\dsty\ibeq{97.5}{(u_{xx} - U_{XX} X_x^2 - 2\:U_{XY} X_x Y_x - U_{YY} Y_x^2)U_X \\{}- (u_{xy} - U_{XX}X_x Y_x - U_{XY}(X_x^2+Y_x^2) - U_{YY} X_x Y_x)U_Y}}{U_X^2-U_Y^2},\\
Y_{xx} = \frac{\dsty\ibeq{97.5}{(u_{xy}-U_{XX} X_x Y_x - U_{XY}(X_x^2+Y_x^2) - U_{YY}X_xY_x)U_X \\{}
-(u_{xx} - U_{XX} X_x^2 - 2\:U_{XY}X_xY_x - U_{YY} Y_x^2)U_Y}}{U_X^2-U_Y^2},}
\end{equation}
%\begin{equation}\label{nf det ex12 eq2}
%\eeq{X_{xx} = \frac{1}{U_X^2-U_Y^2}\big[(u_{xx} - U_{XX} X_x^2 - 2\:U_{XY} X_x Y_x - U_{YY} Y_x^2)U_X \cthn{120}
%- (u_{xy} - U_{XX}X_x Y_x - U_{XY}(X_x^2+Y_x^2) - U_{YY} X_x Y_x)U_Y\big],\\
%%
%Y_{xx} = \frac{1}{U_X^2-U_Y^2}\big[(u_{xy}-U_{XX} X_x Y_x - U_{XY}(X_x^2+Y_x^2) - U_{YY}X_xY_x)U_X \cthn{120}
%-(u_{xx} - U_{XX} X_x^2 - 2U_{XY}X_xY_x - U_{YY} Y_x^2)U_Y\big],
%}
%\end{equation}
%
and
\begin{equation}\label{uyy}
u_{yy} = u_{xx} - \frac{(u_x^2-u_y^2)\:(U_{XX} - U_{YY})}{U_X^2-U_Y^2}.
\end{equation}
Note that to place \eqref{nf det ex12 eq2} in the proper reduced Cartan normal form, we should replace $X_x, X_y, Y_x,Y_y$ by their formulas from \eqref{nf order1 eq ex12}, although the resulting expressions are a bit unwieldy.  We also note that the second order parametric derivatives are $u_{xx}, u_{xy}$.
\pari
We can easily verify that the order two normal form determining equations are involutive.  Indeed, the indices and Cartan characters\fnote{As above, we only need to compute one of these sets to verify involutivity.} are
\[
\index{2}{1} = 4,\quad \index{2}{2} = 3,\qquad
\cartan{2}{1} =2,\quad \cartan{2}{2} =0,
\]
and they satisfy the algebraic involutivity tests
\[
\index{2}{1} + 2\:\index{2}{2} = 10 = \rk_3 \qquad\text{or}\qquad
\cartan{2}{1} + 2\:\cartan{2}{2} = 2 = \d_3.
\]
Since there are no integrability conditions, the equations are involutive.
\pari
On the space 
$$V\ii = \{U_X^2\neq U_Y^2\} \subset \Ji$$
 of regular jets, the prolonged action becomes free at order one\footnote{Every Lie pseudo-group is trivially free at order $n=0$.  Freeness is only of interest when $n\geq 1$, \rf{OP08}.}, and a cross-section is given by
\begin{equation}\label{K pg3}
\mK = \{x=y=0,\ \ u_{x^k} = c_k,\ \ u_{yx^k} = d_k\;|\; k\geq 0,\ \ c_1^2 - d_0^2\neq 0 \}.
\end{equation}
The corresponding cross-section function is
\begin{equation}\label{cs fct ex12}
C(x,y) = c(x) + y\, d(x)\qquad \text{with}\qquad c_x^2(0)-d^2(0) \neq 0,
\end{equation}
and the normal form is
\begin{equation}\label{nf ex12}
u(x,y) = c(x) + y\, d(x) + y^2\: w(x,y).
\end{equation}
In the simplest case, we can take $c(x) = x$ and $d(x)=0$.  Observe that the initial conditions \eqref{cs fct ex12} depend on functions of the same variable $x$, which would not be the case if we were to write the system in the original $\delta $ irregular coordinates.
\pari
According to the general theory, since the action becomes free at order one, the involutivity of the order two normal form determining equations \eqref{nf det ex12 eq1}, \eqref{nf det ex12 eq2}, \eqref{uyy} guarantees the convergence of the normal form \eqref{nf ex12} provided well-posed analytic initial conditions are provided and the target function $U(X,Y)$ is analytic.  On the other hand, the  equations \eqref{nf order1 eq ex12} provide algebraic constraints among the order one jets at the origin.  The desired initial conditions are given by
$$\req{\ceq{X(0,0) = X_0,\qquad Y(0,0) = Y_0,\qquad u(0,0)=C(0,0) = c_0,\\ u_x(x,0) = C(x,0) = c_x(x),\qquad
u_y(x,0) = C_y(x,0) = d(x),}\\c_x^2(0)-d^2(0) \neq 0,}$$
where $c(x)$, $d(x)$ are analytic functions.  This shows that \eqref{K pg3} is a well-posed cross-section satisfying the hypotheses of \th{well-posedK}.  Indeed the cross-section is of minimal order with the set of defining indices of order $>1$ admitting the Rees decomposition
\begin{align*}
\indK^{>1} &= \set{(k+1,0),\; (k,1)}{k\geq 1} \\
&= \set{(k+1,0)}{k\geq 1}\; \biguplus\; \set{(k,1)}{k\geq 1} = \cone(2,0)\;\biguplus\; \cone(1,1).
\end{align*}

\Ex{13}
In the examples considered thus far, the Lie pseudo-group actions were all, in the chosen system of coordinates, quasi-horizontal as defined in \rf{Arnaldsson2}. This property is not necessary for the results of this paper to be valid, and we illustrate this fact by considering the Lie pseudo-group
\begin{equation}\label{pg13}
X = x+a,\qquad Y = y+b,\qquad U=f(u),
\end{equation}
where $a,b \in \mathbb{R}$ and $f\in \D(\mathbb{R})$. Of course, the pseudo-group \eqref{pg13} can be transformed into a quasi-horizontal action via the hodograph transformation $(x,y,u) \to (u,y,x)$, but we will not make this transformation here.
\pari
Provided $U_X\neq 0$, the normal form determining equations of order $1$ are
\begin{equation}\label{order1 ex13}
X_x = Y_y = 1,\qquad X_y = Y_x=0,\qquad u_y = \frac{u_x U_Y}{U_X},
\end{equation}
while at order $2$ we have
\begin{equation}\label{order2 ex13}
\ceq{X_{xx} = X_{xy} = X_{yy} = Y_{xx}=Y_{xy}=Y_{yy} = 0,\\
u_{xy} = \frac{u_x U_{XY}+u_{xx} U_Y}{U_X} - \frac{u_x U_Y U_{XX}}{U_X^2},\qquad
u_{yy} = \frac{u_x U_{YY}}{U_X} + \frac{u_{xx} U_Y^2}{U_X^2} - \frac{u_x U_Y^2U_{XX}}{U_X^3}.}
\end{equation}
The indices and Cartan characters for the order one determining equations \eqref{order1 ex13} are
\[
\index{1}{1} = 2,\quad \index{1}{2} = 3,\qquad \cartan{1}{1} = 1,\quad \cartan{1}{2} = 0
\]
so that the involutivity condition
\[
\index{1}{1} + 2\:\index{1}{2} = 8 = \rk_2,\roq{or, equivalently,}
\cartan{1}{1} + 2\:\cartan{1}{2} = 1 = \d_2
\]
is satisfied. Since there are no integrability conditions, the order one determining equations \eqref{order1 ex13} are involutive.
\pari
The pseudo-group action becomes free at order one.  A well-posed cross-section is given by
\[
\K = \set{x=y=0,\; u_{x^k} = c_k}{k\geq 0\text{ and } c_1\neq 0},
\]
with the set of defining indices of order $>1$ admitting the Rees decomposition
\[
\indK^{>1} = \set{(k,0)}{k\geq 2} = \cone(2,0).
\]
The corresponding cross-section function is
\[
C(x) = c(x)\qquad\text{with}\qquad c_x(0)\neq 0,
\]
and the normal form is given by
\begin{equation}\label{nf ex13}
u(x,y) = c(x) + y\, w(x,y).
\end{equation}
In the simplest case, we could have taken $c(x)=x$.  
\pari
The action being free at order one, the general theory dictates that, assuming analyticity of the function $U(X,Y)$, analyticity of the normal form \eqref{nf ex13} will follow from the involutivity of the order two normal form determining equations \eqref{order2 ex13} along with the equations \eqref{order1 ex13} providing algebraic constraints among the first order jets at the origin.  Formally well-posed initial conditions are given by
\[
X(0,0) = X_0,\quad Y(0,0) = Y_0,\quad u(0,0) =  c(0),\qquad u_x(x,0) = c_x(x),
\]
with $c_x(0) \neq 0$.

\Ex{14}
In the previous examples, the Lie pseudo-group actions considered only involved local diffeomorphisms of the real line.  We now examine the pseudo-group
\begin{equation}\label{pg14}
X = x+a,\qquad Y = g(x,y),\qquad Z=z+b,\qquad U = u,
\end{equation}
where $g(x,y)$ is an analytic function that depends on two variables with $g_y(x,y)\neq 0$, while $a,b\in \mathbb{R}$.  In this example, we assume that $u=u(x,y,z)$ is a function of three variables, \ie we consider the action on three-dimensional submanifolds.
\pari
We can obtain the normal form determining equations by applying the total derivative operators
\Eq{14Dt}
$$\qeq{\Dt_x = \Dt_X + Y_x\: \Dt_Y,\\\Dt_y = Y_y\: \Dt_Y,\\\Dt_z = \Dt_Z,}$$
 to \eqref{pg14}.  Assuming $U_Y\neq 0$, we can rewrite these equations in the form
\Eq{14rnf1}
$$\xeq{X_x = 1,\\ X_y = X_z=0,\\Y_x = \frac{u_x - U_X}{U_Y},\\ Y_y = \frac{u_y}{U_Y}, \\Y_z = Z_x=Z_y=0,\\ Z_z=1,\\u_z = U_Z,}$$
where the parametric derivatives are $u_x,u_y$.  We note that this is compatible with both the theory of moving frames and involutivity.  The second order normal form determining equations can be obtained by applying the differential operators \eq{14Dt} to the first order equations \eq{14rnf1} giving
\Eq{nf eq order2 pg14}
$$\eeq{X_{xx} = X_{xy} = X_{yy} = X_{xz} = X_{yz} = X_{zz} = 0,\qquad
Y_{xz} = Y_{yz} = Y_{zz} = 0,\\
Z_{xx} = Z_{xy} = Z_{yy} = Z_{xz} = Z_{yz} = Z_{zz} = 0,\\
Y_{xx} = \frac{u_{xx}-U_{XX}}{U_Y} - \frac{2\: U_{XY}(u_x - U_X)}{U_Y^2} - \frac{U_{YY}(u_x - U_X)^2}{U_Y^3}, \\
Y_{xy} = \frac{u_{xy}}{U_Y} - \frac{U_{XY}u_y}{U_Y^2}-\frac{U_{YY}u_y(u_x-U_X)}{U_Y^3},\qquad
Y_{yy} = \frac{u_{yy}}{U_Y} - \frac{U_{YY}u_y^2}{U_Y^3},\\
u_{xz} = U_{XZ} + \frac{U_{YZ}(u_x - U_X)}{U_Y},\qquad u_{yz} = \frac{U_{YZ}u_y}{U_Y},\qquad u_{zz} = U_{ZZ},
}$$
with parametric derivatives are $u_{xx}, u_{xy}, u_{yy}$, and similarly for the higher order versions.
The indices and Cartan characters for the order one normal form determining equations \eq{14rnf1} are
\[
\index{1}{1} = 3,\quad \index{1}{2} = 3,\quad \index{1}{3} = 4,\qquad
\cartan{1}{1} = 1,\quad \cartan{1}{2} = 1,\quad \cartan{1}{3}=0,
\]
which satisfy the involutivity condition
\[
\index{1}{1} + 2\:\index{1}{2} + 3\:\index{1}{3} = 21 = \rk_2\:\qquad \text{or, equivalently,}\qquad
\cartan{1}{1} + 2\:\cartan{1}{2} + 3\:\cartan{1}{3} = 3 = \d_2.
\]
Since there are no integrability constraints, the order one normal form determining equations are involutive.
\pari
The pseudo-group action becomes free at order one and a well-posed cross-section is given by
\[
\K=\set{x=y=0,\; u_{x^{k+1}} = c_k,\; u_{x^jy^{k+1}} = d_{j,k}}{j,k\geq 0 \text{ and } d_{0,0} \neq 0}
\]
with the defining indices of order $>1$ admitting the Rees decomposition
\begin{align*}
\indK^{>1} &= \set{(i+2,0),\;(j,k+1)}{i\geq 0,\; j+k\geq 1} \\
&= \set{(i+2,0)}{i\geq 0}\;\biguplus\; \set{(j+1,1)}{j\geq 0} \;\biguplus\;
\set{(j,k+2)}{j,k\geq 0} \\
&= \cone(2,0)\;\biguplus\; \cone(1,1)\;\biguplus\; \cone(0,2).
\end{align*}
The corresponding cross-section function $C(x,y)$ satisfies the constraints
\[
C(0,0) = 0,\qquad C_x(x,0)=c(x),\qquad C_y(x,y) = d(x,y)\quad \text{with}\quad C_y(0,0)=d(0,0)\neq 0.
\]
In the simplest case, we could let $C(x,y) = y$.
In general, the normal form is given by
\begin{equation}\label{nf pg14}
u(x,y,z) = U_0 + C(x,y) + z\, w(x,y,z),
\end{equation}
where $U_0=U(X_0,Y_0,Z_0)$ is a constant, fixed by the prescribed submanifold.  
\pari
Since the prolonged action becomes free at order one, the convergence of the normal form \eqref{nf pg14} follows from the involutivity of the order two normal form determining equations \eq{nf eq order2 pg14}, with the equations \eq{14rnf1} providing algebraic constraints on the order one jets at the origin.  Since the pseudo-group action \eqref{pg14} is intransitive, we also have the order zero normal form determining equation $u = U$, which  needs to be evaluated at the origin.  Well-posed initial conditions are given by
\begin{gather*}
X(0,0,0) = X_0,\qquad Y(0,0,0) = Y_0,\qquad Z(0,0,0) = Z_0,\\
u(0,0,0) = U_0,\qquad u_x(x,0,0) = C_x(x,0) = c(x),\qquad  u_y(x,y,0) = C_y(x,y) = d(x,y).
\end{gather*}

\Ex{15}
As our next example, we consider the Lie pseudo-group
\Eq{15}
$$X = x+a,\qquad Y=y+b,\qquad Z=z+f(x,y),\qquad U=u+g(x,y),$$
where $f,g$ satisfy the Cauchy--Riemann equations
\Eq{CR}
$$f_x = g_y,\qquad f_y = -\:g_x.$$
As in \ex{14}, we obtain the normal form determining equations by recursively applying the total differential operators
\Eq{15Dt}
$$\qeq{\Dt_x = \Dt_X + Z_x\: \Dt_Z,\\\Dt_y = \Dt_Y+ Z_y\: \Dt_Z,\\\Dt_z = \Dt_Z,}$$
to the pseudo-group transformations \eq{15}.  At first order, we have
\Eq{15rnf1a}
$$\qeq{X_x=1,\\ X_y = X_z = 0,\\ Y_y=1,\\ Y_x=Y_z=0,\\ Z_z = 1,}$$
along with
\Eq{15rnf1b}
$$\weq{Z_x=f_x,\\ Z_y = f_y,\\ U_X + Z_x U_Z= u_x + g_x,\\  U_Y + Z_y U_Z= u_y + g_y,\\ U_Z = u_z.}$$
Eliminating the derivatives of $f,g$ from the latter equations using \eq{CR} produces
\Eq{15rnf1c}
$$\qeq{Z_x = \frac{U_Z(u_x-U_X)-(u_y-U_Y) }{1+U_Z^2},\\
Z_y = \frac{U_Z(u_y-U_Y)+u_x - U_X}{1+U_Z^2},\\ u_z = U_Z,}$$
where the parametric derivatives are $u_x,u_y$. As in \ex{12}, this is compatible with the moving frame construction but not with involutivity, which would require solving for $u_y$ in the first equation of \eq{15rnf1c}, assuming the ordering $x\prec y \prec z$.  The second order normal form determining equations can be obtained by using \eq{15Dt} to differentiate \eqs{15rnf1a}{15rnf1c}.  We find
\Eq{15rnf2}
$$\eeq{X_{xx} = X_{xy} = X_{yy} = X_{xz} = X_{yz} = X_{zz} = 0,\\
Y_{xx} = Y_{xy} = Y_{yy} = Y_{xz} = Y_{yz} = Y_{zz} = 0,\qquad
Z_{xz} = Z_{yz} = Z_{zz} = 0,\\
Z_{xx} = -\:Z_{yy} = [\:-u_{xy}+U_{XY} + (u_{xx}-U_{XX})\:U_Z + (U_{YZ} - 2U_{XZ}U_Z)Z_x +U_{XZ}Z_y\cthn{210}-U_ZU_{ZZ}Z_x^2 + U_{ZZ}Z_xZ_y\:]/\pa{1+U_Z^2},\\
Z_{xy} = [\:u_{xx}-U_{XX} + (u_{xy}-U_{XY})\:U_Z - (2\:U_{XZ}+U_{YZ}U_Z)Z_x - U_{XZ}U_Z Z_y \cthn{210} -U_{ZZ}Z_x^2 - U_Z U_{ZZ}Z_xZ_y\:]/\pa{1+U_Z^2},\\
u_{yy} = -\:u_{xx} + U_{XX} + U_{YY} + U_{ZZ} (Z_x^2+Z_y^2) + 2 \:U_{XZ} Z_x + 2\:U_{YZ} Z_y,\\
u_{xz} = U_{XZ} + U_{ZZ}Z_x,\qquad u_{yz}=U_{YZ}+U_{ZZ}Z_y,\qquad u_{zz} = U_{ZZ}.
}$$
To write the equations in reduced Cartan normal form, one should replace $Z_x,Z_y$ by their expressions in \eq{15rnf1c} to express the right hand sides in terms of only the parametric derivatives $u_x,u_y,u_{xx},u_{xy}$; however, the resulting formulas are too unwieldy to display.
\pari
The indices and Cartan characters for the order two normal form determining equations \eq{15rnf2} are
\[
\index{2}{1} = 10,\quad \index{2}{2} = 8,\quad \index{2}{3} = 4,\qquad
\cartan{2}{1}=2,\quad \cartan{2}{2} = 0, \quad \cartan{2}{3} = 0.
\]
Omitting the computational details, the only third order parametric derivatives are $u_{xxx} $ and $u_{xxy}$, and hence 
\[
\index{2}{1} + 2\:\index{2}{2} + 3\:\index{2}{3} = 38 = \rk_3\qquad\text{or, equivalently,}\qquad
\cartan{2}{1} + 2\:\cartan{2}{2} + 3\:\cartan{2}{3} = 2 = \d_3.
\]
Since there are no integrability conditions, the order two normal form equations are involutive.
\pari
The prolonged pseudo-group  action becomes free at order one and a well-posed cross-section is given by
\[
\K = \set{x=y=z=0,\ \ u_{x^k} = c_k,\ \ u_{x^k y} = d_k}{k\geq 0}
\]
so that the set of defining indices of order $>1$ admits the Rees decomposition
\[
\indK^{>1} = \cone(2,0,0) \;\biguplus\; \cone(1,1,0).
\]
The cross-section function is given by
\[
C(x,y) = c(x) + y\: d(x)
\]
and the normal form is
\[
u(x,y,z) = c(x) + y\: d(x) + y^2\, v(x) + z\, w(x,y,z).
\]
The convergence of the normal form follows from the involutivity of the order two normal form determining equations \eq{15rnf2}, combined with the algebraic constraints obtained by evaluating the order one equations \eq{15rnf1a}, \eq{15rnf1c} at the origin.  Formally well-posed initial conditions are given by
\begin{gather*}
X(0) = X_0,\qquad Y(0)=Y_0,\qquad Z(0)=Z_0,\qquad u(0,0,0)=C(0,0)=c_0,\\
u_x(x,0,0) = C_x(x,0) = c_x(x),\qquad u_y(x,0,0) = C_y(x,0) = d(x).
\end{gather*}

\Ex{CM-exm}
In \cite{OSV} we revisited the Chern--Moser normal form problem, \cite{CM74}, for nondegenerate real hypersurfaces in $\mathbb C^2$ under the action of the pseudo-group of holomorphic transformations, obtaining five inequivalent classes of normal forms termed locally umbilic, non-umbilic, generic, circular, and semi-circular.  The convergence of these normal forms relied on results from \cite{CM74}.  We now use Theorems \ref{well-posedK} and \ref{main} to give an alternative argument.
\pari
Let $z = x + \i y$, $w=u+\i v$ be local coordinates on $\mathbb C^2$.  Accordingly, the pseudo-group of holomorphic transformations $(z,w) \mapsto (Z(z,w), W(z,w))$ of $\mathbb{C}^2$, with $W=U+\i V$, is determined by the differential equations
\begin{equation}\label{PDE-real}
Z_{\oz}=0, \qquad Z_v=\i Z_u, \qquad V_{\oz}=\i U_{\oz}, \qquad V_u=-U_v, \qquad V_v=U_u.
\end{equation}
We consider real hypersurfaces $S \subset \mathbb{C}^2$ that are locally parametrized as the graph of a real-valued function
\begin{equation}\label{v}
v = v(z,\oz,u).
\end{equation}
A partial cross-section to the prolonged action was found in \cite[eq.\ (3.14)]{OSV} and is given by
\begin{multline}\label{completeK}
\{v_{z\oz}=1,\, z = \oz = u = v = v_{z^ku^\ell} = v_{\oz^ku^\ell} = v_{z\oz u^{\ell+1}} = v_{z^{k+2}\oz u^\ell} \\
= v_{z\oz^{k+2}u^\ell} = v_{z^2\oz^2 u^\ell} = v_{z^3 \oz^2 u^\ell} = v_{z^2\oz^3u^\ell} = v_{z^3\oz^3 u^\ell} = 0\,|\, k,\ell \geq 0\}.
\end{multline}
Depending on the class of the normal form, only a finite number of normalizations must be added to \eqref{completeK} to obtain a complete cross-section.  These normalizations do not affect the convergence argument, and we therefore work with the partial cross-section \eqref{completeK}.  
\pari
The normal form for locally umbilic hypersurfaces is given by the Heisenberg sphere $v=z\:\oz$, which is obviously analytic.   We thus focus on the remaining four classes of normal forms.  Since the equations
\[
v_{\oz^k u^\ell}=v_{z\oz^{k+2} u^\ell}=v_{z^2 \oz^3 u^\ell}=0
\]
can be obtained by conjugating $v_{z^k u^\ell}=v_{z^{k+2}\oz u^\ell}=v_{z^3 \oz^2 u^\ell}=0$, they can be omitted from \eqref{completeK}.  No information is lost as, for example, the pseudo-group normalization originating from the normalization $v_{\oz^k u^\ell}=0$ is recovered by taking the conjugate of the pseudo-group normalization obtained by solving $v_{z^k u^\ell} = 0$.  Said differently, the normalization of a Taylor coefficient of the real-valued function \eqref{v} induces a normalization of its conjugated Taylor coefficient.  We thus focus on the reduced partial cross-section
\begin{multline}
\K = \big\{\, v_{z\oz}=1,\; z= u = v_{z^k u^\ell} = v_{z\oz u^{\ell+1}} = v_{z^{k+2}\oz u^\ell}\\
= v_{z^2\oz^2 u^\ell} = v_{z^3 \oz^2 u^\ell}  = v_{z^3\oz^3 u^\ell}=0 \;|\; k,\ell \geq 0 \,\big\}.
\end{multline}
\pari
Since the given coordinates turn out to be $\delta $ irregular, as in \ex{12}, we need to make a change of variables in order for the pseudo-group determining equations \eqref{PDE-real} to become involutive.  Reverting back to complex variables, let
\begin{equation}\label{uv}
u = \frac{w+\ow}{2},\qquad v = \frac{w-\ow}{2\i}.
\end{equation}
The determining equations of the pseudo-group then become
\begin{equation}\label{PDE-complex}
Z_{\oz}=Z_{\ow}=W_{\oz}=W_{\ow}=0.
\end{equation}
Introducing the  ordering $w\prec z\prec \oz\prec \ow$, the indices and Cartan characters of \eqref{PDE-complex} are
\[
\index{1}{1}=\index{1}{2}=0, \quad \index{1}{3}=\index{1}{4}=2\qquad\text{and}\qquad
\cartan{1}{1} = \cartan{1}{2} = 2,\quad \cartan{1}{3}=\cartan{1}{4}=0.
\]
Since the second order determining equations are
\[
\aligned
Z_{z\oz}&=Z_{w\oz}=Z_{\oz\oz}=Z_{\oz\ow}=Z_{z\ow}=Z_{w\ow}=Z_{\ow\ow}=0,
\\
W_{z\oz}&=W_{w\oz}=W_{\oz\oz}=W_{\oz\ow}=Z_{z\ow}=W_{w\ow}=W_{\ow\ow}=0,
\endaligned
\]
the involutivity test $\index{1}{1}+2 \index{1}{2}+3 \index{1}{3}+ 4 \index{1}{4} = 14 = \rk_2$ is satisfied, and there are clearly no integrability conditions.  Therefore, the determining equations \eqref{PDE-complex} are involutive.
\pari
Substituting the change of variables \eqref{uv} into the hypersurface defining equation \eqref{v} and solving for $\ow$ using the Implicit Function Theorem, we obtain the complex defining equation\footnote{
Kossovskiy and Zaitsev also used the complex defining equation \eqref{def-complex} in their convergence argument; see the acknowledgments in their paper \cite{KZ2015}.}
\begin{equation}\label{def-complex}
\ow=\ow(z,\oz,w)
\end{equation}
of the hypersurface $S$.  Thus, in the new coordinates, the jet variables are $\ow_{z^j\oz{}^k w^\ell}$ with $j,k,\ell\geq 0$.  To find the cross-section in these new jet variables, we substitute the real and complex defining equations \eqref{v}, \eqref{def-complex} into the second equation of \eqref{uv} to obtain the relationship
\begin{equation}\label{def-re-com}
\ow(z,\oz,w) = w -2\i v\bigg(z,\oz,\frac{w+\ow(z,\oz,w)}{2}\bigg).
\end{equation}
Implicitly differentiating \eqref{def-re-com} produces the expressions for the new jet coordinates $\ow_J$ in terms of the original ones $v_K$. For example, at order one, we have
\[
\aligned
\ow_z=-2\i v_z-\i v_u\ow_z, \qquad \ow_{\oz}=-2\i v_{\oz}-\i v_u \ow_{\oz}, \qquad \ow_w=1-\i v_u (1+\ow_w).
\endaligned
\]
These equations can be solved for $\ow_z, \ow_{\oz}, \ow_w$ and the result will depend on whether $v_u$ is zero or not. For orders $\geq 2$, one finds, using induction, that
\begin{equation}\label{expression}
\ow_{z^j \oz{}^k w^\ell}= -2\i v_{z^j \oz{}^k u^\ell} + \mathcal S_{j,k,\ell}(\ow_J, v_{K}),
\end{equation}
where $\mathcal S_{j,k,\ell}$ is a polynomial involving $\ow_J$, with $|J| \leq j+k+\ell$, and $v_K = v_{z^\alpha \oz{}^\beta u^\gamma}$, with $\alpha \leq j$, $\beta \leq k$, $\gamma \geq 1$ and $|K| \leq j+k+\ell$.  Moreover,
\[
\mathcal S_{j,k,\ell}(\ow_J, 0)=0.
\]
Using \eqref{expression} and induction, the partial cross-section in the new complex jet coordinates is
\begin{multline*}
\widetilde{\K} = \big\{\, \ow_w=1,\;\ow_{z\oz}=-2\i, \; z= w = \ow_{z^k w^\ell} = \ow_{z\oz w^{\ell+1}} = \ow_{z^{k+2}\oz w^\ell}\\
=\ow_{z^2\oz^2 w^\ell} = \ow_{z^3 \oz^2 w^\ell}  = \ow_{z^3\oz^3 w^\ell}=0 \;|\; k,\ell \geq 0 \,\big\}.
\end{multline*}
\pari
As shown in \cite[Section 4]{OSV}, the prolonged action of the holomorphic pseudo-group becomes free at some order $n_0\geq 7$ for generic, non-umbilic, and semi-circular hypersurfaces.  Circular hypersurfaces retain a one-dimensional isotropy group, but the convergence argument remains valid at some order $n_0 \geq 8$. In this case, the isotropy pseudo-group parameter is added to the order zero jet $\rX$ on the right hand side of the normal form determining equations \eqref{Cartan nf det eq1}, \eqref{Cartan nf det eq2}, and its value at the origin is an extra initial condition included on the last line of \eqref{alg eq}. In all cases, it is possible to construct a minimal cross-section and at the appropriate order $n_0$, one observes that $\mI_{\widetilde{\K}}^{> n_0}$ admits the following Rees decomposition with respect to the ordering $w\prec z\prec\oz$
\[
\biguplus_{j=0}^{n_0+1} \cone(n_0+1-j, j, 0)\biguplus_{j=1}^{n_0}  \cone(n_0-j, j, 1) \biguplus \cone(n_0-3,2,2) \biguplus \cone(n_0-4,3,2) \biguplus \cone(n_0-5,3,3).
\]
By \th{well-posedK}, the cross-section is well-posed and thus \th{main} implies that the normal form of a nondegenerate analytic hypersurface converges, thus reproducing Chern and Moser's celebrated convergence result.

%%%%%
\Section{z} Final Comments.
%%%%%

In this paper, we have proven a fundamental result establishing the convergence of normal form power series for suitably regular submanifolds under a large class of Lie pseudo-group actions, which includes, in particular, all those for which the equivariant moving frame methods developed in \rf{OP08,OP09} can be applied.  To do so, we introduced the normal form determining equations \eqref{nf det eq}, whose solution includes the normal form, and proved their involutivity when the submanifold is reducible.  In \sect{imf}, we showed that, beyond the order of freeness, the involutivity of the normal form determining equations is compatible with the moving frame construction, and that a well-posed cross-section provides suitable analytic initial conditions.  The convergence of the normal form is then guaranteed by an application of the Cartan--K\"ahler Theorem.

The results of the paper have been obtained under the assumption that the prolonged pseudo-group action eventually acts freely, which is a necessary requirement for the construction of a moving frame.  That said, there are many circumstances where the prolonged pseudo-group action never becomes free, in which case the geometric problem admits a non-trivial isotropy groups.  In these situations one can construct a partial moving frame, \rf{O2011,V2013}. As indicated in \ex{CM-exm}, if the isotropy group is finite-dimensional, then the constructions and results of the paper can be adapted to encompass this setting.  In this case, the isotropy pseudo-group jet coordinates are parameters that can be added to the right hand side of the normal form determining equations \eqref{Cartan nf det eq1}, \eqref{Cartan nf det eq2}.  At a sufficiently high order, the isotropy pseudo-group jet coordinates do not influence the involutivity of the normal form determining equations.  These parametric variables can take any value and those are appended to the initial conditions \eqref{alg eq}.  Extending the result of the paper to Lie pseudo-groups that do not eventually act freely, and to singular submanifolds that admit infinite-dimensional isotropy pseudo-groups will be the subject of future research.

We anticipate that our general convergence result will find a wide range of applications in the construction of normal forms.  This include, for example, the investigation of Bishop surfaces in CR geometry, \rf{XW2010}, the construction of Poincar\'e--Dulac normal forms, \rf{G2002,KK2004}, as well as normal forms in control theory, dynamical systems, partial differential equations, and so on.

%%%%%
\subsection*{Acknowledgments}
%%%%%

The first author would like to thank Niky Kamran for his prescient suggestion, made a long time ago, to apply the equivariant method of moving frames for Lie pseudo-groups to the Chern--Moser example.  The research of the second author was supported, in part, by the Iran National Science Foundation (INSF), under the project No.\ 4031893, and the Institute for Research in Fundamental Science (IPM), grant No.\ 1403320417.  The third author would like to thank Jo\"el Merker for his discussion of Chern--Moser chains during his visit at l'Universit\'e Paris--Saclay.

\newpage

%%%%%

\bigskip
\noindent
{}\textsuperscript{$\dagger$}\footnotesize{\textsc{School of Mathematics, University of Minnesota, Minneapolis, MN \  55455}
\\
{\it Email address}}: {\tt olver@umn\hspace{-0.075em}.\!edu}

\medskip
\noindent
{}\textsuperscript{$\ddagger$}\footnotesize{\textsc{Department of Mathematics, Shahrekord University, Shahrekord, Iran \ 88186 and School of Mathematics, Institute for Research in Fundamental Sciences (IPM), Tehran, Iran \ 19395-5746}
\\
{\it Email address}}: {\tt sabzevari@ipm\hspace{-0.075em}.\!ir}

\medskip
\noindent
{}\textsuperscript{$\star$}\footnotesize{\textsc{Department of Mathematics, Monmouth University, West Long Branch, NJ \ 07764}
\\
{\it Email address}}: {\tt fvalique@monmouth\hspace{-0.075em}.edu}

\bigskip
\noindent \today

\end{document}